\documentclass[12pt,fleqn]{ucithesis}

\usepackage{amsmath}
\usepackage{amsthm}
\usepackage{array}
\usepackage[pdftex]{graphicx}
\usepackage{relsize}
\usepackage[titletoc]{appendix}
\usepackage{xcolor}
\usepackage[utf8]{inputenc}
\usepackage[utf8]{inputenc}

\usepackage{caption}
\usepackage{subcaption}  
\usepackage{multirow}
\usepackage{tabularx}

\usepackage[plainpages=false,pdfborder={0 0 0}]{hyperref}

\thesistitle{Efficient ICBased Solutions for Medical Devices and 
Automotive Radars}

\documenttitle{Dissertation}

\degreename{Doctor of Philosophy}

\degreefield{Electrical and Computer Engineering}

\authorname{Behnam Moradi Shahrbabak}

\committeechair{Professor Michael Green}
\othercommitteemembers
{
  Professor Ozdal Boyraz\\
  Professor Hamidreza Aghasi
}

\degreeyear{2024}

\copyrightdeclaration
{
  {\copyright} {\Degreeyear} \Authorname
}

\dedications
{

Dedicated to my family and my wife, whose love, support, and encouragement have been the foundation of my growth and success.
}

\acknowledgments
{
  First and foremost, I would like to extend my deepest gratitude to my supervisors, Professor Michael Green and Professor Hamidreza Aghasi. Their guidance, support, and invaluable insights over the past years have profoundly shaped my academic journey and prepared me to tackle the challenges I faced. Their impact on my personal and professional development is immeasurable, and I will forever be grateful for your mentorship and encouragement.

I would also like to express my sincere thanks to all those whose support was crucial in completing my project and design. I am deeply grateful to Professor Hamid Djalilian, Professor FanGang Zeng, and Rabih Nassif for their invaluable assistance, insights, and encouragement throughout the course of this project.

Thirdly, I thank my PhD committee members, Professor Ozdal Boyraz and Professor Maxim Shcherbakov. Your thoughtful comments and insightful support were instrumental in helping me navigate and overcome numerous challenges throughout the project. Your guidance has been invaluable.

Lastly, I want to take a moment to thank all the colleagues who have supported me throughout this journey. I am particularly grateful to Alireza Nikzamir, whose help on the project and friendship have been invaluable since we first met. My thanks also go to Dr. Yaoyu Cao and Dr. Hang Zhao for providing crucial guidance when I started the project a few years ago. I would also like to acknowledge Xuyang Liu, with whom I spent considerable time tackling RFIC design challenges and from whom I learned a great deal.

}

\newcommand{\mypubentry}[3]{
  \begin{tabular*}{1\textwidth}{@{\extracolsep{\fill}}p{4.5in}r}
    \textbf{#1} & \textbf{#2} \\ 
    \multicolumn{2}{@{\extracolsep{\fill}}p{.95\textwidth}}{#3}\vspace{6pt} \\
  \end{tabular*}
}

\curriculumvitae
{

\textbf{EDUCATION}
  
  \begin{tabular*}{1\textwidth}{@{\extracolsep{\fill}}lr}
    \textbf{Doctor of Philosophy in Electrical and Computer Engineering} & \textbf{2024} \\
    \vspace{6pt}
    University of California, Irvine & \emph{Irvine, CA} \\
    \textbf{Master of Science in Electrical Engineering} & \textbf{2021} \\
    \vspace{6pt}
    University of California, Irvine & \emph{Irvine, CA} \\
    \textbf{Bachelor of Science in Electrical Engineering} & \textbf{2018} \\
    \vspace{6pt}
    Sharif University of technology & \emph{Tehran, Tehran} \\
  \end{tabular*}

\vspace{12pt}
\textbf{RESEARCH EXPERIENCE}

  \begin{tabular*}{1\textwidth}{@{\extracolsep{\fill}}lr}
    \textbf{Graduate Research Assistant} & \textbf{20192024} \\
    \vspace{6pt}
    University of California, Irvine & \emph{Irvine, California} \\
  \end{tabular*}

\vspace{12pt}
\textbf{TEACHING EXPERIENCE}

  \begin{tabular*}{1\textwidth}{@{\extracolsep{\fill}}lr}
    \textbf{Teaching Assistant} & \textbf{2020} \\
    \vspace{6pt}
    University of California, Irvine & \emph{Irvine, California} \\
  \end{tabular*}

\pagebreak

\textbf{REFEREED JOURNAL PUBLICATIONS}

  \mypubentry{A Single Switch 3.1 4.7 GHz 194.52dB FoM Class D VCO With 495 W Power Consumption}{May 2024}{IEEE Transactions on Circuits and Systems II: Express Briefs}

  \mypubentry{A 76–82 GHz VCO in 65 nm CMOS With 189.3 dBc/Hz PN FOM and -0.6 dBm Harmonic Power for mmWave FMCW Applications}{October 2023}{IEEE Transactions on Circuits and Systems II: Express Briefs}

\vspace{12pt}
\textbf{REFEREED CONFERENCE PUBLICATIONS}

  \mypubentry{A Digitally Controlled Integrated Circuit Solution for Tinnitus Treatment with Charge Balancing}{July 2023}{Engineering in Medicine \& Biology Society}

 \mypubentry{ A Compact CMOS 76–82 GHz Super Harmonic VCO with 189 dBc/Hz FoM Operating based on Harmonic Assisted ISF Manipulation}{June 2022}{ IEEE Radio Frequency Integrated Circuits Symposium (RFIC)}

\vspace{12pt}

}

\thesisabstract
{
This thesis focuses on developing integrated circuit (IC) solutions for medical devices and automotive radars, and is divided into two main parts.

Part One presents the design and evaluation of a miniaturized multi chip module (MCM) solution intended to deliver welldefined, charge balanced current stimuli directly to the inner ear. This section emphasizes the design of the supply chip, which includes a DC DC converter. It involves a comprehensive study aimed at optimizing and enhancing the efficiency of the design.

Part Two investigates the fundamental principles of designing millimeter wave (mmWave) voltagecontrolled oscillators (VCOs). This section introduces a VCO with stateoftheart performance, showcasing advancements in mmWave technology.

Overall, this thesis contributes to both the medical device field and automotive radar technology through innovative IC solutions.
}

\hypersetup{
	pdftitle={\Thesistitle},
	pdfauthor={\Authorname},
	pdfsubject={\Degreefield},
}

\begin{document}

\preliminarypages
\chapter{Integrated Solution for Tinnitus Treatment - Introduction}
\newpage

\section{What is Tinnitus? An In-Depth Exploration of Its Symptoms and Variants}

Tinnitus, often described as a ringing in the ears, is a prevalent condition characterized by the perception of noise or sound in the ears or head in the absence of any external acoustic stimulus. This condition can manifest in various forms, presenting a range of sounds from a subtle background noise to a potentially overwhelming sound that can significantly impact daily functioning and quality of life \cite{EsmailiRenton2018}.Tinnitus is a symptom rather than a disease itself, reflecting an underlying condition or dysfunction within the auditory system or related systems in the body. The experience of tinnitus is highly individual, with the sound perceived varying greatly among those affected. Common descriptions of the sounds heard include ringing, buzzing, whistling, hissing, humming, roaring, and clicking \cite{han2009tinnitus}, and \cite{tinnituskorea}. Tinnitus can affect one or both ears and can fluctuate in intensity based on factors such as stress, fatigue, and changes in the environment\cite{kleinjung2020avenue}.
Tinnitus is categorized into two main types: subjective and objective. Subjective is the most common form, where the sound is heard only by the individual. This type can be caused by auditory and neurological reactions to hearing loss, but it can also be influenced by various other factors, including ototoxic medications, metabolic disorders, and stress. Objective, the rarer form of tinnitus, can be heard by an external observer, such as a physician during an examination. Objective tinnitus may be caused by vascular anomalies, muscle contractions, or inner ear bone conditions, and its presence often points to a specific underlying condition that may be treatable \cite{heller2003classification}.

\begin{figure}
    \centering
  \includegraphics[width=0.8\columnwidth,height=0.5\columnwidth]{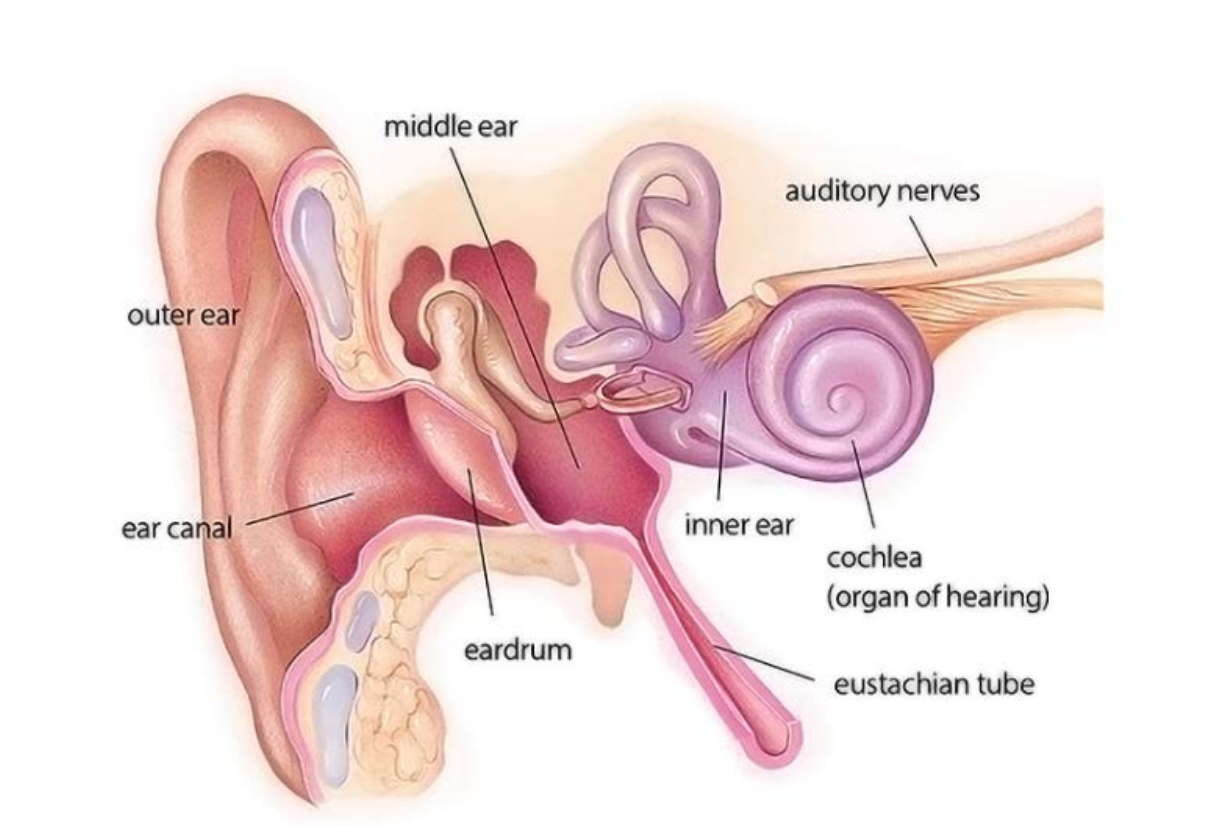}
    \caption{Human ear anatomy}
    \label{Tinnitus}
\end{figure}

Several risk factors can heighten the likelihood of developing tinnitus. Regular exposure to loud environments, common in certain occupations or recreational activities, can cause irreversible hearing damage and tinnitus. The risk of tinnitus increases with age as the cochlea or other ear components naturally deteriorate. Men are more prone to tinnitus, possibly due to more frequent loud noise exposure. Smoking, by affecting blood flow to the auditory system, can exacerbate tinnitus. Cardiovascular issues that impede blood flow, such as hypertension or arteriosclerosis, are also associated with an increased risk of tinnitus. The impact of tinnitus on an individual's life can vary widely. For some, it is a minor nuisance that can be ignored or masked with background noise. For others, it can interfere with concentration, sleep, and normal daily activities, leading to frustration, anxiety, and depression. The variability in the experience of tinnitus underscores the complexity of the condition and the need for personalized approaches to understanding and managing it. Despite its prevalence and the significant impact it can have on individuals' lives, tinnitus remains a condition shrouded in mystery, with its exact mechanisms not fully understood. It is thought to result from alterations in the neural circuits in the brain that process sound, but why these changes occur in some individuals and not in others is still being investigated. The subjective nature of the condition, the variety of sounds experienced, and the range of underlying causes make tinnitus a challenging to study and understand.

\newpage
\section{A Review of Available Assessment Techniques}

Diagnosing tinnitus is a multifaceted process that integrates patient-reported experiences with clinical evaluations and specialized testing. The inherently subjective nature of tinnitus, where the sounds perceived do not originate from an external source, necessitates a thorough and individualized assessment approach. 

The diagnostic journey typically begins with a detailed patient history, where individuals describe their tinnitus perceptions, including the onset, duration, and any associated symptoms such as hearing loss or vertigo. This initial conversation aims to gather comprehensive information about the patient's experience, potential triggers, and the overall impact on their daily life.

Following the patient history, a clinical examination focuses on identifying any observable causes of tinnitus. This examination often includes an otoscopic evaluation to check for abnormalities within the ear canal and tympanic membrane. The presence of earwax buildup, infections, or structural anomalies can provide valuable clues to the underlying cause of tinnitus. A cornerstone of tinnitus diagnosis is the audiological assessment, which evaluates the individual's hearing capabilities. This assessment typically involves pure-tone audiometry to determine the softest sounds a person can hear at various frequencies, thereby identifying hearing loss patterns that might be associated with tinnitus. Additional tests may include speech recognition measures and otoacoustic emissions testing to assess the function of the inner ear's hair cells. To directly assess tinnitus, several specialized techniques are employed. Tinnitus pitch and loudness matching tests aim to quantify the frequency and intensity of the perceived tinnitus sound. Furthermore, tinnitus masking and residual inhibition tests explore the ability of external sounds to mask tinnitus or reduce its intensity after the external sound is removed. Questionnaires play a crucial role in evaluating the subjective impact of tinnitus on an individual's quality of life. Instruments like the Tinnitus Handicap Inventory or the Tinnitus Functional Index help quantify the psychological and functional burdens of tinnitus, guiding treatment planning and outcome measurement. In cases where the clinical assessment suggests a vascular anomaly or a structural issue, imaging techniques such as magnetic resonance imaging (MRI) or computed tomography (CT) scans may be utilized. These imaging modalities can reveal tumors, vascular malformations, or other abnormalities that might contribute to tinnitus symptoms. The diagnosis of tinnitus is comprehensive, incorporating patient narratives, clinical evaluations, audiological testing, and, when necessary, advanced imaging techniques. This multidisciplinary approach ensures a thorough understanding of the tinnitus experience, facilitating targeted management strategies tailored to the individual's needs.

\newpage

\section{What Are the Shortcomings of Current Tinnitus Therapies?}

\begin{figure*}[!htb]
\centering
\includegraphics[width=0.9\columnwidth]{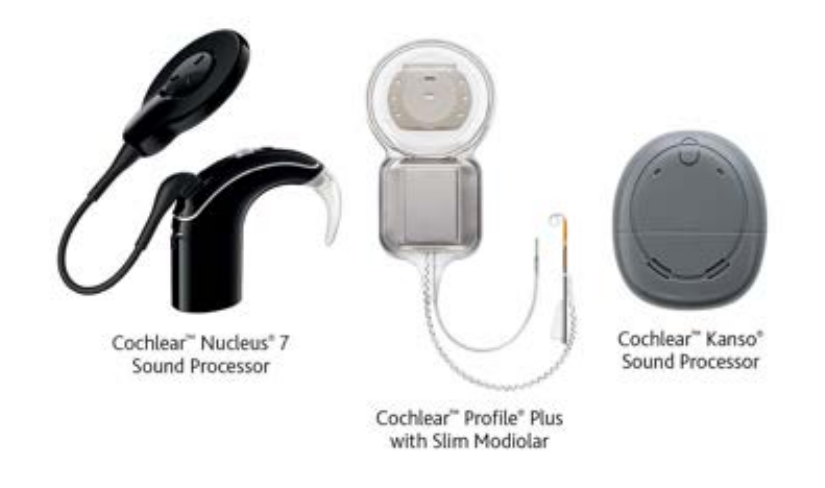}
\caption{Nucleus system. (Image courtesy of Cochlear \cite{Cochlear_website})}
\label{Nucleus system}
\end{figure*}

\begin{figure*}[!htb]
\centering
\includegraphics[width=0.5\columnwidth]{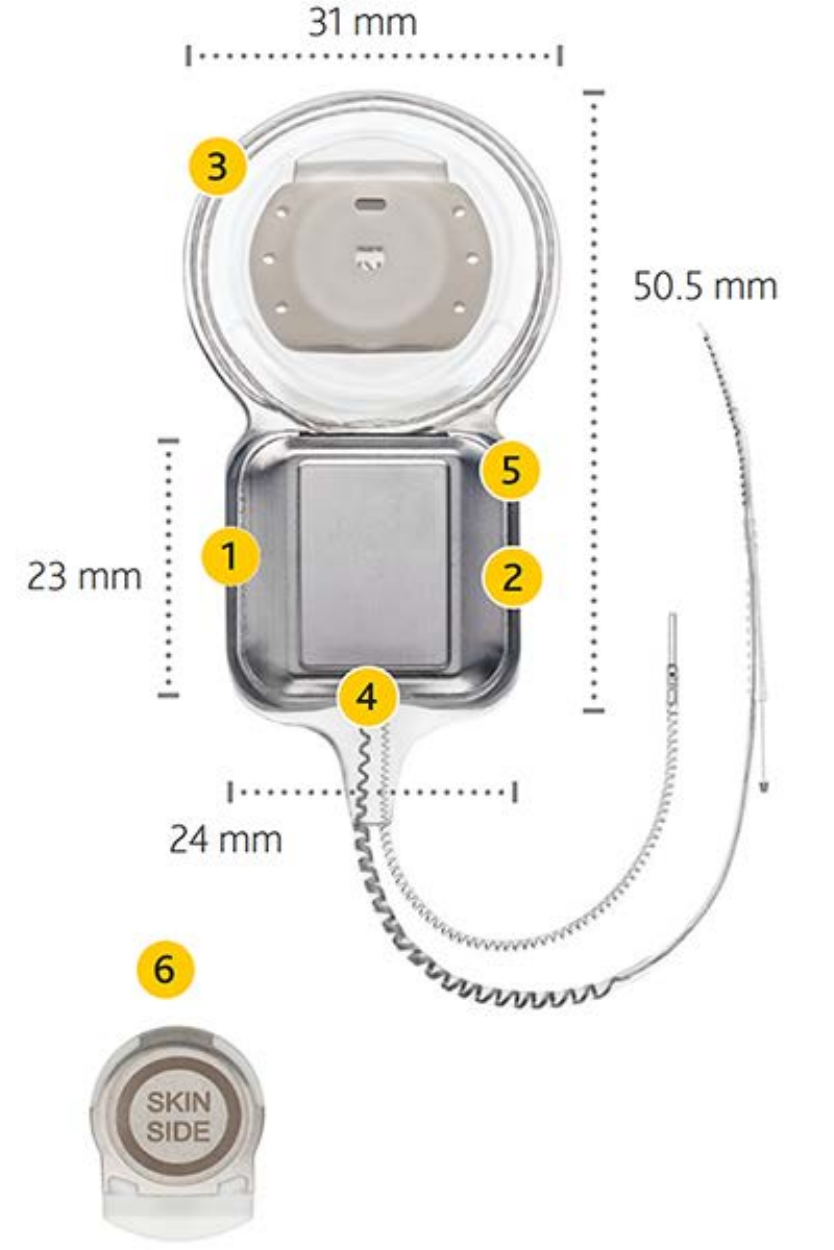}
\caption{Implant size. (Image courtesy of Cochlear \cite{Cochlear_website})}
\label{Implant size}
\end{figure*}

Current tinnitus therapies encompass a broad spectrum of approaches, each aimed at mitigating the symptoms and improving the quality of life for those affected by this condition as it is shown in Fig.~\ref{Nucleus system}. Despite the diversity of treatment options, many individuals with tinnitus continue to face significant hurdles, highlighting notable shortcomings in the efficacy and accessibility of these therapies. Tinnitus is a highly individualized condition, with symptoms and underlying causes varying greatly among patients. This diversity presents a major challenge in developing a one-size-fits-all treatment approach as it is shown in Fig.~\ref{Implant size}. Consequently, the absence of a standardized treatment protocol leads to inconsistent treatment experiences and outcomes, making it difficult for patients and clinicians to identify the most effective therapy. Moreover, the precise mechanisms underlying tinnitus are not fully understood, which complicates the development of targeted therapies. Without a clear understanding of the condition's pathophysiology, treatments often focus on symptom management rather than addressing the root causes, potentially limiting their long-term efficacy.

\section*{Conclusion}

While current tinnitus therapies offer valuable support to many individuals, significant shortcomings remain. These include the lack of a standardized treatment approach, limited understanding of tinnitus pathophysiology, variable treatment efficacy, access and affordability issues, and the absence of a definitive cure. Addressing these gaps through research and clinical practice is essential for advancing tinnitus care and improving patient outcomes.

\newpage

 \chapter{Proposed Miniaturized MCM Solution}

\clearpage

\section{Introduction of inner-ear tissue model under biphasic stimulation}

Creating a device capable of delivering precise electrical stimulation to inner-ear tissue demands an accurate yet straightforward electrical model. This model is indispensable for effectively representing the intricate electrical properties of the inner-ear tissue without unnecessary complexity. However, crafting such a model can be challenging. A thorough investigation and consolidation of the impacts of electrical stimulation on neural tissue are detailed in \cite{earmodel1}.

\begin{figure*}[!htb]
\centering
\includegraphics[width=0.7\columnwidth]{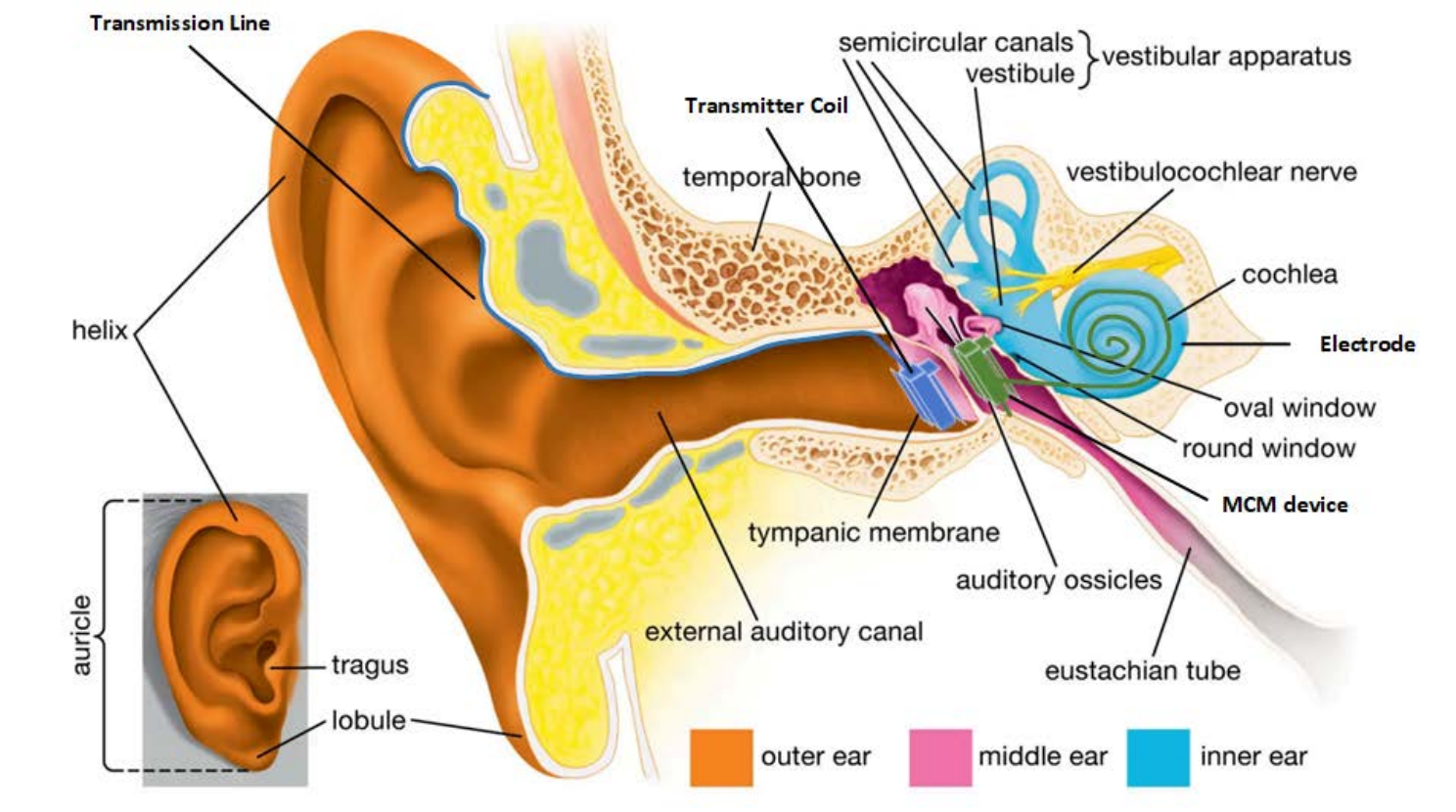}
\caption{Proposed MCM. (Image courtesy of NIH/NIDCD \cite{Proposed_MCM})}
\label{Proposed MCM}
\end{figure*}

\begin{figure*}[!htb]
\centering
\includegraphics[width=1\columnwidth]{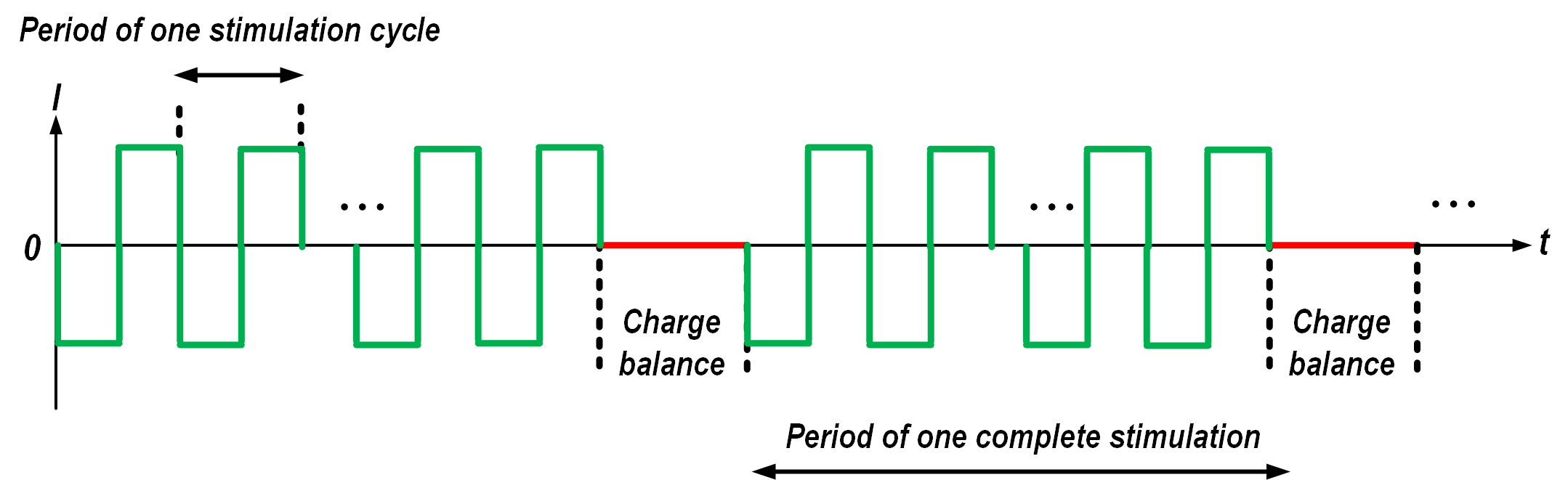}
\caption{Biphasic current waveform}
\label{Biphasic current waveform}
\end{figure*}

Ear model with a simple representation of proposed module is shown in Fig.~\ref{Proposed MCM}. One of the fundamental and widely adopted approaches to stimulation of inner-ear tissue involves the application of biphasic pulse-shaped current stimulation signals. These signals, depicted in Fig.~\ref{Biphasic current waveform}, serve as the basis for many electrical models. Each stimulation cycle comprises two phases: an initial negative phase followed by a positive phase. Ideally, the amplitude and duration of these phases should be precisely matched to prevent any net charge delivery to the tissue. However, practical stimulation setups often exhibit discrepancies between these phases. To address this, a charge-balancing period, illustrated in Fig.~\ref{Biphasic current waveform}, is introduced to counteract any residual charge resulting from these disparities.

\begin{figure}[h]
\centering
\includegraphics[width=0.5\textwidth]{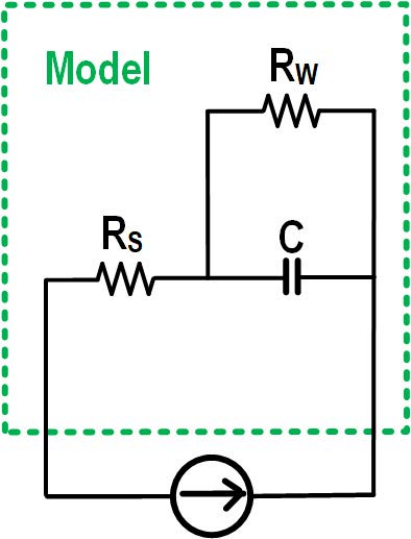}
\caption{Illustration of the Randles R-C-R model depicting electrical components.}
\label{R-C-R model}
\end{figure}

To deliver current stimulation signals to the inner ear, electrodes are employed. Upon surgical implantation within the inner-ear tissue, these electrodes establish a capacitor-like interface with the tissue. Here, the electrodes function as the capacitor plates, while the inner-ear tissue, acting as an electrolyte, serves as the dielectric material \cite{model_cap}. The overall resistance encountered by the current flowing through the inner-ear tissue, encompassing resistance from wires and electrode contacts, is denoted as spreading resistance ($R_s$). Additionally, the current flow arising from oxidation-reduction reactions at the electrode-electrolyte interface is represented as charge-transfer resistance ($R_{ct}$). Hindrances caused by ion movement within the electrolyte are reflected by the Warburg impedance \cite{model_warburg}. These components, namely the charge-transfer resistance and Warburg impedance, collectively constitute the Faradaic impedance, identified as $R_w$ \cite{model_faradaic}. Together, these elements form the basis of the Randles R-C-R model, an electrical model commonly utilized, as depicted in Fig.~\ref{R-C-R model} \cite{earmodel1}.

Another prevalent model utilized for the electrode-tissue interface in stimulators is the series R-C model, as depicted in Fig.~\ref{R-C-R model} \cite{inner_ear_rc_model}. Unlike the R-C-R model showcased in Fig.~\ref{R-C-R model}, this model consolidates the combined effects of spreading resistance, \(R_s\), charge-transfer resistance, \(R_{ct}\), and the Warburg impedance into a single resistor, \(R\). The capacitance formed between the metal electrodes and the inner-ear tissue's electrolyte is encapsulated by a solitary capacitor, \(C\). This simplified model, compared to the R-C-R model illustrated in the model shown in Fig.~\ref{R-C-R model}, is more frequently employed in the circuit design and simulation phase. This preference arises from its enhanced convergence and reduced simulation time during transient simulations when designing circuits using modern CAD software.

\clearpage

\section{Miniaturized MCM solution in detail}

\begin{figure*}[!htb]
\centering
\includegraphics[width=0.8\columnwidth]{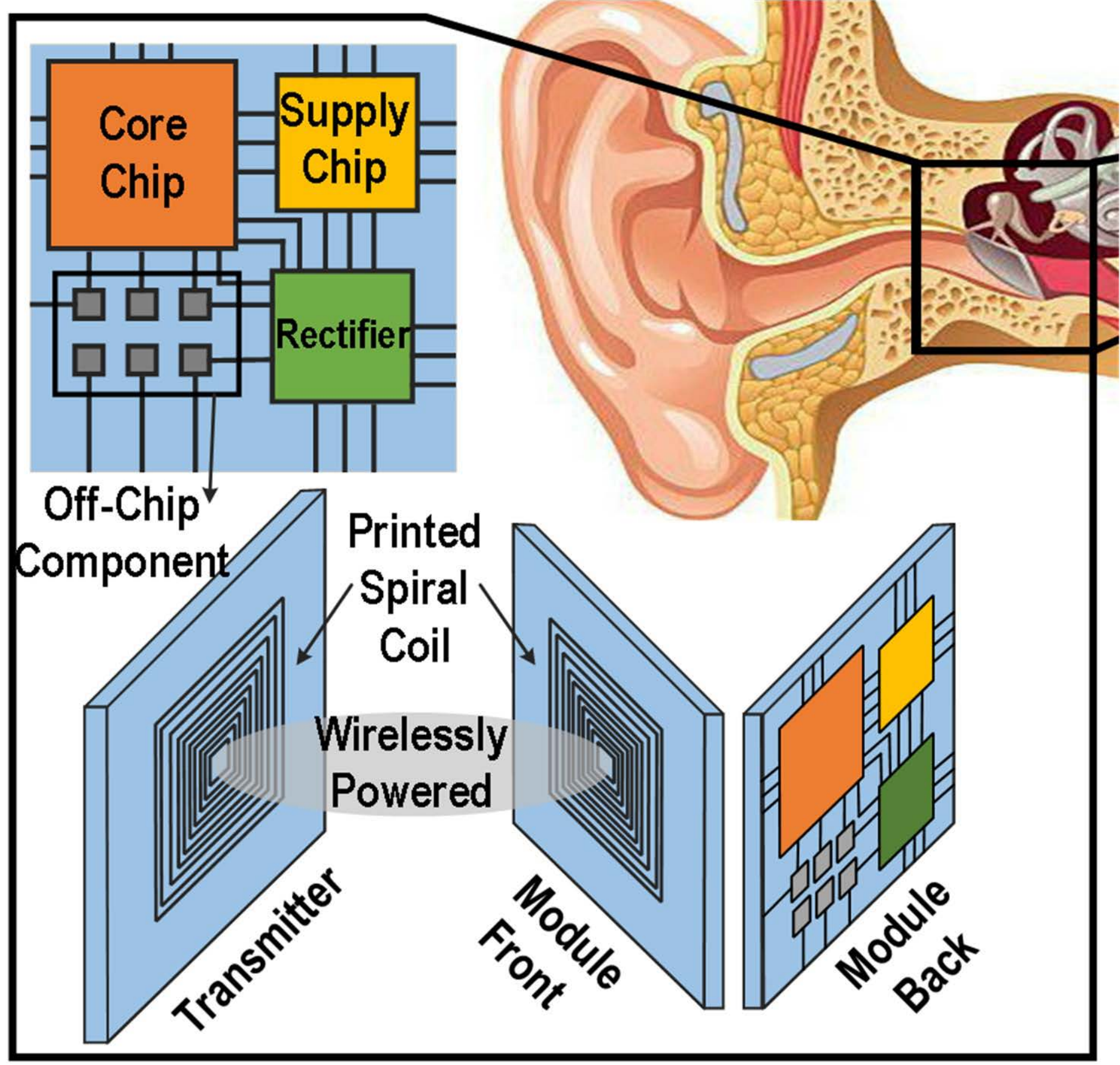}
\caption{Multi chip module.}
\label{Multi chip module}
\end{figure*}

For the design of a non-implantable transmitter that sends waveform data to a module, it is practical to leverage standard commercial chips and available development boards. However, there are a few reasons that leads us to a multi-chip system when it comes to a implantable design. Firstly, the use of an implantable module requires wireless reliable transmission of data and power which requires a more sophisticated approach to power delivery and data transmission. Secondly, the form factor needed for an implantable transmitter capable of providing the required stimulation's is only possible throughout a multi-chip system design.
To better capture and characterize the specification for the multi-chip system, a simplified model of the inner ear, adapted from existing research \cite{inner_ear_rc_model}, that is central to the design of our stimulation device, is proposed. The model is essentially a resistor-capacitor (RC) circuit, reflecting the electrical characteristics of the inner ear environment. At a stimulation frequency of 150 Hz, the impedance is approximately 11k$\Omega$. On the other hand, the reported maximum current stimulation value for a effective treatment is 1.25mA. To accommodate for potential variations under different process, voltage, and temperature (PVT) conditions, a robust 20-V power supply is utilized. The choice of a high-voltage bipolar-CMOS-DMOS (BCD) process for the circuit design allows for an optimal blend of analog precision, digital efficiency, and high-voltage capability.
The schematic shown in Fig. e~\ref{Multi chip module} shows the multi-chip module, which is engineered to be compact enough to fit within the restricted space of the inner ear, maintaining a maximum dimension of 4.5 mm on each side. This module comprises several critical components mounted on a single laminate: The core chip, a DC-to-DC converter for voltage regulation, a rectifier for AC to DC conversion, the necessary receiving coil for energy and data reception, and various discrete components like capacitors which ensure the stability and efficiency of the power conversion process.

The integration of these components into a cohesive unit not only challenges traditional design paradigms but also pushes the boundaries of what is technically feasible in biomedical implants, promising enhanced outcomes for auditory stimulation therapies.

In any integrated circuits, random mismatches invariably lead to asymmetries between the stimulation's positive and negative phases. Such asymmetries, though subtle, can cause an incremental accumulation of charge in the tissues of the inner ear, which, if unchecked, may lead to tissue damage. To mitigate these effects, various strategies for charge equalization are deployed in stimulation devices. Commonly, sophisticated feedback systems are implemented to dynamically regulate and correct charge distribution, ensuring operational balance as detailed in \cite{dac}. However, these solutions can introduce additional complexity, increasing power usage and chip footprint.

\begin{figure*}[!htb]
\centering
\includegraphics[width=1\columnwidth]{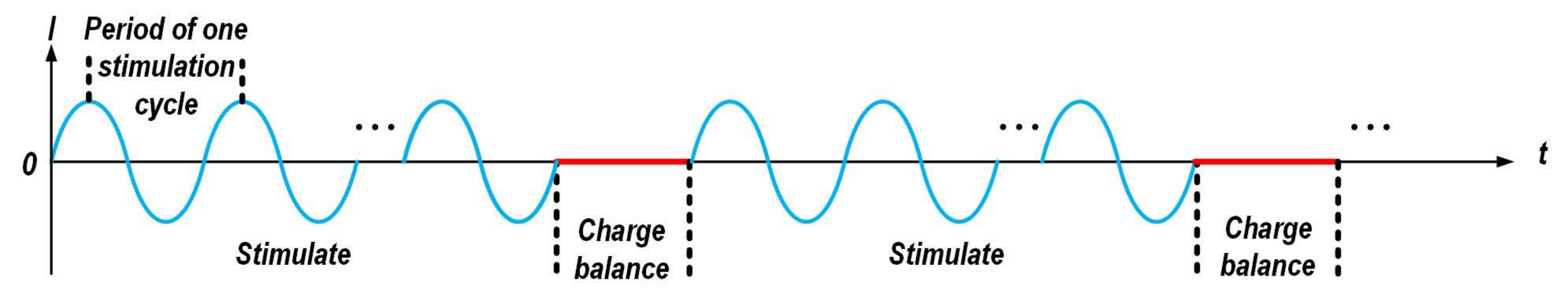}
\caption{Charge balanced waveform.}
\label{Charge balanced waveform}
\end{figure*}

This project explores a streamlined alternative to manage charge as shown in Fig.~\ref{Charge balanced waveform}. Our approach simplifies the process by directly short-circuiting the electrodes in contact with the ear tissue through an on-chip mechanism, promptly dissipating any residual charges. We suggest a new circuit design for this purpose, which is feasible within the TSMC 180nm BCD fabrication technology. This method, which is described in detail in the next chapter, not only simplifies the design but also reduces the power requirements and chip area, making it an efficient solution for charge management in medical devices.

\section{System Diagram}

\begin{figure*}[!htb]
\centering
\includegraphics[width=1\columnwidth]{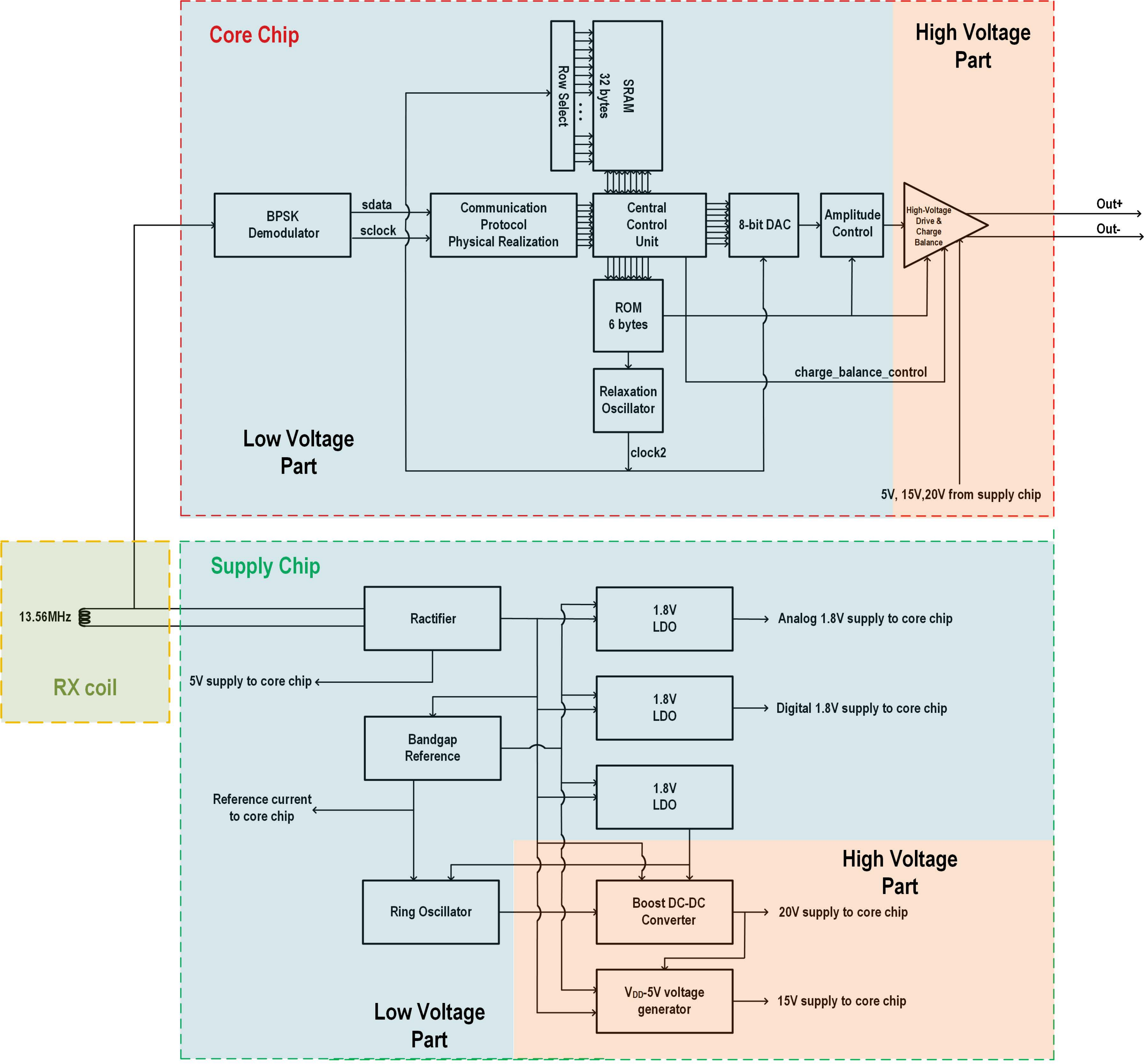}
\caption{System diagram.}
\label{System diagram}
\end{figure*}

The system block-level diagram is illustrated in Fig.~\ref{System diagram}, with the circuit blocks enclosed in the red box representing the circuit implementation of the core chip, and those within the green box representing the circuit implementation of the supply chip.

The core chip's operations are structured sequentially, initiating with the transmission of a digital control sequence. This sequence is modulated onto a 13.56-MHz AC signal facilitating energy transfer between two inductively coupled coils. Post reception, the signal is demodulated by a BPSK demodulator to extract digital data, which is then forwarded to both the communication protocol hardware and the main control unit for further processing. The process commences with a 12-bit synchronization code with the receiver chip, followed by uploading signal parameters into the chip’s memory units—SRAM and ROM. The transfer concludes with an additional 12-bit code signaling the end of transmission, subsequently leading the chip to route stimulation signals through nodes \textit{out+} and \textit{out-}, connected to stimulation electrodes. During the implantation procedure, one electrode is strategically positioned at the round window and the other on the bone surface of the cochlea.

The system's flexibility allows it to be adjusted to accommodate various severities of tinnitus experienced by patients. It is capable of altering the number of stimulation cycles, ranging from 7 to 2047, and the audio stimulation frequency, adjustable between 150 Hz and 20 kHz. The duration of the charge balance can also be adjusted, from 16 ms to 1.28 seconds.

To support these functions, the core chip depends on two separate 1.8-V power sources—one allocated for digital circuits and the other for analog circuits. Additionally, the chip's BPSK demodulator, relaxation oscillator, and digital-to-analog converter (DAC) require a stable 5-$\mu$A reference current. Higher voltages of 5 V, 15 V, and 20 V, necessary for driving high-voltage circuits and managing charge balance, are provided by an external supply chip. This supply chip is outfitted with various stages of noisy large-swing switch-capacitor circuits, strategically isolated to mitigate any potential switching noise impacts on the core chip.


\chapter{Integrated Solution for Tinnitus Treatment - Design of the Core Chip}

This chapter is primarily contributed by my colleague, detailing the comprehensive design of the core chip used in our tinnitus treatment project. Here, I provide a summarized description to elucidate the integration and functionality of the system components.

\section{Core Chip Overview}

\begin{figure*}[!htb]
\centering
\includegraphics[width=1\columnwidth]{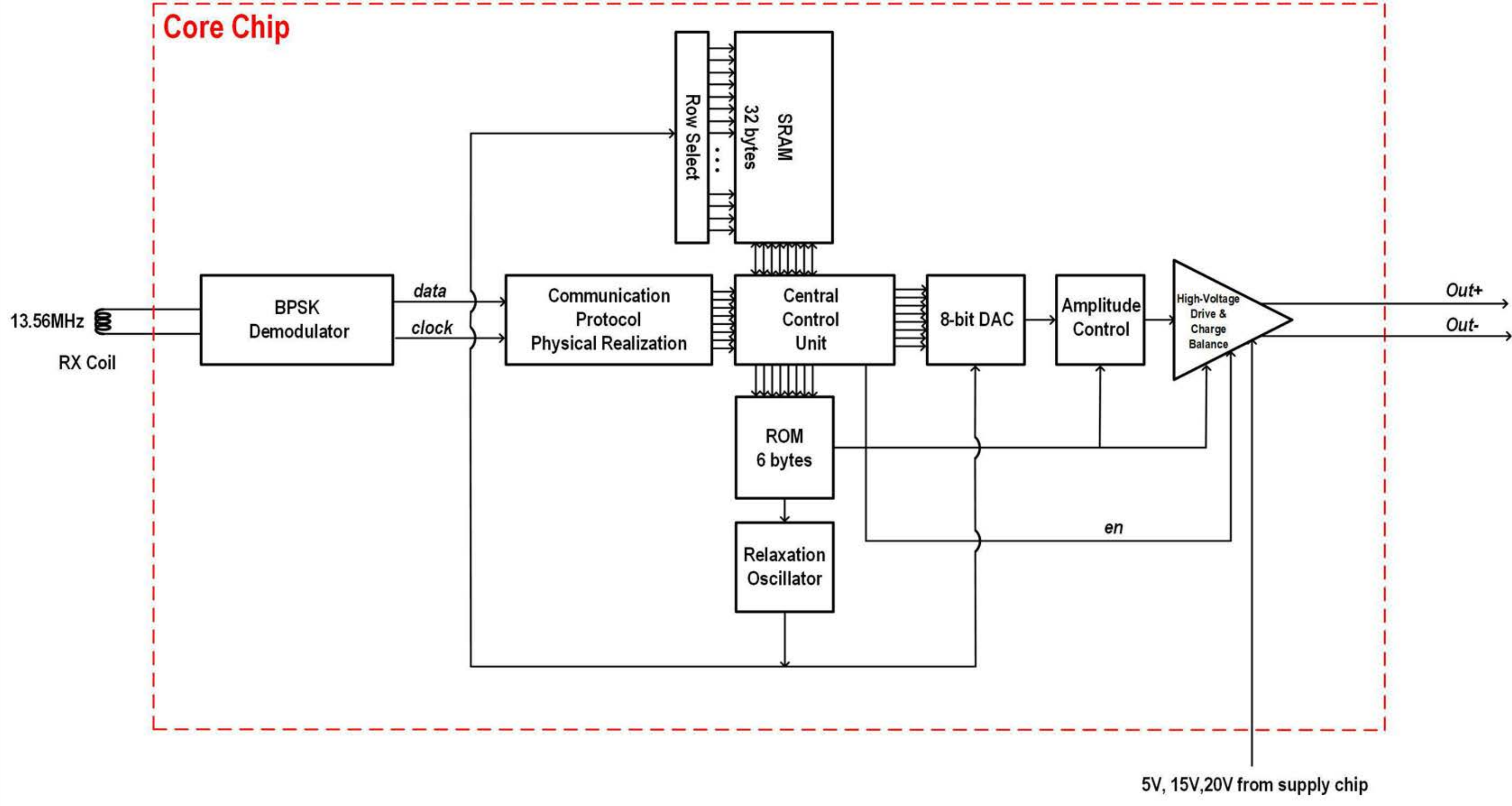}
\caption{Block diagram of core chip.}
\label{Block diagram of core chip}
\end{figure*}

The core chip is fabricated using the TSMC180nm BCD G2 process. It incorporates low-voltage transistors and high-voltage DMOS transistors, facilitating various operational voltages crucial for the system's diverse functions, such as BPSK demodulation, digital-to-analog conversion, and high-voltage driving of the load. the overview is shown in Fig. \ref{Block diagram of core chip}.

\section{BPSK Demodulator and Clock Recovery Circuit}

\begin{figure*}[!htb]
\centering
\includegraphics[width=0.5\columnwidth]{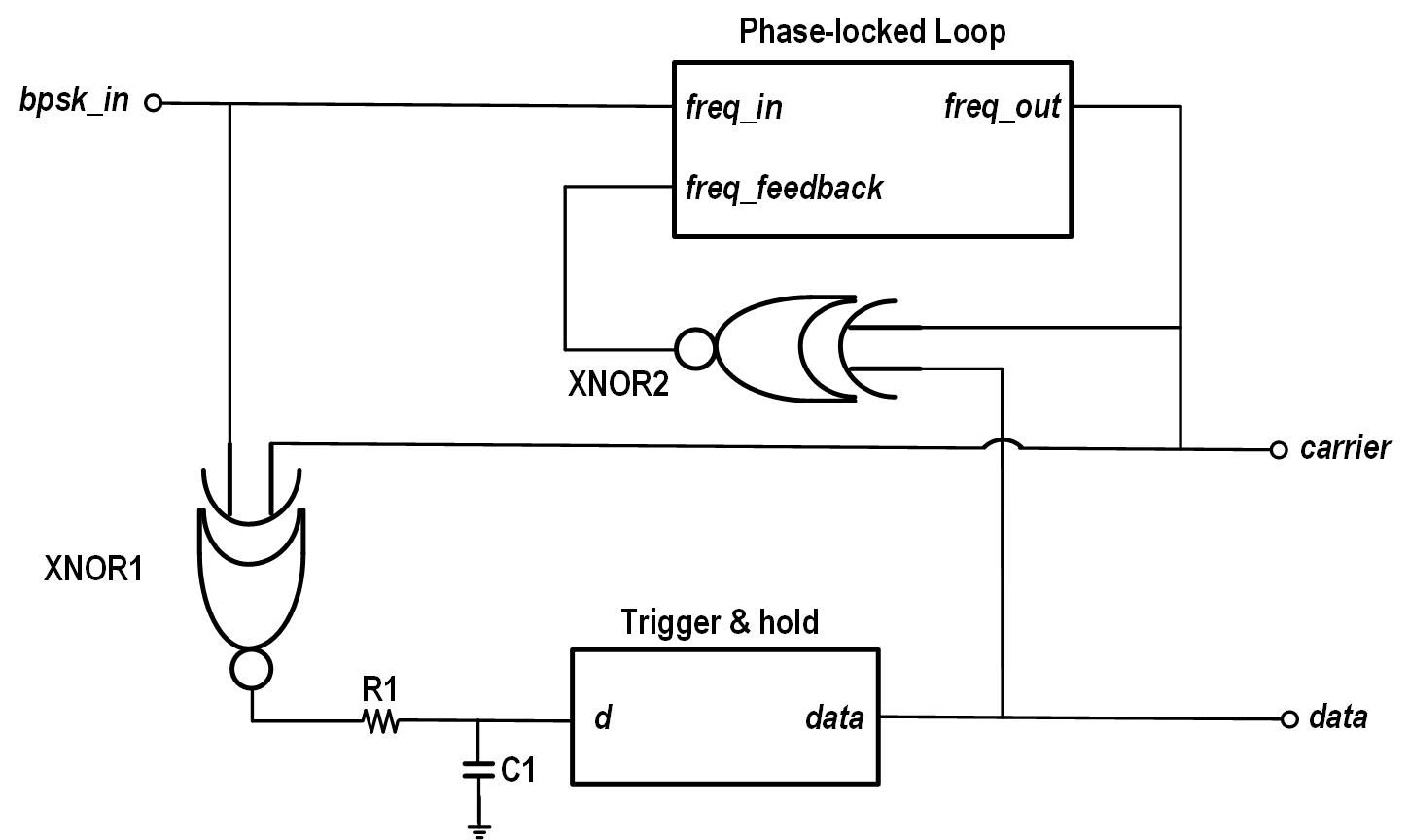}
\caption{Block diagram of BPSK demodulator.}
\label{bpsk_demodulator_core_top}
\end{figure*}

The BPSK demodulator is designed to function independently of an external clock source. It employs a phase-locked loop (PLL) to ensure synchronization, effectively compensating for phase shifts that may occur due to patient movements. A clock recovery mechanism, incorporating a frequency divider, is utilized to maintain the accuracy and stability of the system clocks. An overview of the BPSK demodulator is depicted in Fig.~\ref{bpsk_demodulator_core_top}.

The operational principle of this circuit is outlined as follows. In the absence of input data transitions, the PLL, as shown in Fig.~\ref{bpsk_demodulator_core_top}, maintains synchronization between the \texttt{bpsk\_in} and \texttt{freq\_feedback} inputs. In this steady state, the output of XNOR1 remains consistently high or low. When a sudden $180^\circ$ phase shift occurs in the \texttt{bpsk\_in} signal, the PLL output \texttt{freq\_out} initially continues to produce the previously synchronized carrier signal without any phase alteration, due to the PLL's designed response time. Consequently, the output state of XNOR1 toggles from high to low, or vice versa. This state change triggers the action in the trigger \& hold circuit, shown in Fig.~\ref{bpsk_demodulator_core_top}, which then outputs the demodulated data.

\section{Communication Protocol}

\begin{figure}[!htb]
\centering
\includegraphics[width=1\columnwidth]{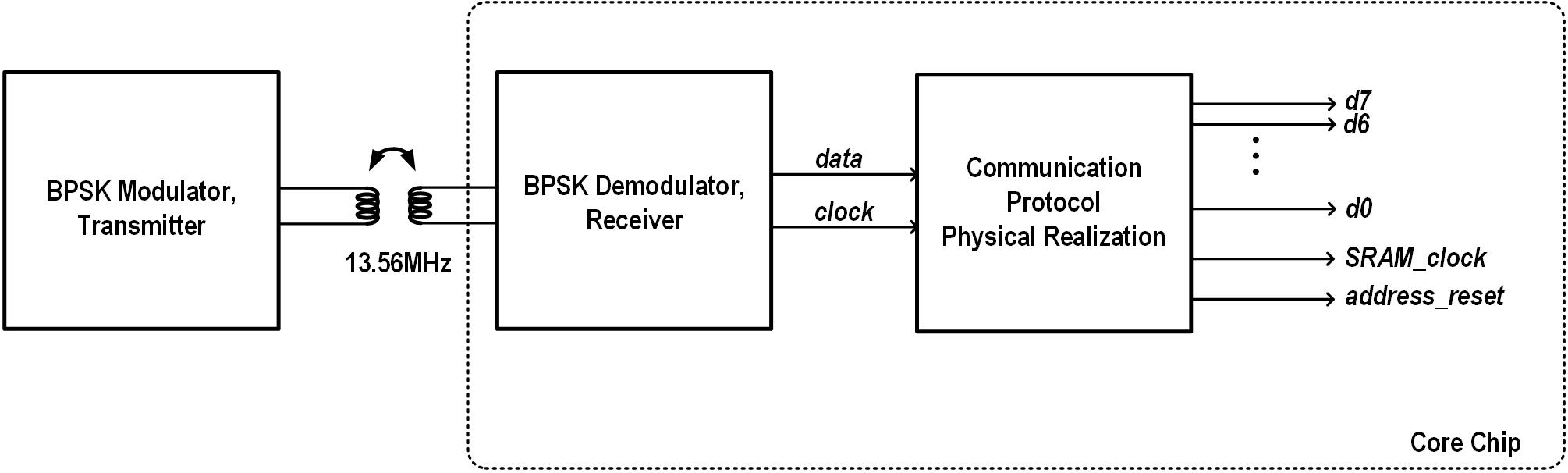}
\caption{Communication system.}
\label{communication_system}
\end{figure}

\begin{figure}[!htb]
\centering
\includegraphics[width=0.7\columnwidth]{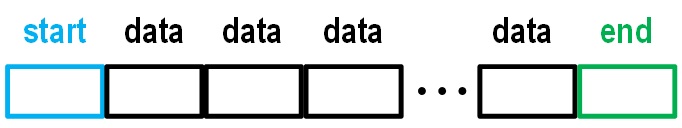}
\caption{Entire data transfer process.}
\label{entire_data_transfer_process}
\end{figure}

A customized communication protocol has been developed to ensure efficient data transfer within the system. This protocol is designed for low-power operation and incorporates a serial-to-parallel conversion mechanism tailored specifically to the device's specifications. The communication system, as illustrated in Fig.~\ref{communication_system}, continuously operates the BPSK demodulator, feeding its outputs, \texttt{data} and \texttt{clock}, to the communication protocol physical realization circuit block. This block then outputs 8-bit parallel data (\texttt{d7} to \texttt{d0}), an associated clock signal (\texttt{SRAM\_clock}), and a control signal (\texttt{address\_reset}). The voltage levels for the digital low of these inputs range from -0.3V to 0.5V, and for the digital high from 1.3V to 2.1V. The clock signal \texttt{clock} samples the serial data signal \texttt{data} at its rising edges. The SRAM holds 32 bytes of data and the ROM contains 6 bytes. Each data sequence transfers 1 byte of data into the core chip, requiring a total of 38 data sequences to transfer all necessary data. The complete data transfer process includes one start sequence, 38 data sequences, and one end sequence, as depicted in Fig.~\ref{entire_data_transfer_process}.

\section{Relaxation Oscillator}

Each cycle of the stimulation waveform consists of 32 sampling points, covering the audio frequency range of 150Hz to 20kHz. Consequently, the clock frequency for the digital-to-analog converter (DAC), which generates this waveform, should range from 4.8kHz to 640kHz. Generating such a low frequency range using a simple on-chip ring oscillator is not feasible. Instead, an on-chip relaxation oscillator, which is specifically designed for low-frequency operation, is utilized.

\begin{figure}[!htb]
    \centering
    \includegraphics[width=0.8\columnwidth]{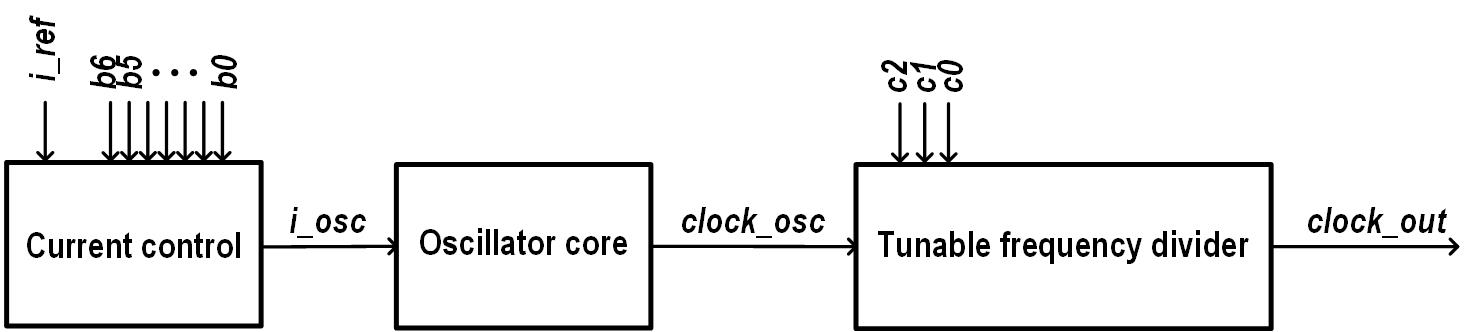}
    \caption{Clock generator.}
    \label{fig:Clock_generator}
\end{figure}

Figure~\ref{fig:Clock_generator} presents the block diagram of the entire clock generator, which comprises three main internal sub-blocks: the current control, the oscillator core, and the frequency divider. The current control block produces a tunable DC current $i\_osc$ for the oscillator core. The oscillator core, detailed later in this chapter, outputs the signal $clock\_osc$ with a frequency range of 65kHz to 3MHz. The tunable frequency divider block then reduces the $clock\_osc$ signal to yield the $clock\_out$ signal, achieving frequencies between 4.8kHz and 640kHz, as required.

\begin{figure}[!htb]
\centering
\includegraphics[width=1\columnwidth]
{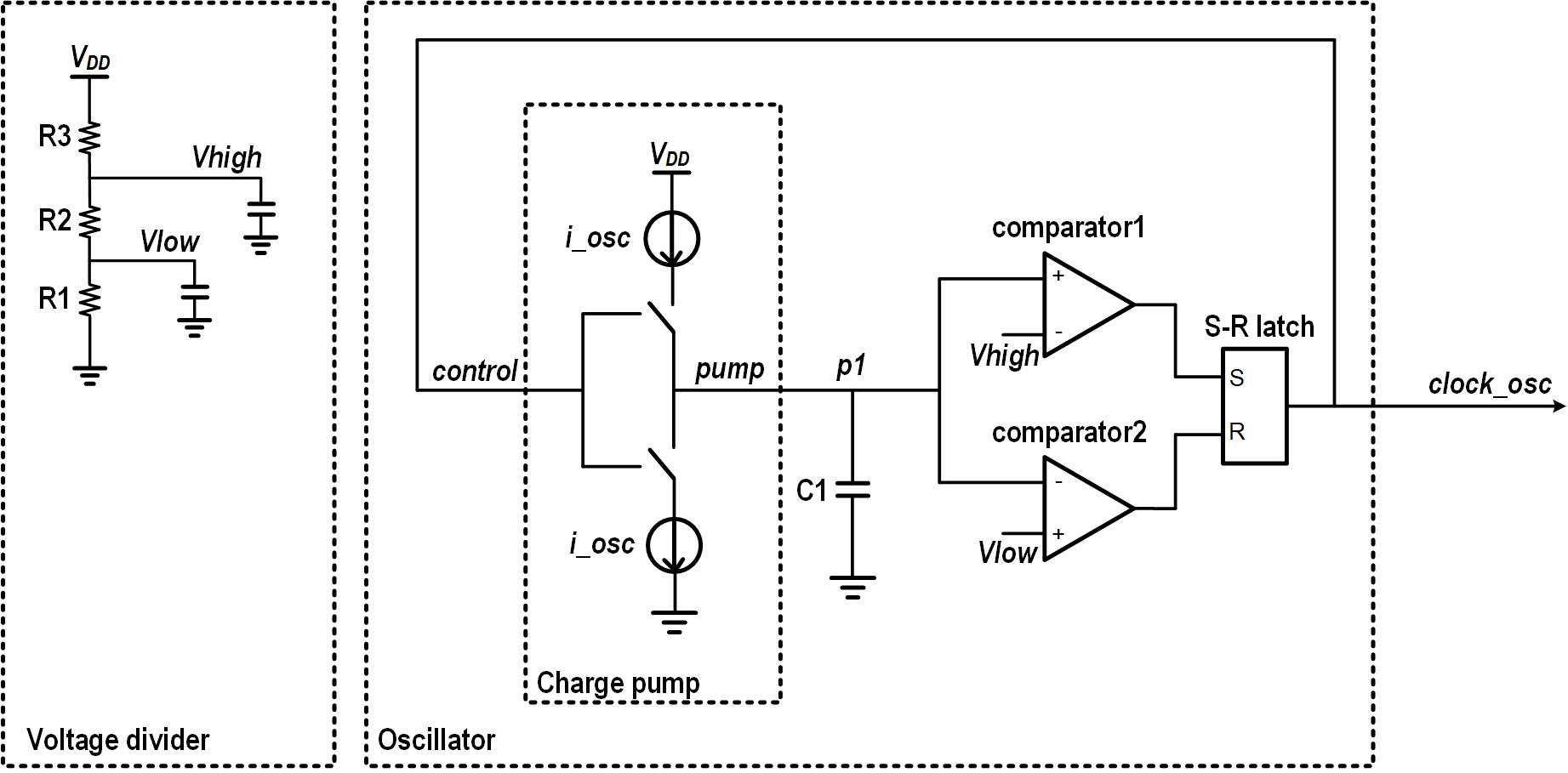}
\caption{Oscillator core.}
\label{Oscillator_core}
\end{figure}

Fig. ~\ref{Oscillator_core} illustrates the schematic of the well-known relaxation oscillator, which includes two comparators, a charge pump, and an S-R latch. The reference voltages $V_{\text{high}}$ and $V_{\text{low}}$ are established by a resistor voltage divider, with decoupling capacitors ensuring stable small-signal grounds for these voltages. The block's supply voltage is sourced from a regulated Low-Dropout (LDO) regulator, fixed at 1.8V and designed to be insensitive to Process, Voltage, and Temperature (PVT) variations.

Upon powering up the chip, the initial charge in capacitor C1 is zero. The S-R latch outputs a digital low control signal, activating the top switch in the charge pump and deactivating the bottom switch. Consequently, the top current source begins charging capacitor C1, and the voltage at node p1 increases steadily until it reaches $V_{\text{high}}$. At this threshold, the S-R latch toggles, outputting a digital high signal, deactivating the top switch and activating the bottom switch. This action causes the bottom current source to discharge capacitor C1, with the voltage at p1 decreasing until it falls below $V_{\text{low}}$. This cycle repeats, causing the circuit to oscillate at a frequency determined by (~\ref{osc_freq_equation}).

\begin{align}
\label{osc_freq_equation}
    {clock\_out}=\frac{i\_osc}{2\cdot{C1}\cdot{(Vhigh-Vlow)}}
\end{align}

\section{DAC and Amplitude Control}
The 8-bit DAC, crucial for generating precise stimulation signals, is augmented by an amplitude control circuit. This system enables dynamic adjustment of output signals, essential for addressing patient-specific treatment needs.

\begin{figure}[htb]
    \centering
    \includegraphics[width=0.8\columnwidth]{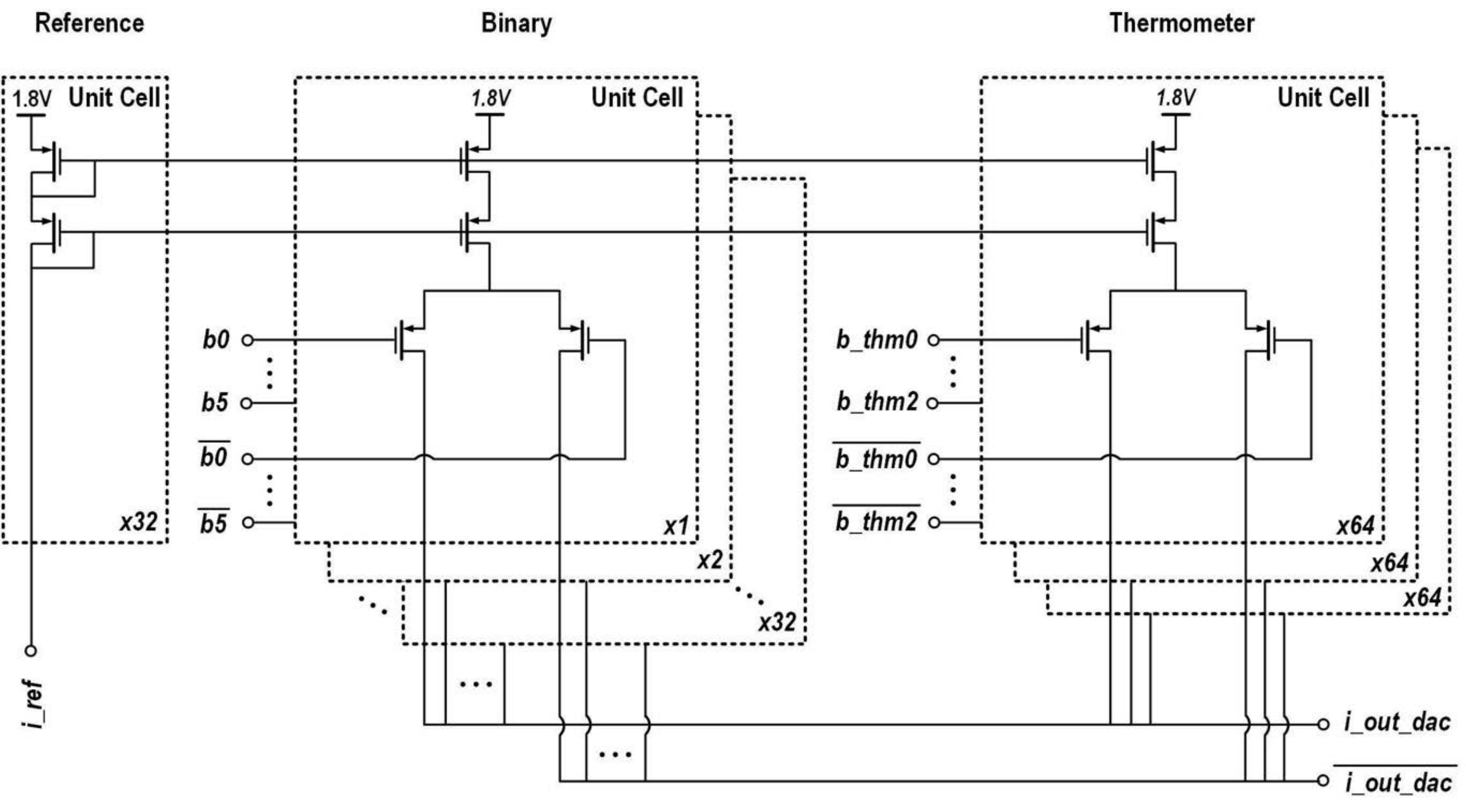}
    \caption{Schematic of the DAC core, highlighting its structured branches.}
    \label{fig:Digital_to_analog_converter}
\end{figure}

Fig.~\ref{fig:Digital_to_analog_converter} shows the DAC core schematic, which includes the reference, binary, and thermometer-code weighted branches. The reference branch is composed of 32 unit cells. The binary branches vary, containing from 1 to 32 unit cells corresponding to control signals $b_0$ through $b_5$, while each thermometer-code weighted branch incorporates 64 unit cells. This DAC is meticulously designed using the common-centroid technique to minimize mismatches among unit cells.

In the context of inner-ear tissue stimulation, this DAC operates within stimulation cycles in stimulation mode, with each cycle comprising 32 samples. The clock frequency controlling the DAC, which dictates the rate at which these samples are processed and converted into analog signals, ranges from 4.8 kHz to 640 kHz.

Post-layout simulations confirm that the DAC functions effectively with sampling frequencies up to 1 MHz. The Integral Nonlinearity (INL) of this DAC is less than $\pm 0.5$ LSB, and the Differential Nonlinearity (DNL) is less than $\pm 1$ LSB, indicating that the DAC is adequately monotonic and linear for use in inner-ear tissue stimulation applications.

\begin{figure}[!htb]
\centering
\includegraphics[width=1\columnwidth]
{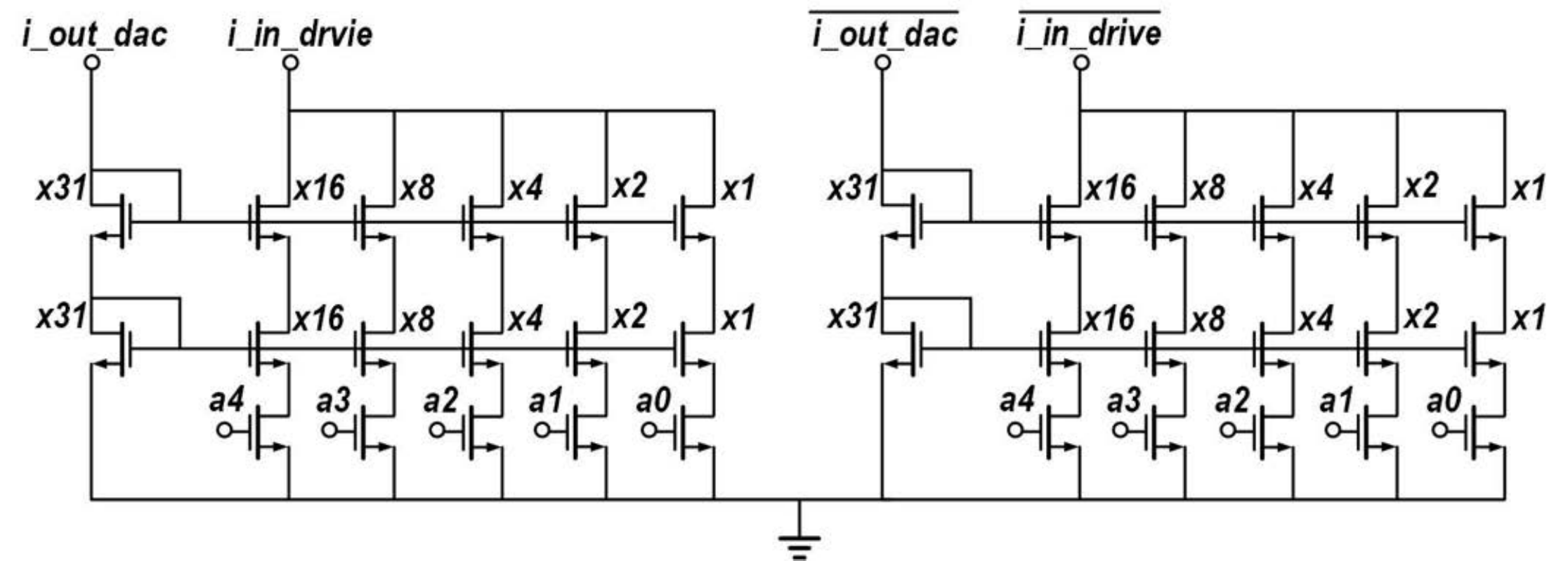}
\caption{Amplitude control.}
\label{Amplitude control}
\end{figure}

Since the DAC is designed to always operate with its full-scale amplitude, in order to generate differential current signals with tunable amplitudes, a 5-bit amplitude control block is designed as shown in Fig.~\ref{Amplitude control}. The output currents from the DAC, $i\_out\_dac$ and $\overline{i\_out\_dac}$, are fed into this block. The generated signals are then mirrored to $i\_drive1$, $i\_drive2$, $\overline{i\_drive1}$, $\overline{i\_drive2}$ and subsequently fed into the high-voltage drive and charge balance block.

\section{High-Voltage Drive and Charge-Balance}

The design includes a high-voltage drive capable of up to 20V operation. Charge balancing mechanisms are implemented to ensure safe and efficient performance.

\begin{figure}[!htb]
\centering
\includegraphics[width=1\columnwidth]
{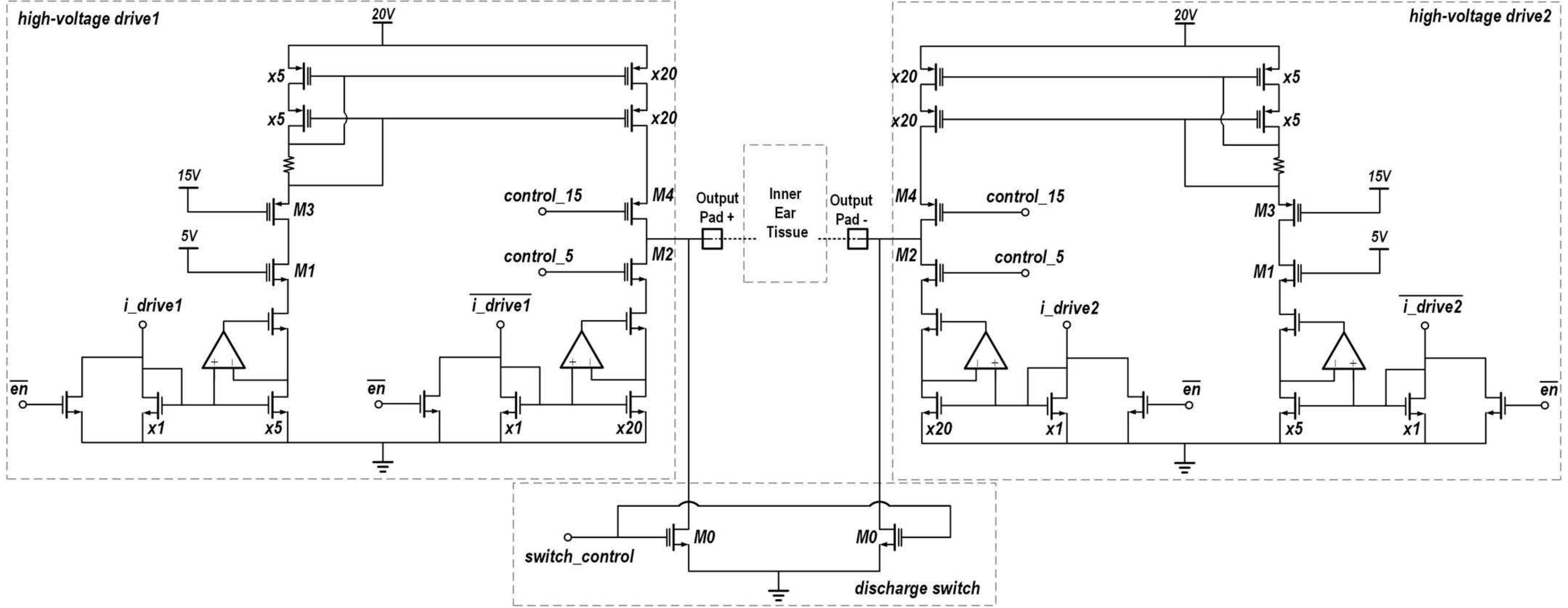}
\caption{High voltage drive.}
\label{High voltage drive}
\end{figure}

In the provided schematics, symbols with double gates represent either high-voltage transistors or low-voltage transistors situated in high-voltage wells, while symbols with single gates denote standard 5-V transistors. Transistors M1-M4, as depicted in Fig.~\ref{High voltage drive}, are 20-V high-voltage LDMOS transistors; all other transistors in this schematic featuring double-gate symbols are 5-V transistors located in high-voltage wells. According to the electrical rules for this process, the maximum voltage difference between the gate and source of any high-voltage transistor must not exceed 5 V to avoid breakdown. Consequently, the gate voltages of M1 and M2 are capped at 5 V, whereas the gates of M3 and M4 are maintained no lower than 15 V.

To ensure fully differential outputs, both high-voltage drive circuits are designed to be identical using precise layout techniques to minimize random mismatches. Nevertheless, even minor mismatches between the two drive circuits can lead to charge accumulation; hence, charge balancing becomes necessary. Ideally, balancing the residual charge accumulating in the load could be achieved by activating a transmission gate linking the two outputs. However, creating a transmission gate capable of withstanding a 20-V swing is impractical in this process due to the limited 5-V breakdown voltage across the gate-source junctions of high-voltage transistors. Instead, a high-voltage transistor M0 is connected between each output node and the ground; these transistors are operable via 5-V or 0-V gate control voltages, allowing for efficient switching on or off as required.

\clearpage

\section{Core Chip Layout}

The layout of the core chip was meticulously crafted to meet rigorous specifications, ensuring optimal performance of all integrated circuits. The size of fabricated chip is 3mm by 2.33mm.

\begin{figure*}[!htb]
\centering
\includegraphics[width=0.7\columnwidth]{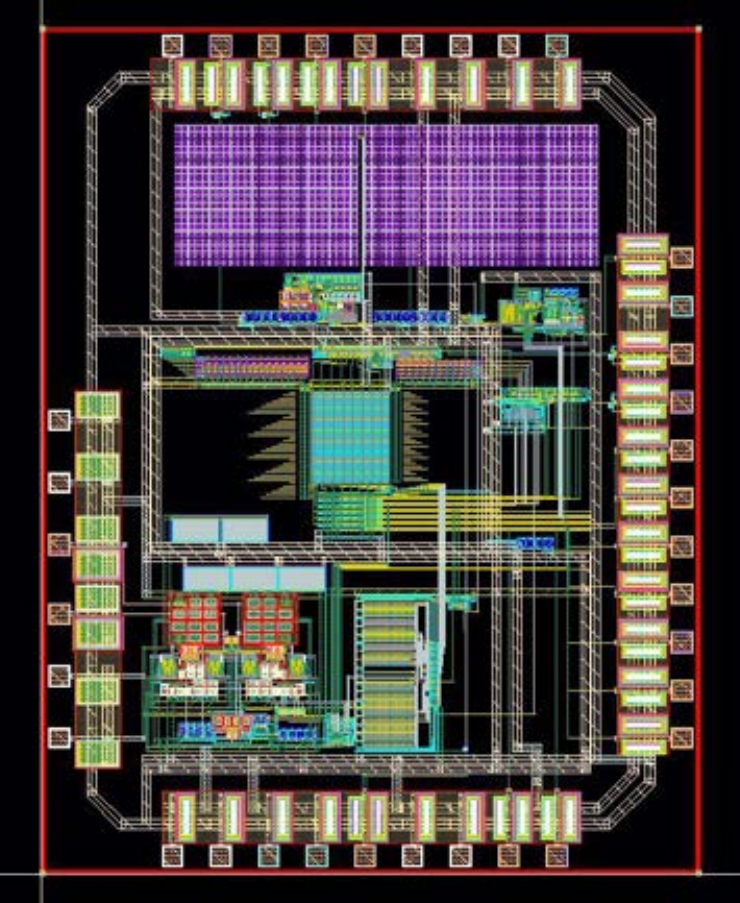}
\caption{Layout of core chip.}
\label{Layout of core chip}
\end{figure*}

\chapter{Integrated Solution for Tinnitus Treatment - Design of the Supply Chip}

\clearpage

\section{Supply chip overview}

This chapter outlines the design of the supply chip, which is crucial for generating five distinct supply voltages and a reference current to power the core chip. This chip was also designed and manufactured using the TSMC 180nm BCD G2 process. The efficiency of the supply chip is extremely significant, as it directly influences the system's overall efficiency. A full bridge on-chip rectifier, consisting of four on-chip diodes with a bypass option, has been designed. Due to the coil design parameters and, more specifically, current handling capacity, the on-chip rectifier has not been used later at the time of the measurement and has been bypassed. 
\begin{figure*}[!htb]
\centering
\includegraphics[width=0.9\columnwidth]{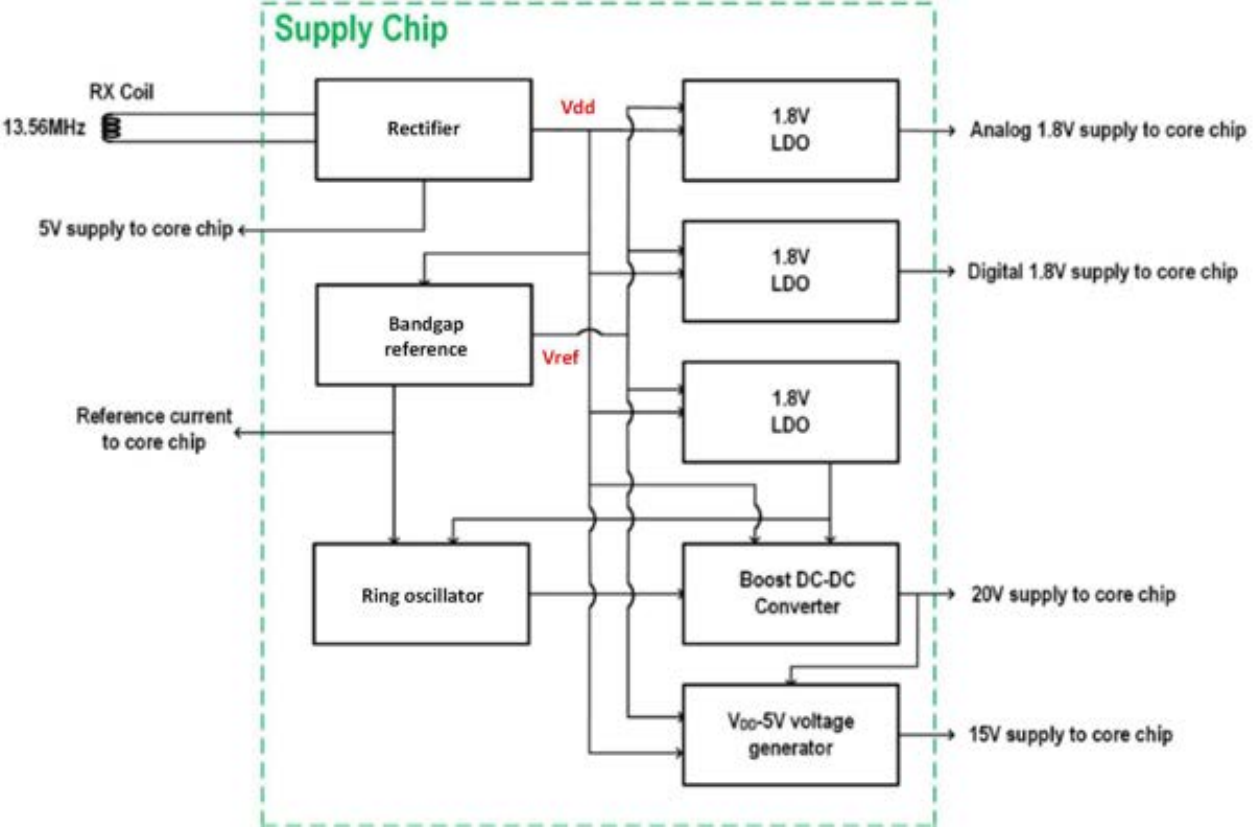}
\caption{Block diagram of supply chip.}
\label{DCDCconv_Diagram}
\end{figure*}

Figure~\ref{DCDCconv_Diagram} shows the block diagram of the supply chip demonstrating each block and the connections between blocks.

\section{Design of the supply chip}

The Supply chip is designed to provide both low and high voltages necessary for the core chip's operation. Consequently, the design of the supply chip is divided into two sections. The following section discusses the design of the low-voltage components.

\subsection{Low-voltage design}

\subsubsection{LDO design}

To guarantee the stable functioning of the core chip's circuits, which primarily need a steady 1.8-V power supply, the best strategy involves incorporating two low-dropout regulators (LDOs) on the supply chip. One LDO is dedicated to analog and the other to digital. This dual configuration is crucial for shielding the noise-sensitive analog circuits from the switching noise generated by the digital circuits.

\begin{figure*}[!htb]
\centering
\includegraphics[width=0.55\columnwidth]{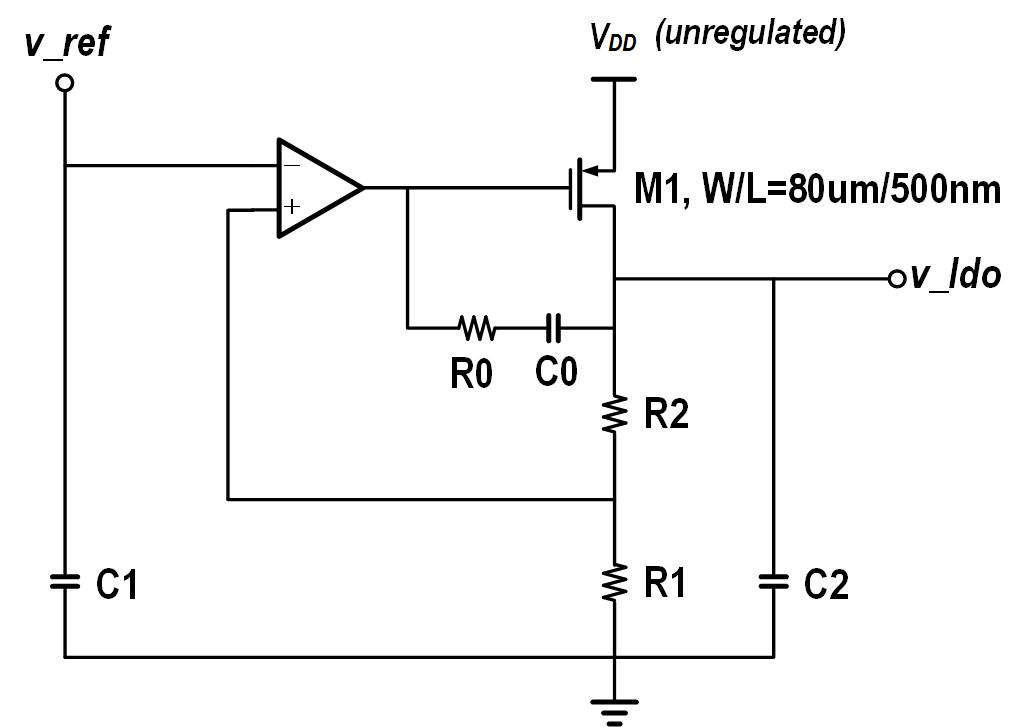}
\caption{Diagram of the LDO Regulators.}
\label{fig:LDO}
\end{figure*}

\begin{figure*}[!htb]
\centering
\includegraphics[width=0.9\columnwidth]{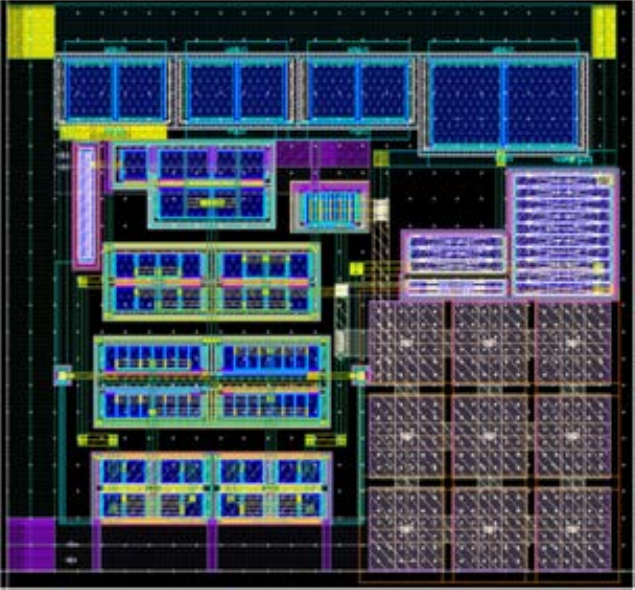}
\caption{Layout of the proposed LDO}
\label{fig:LDOlayout}
\end{figure*}

Fig.~\ref{fig:LDO} displays the configuration of the two identical LDO regulators integrated into the supply chip. The input to the LDOs, marked as $V_{DD}$, comes from the rectifier’s output, which varies between 3.5V and 5.5V depending on the coupling efficiency of the wireless power transfer coils. Transistor M1, serving as the secondary gain stage of the LDO, is designed to provide a consistent 10-mA DC output current. The layout of LDO is shown in Fig.~\ref{fig:LDOlayout}.

\subsubsection{Design of the bandgap current reference}

\begin{figure*}[!htb]
\centering
\includegraphics[width=0.85\columnwidth]{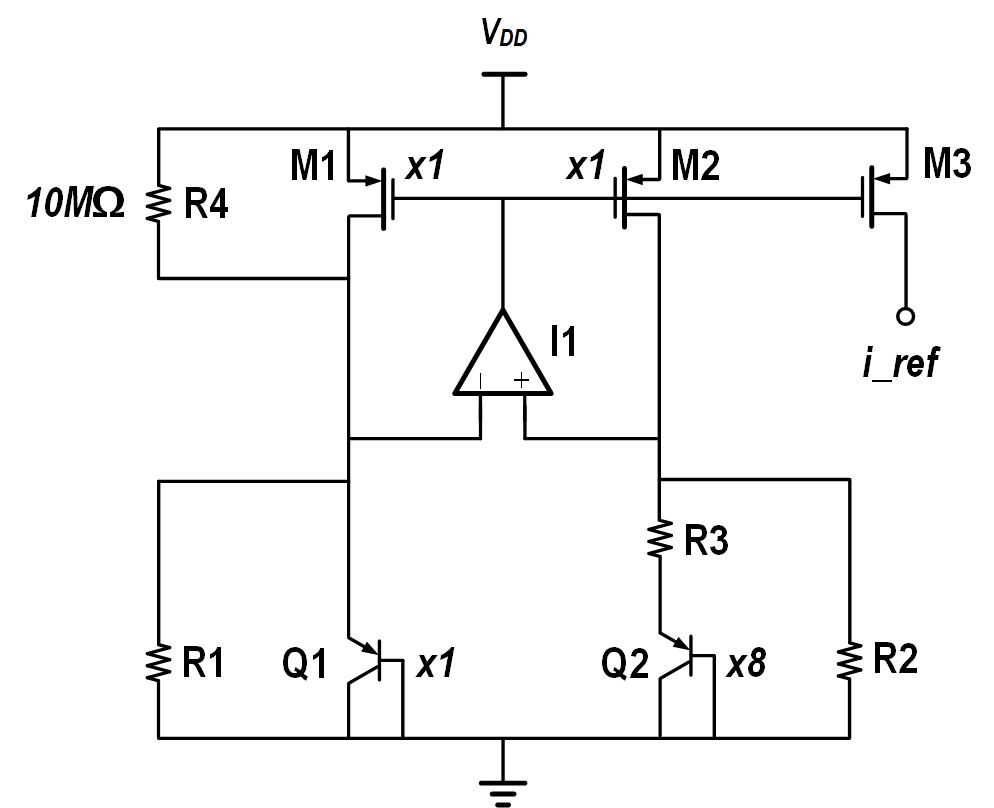}
\caption{Schematic of the Bandgap Current Reference Circuit.}
\label{fig:BandgapCurrentReference}
\end{figure*}

In addition to the bandgap voltage reference, the system includes a bandgap current reference circuit, shown in Fig.~\ref{fig:BandgapCurrentReference}. Value of $R_{1}$, and $R_{2}$ are identical. This circuit is essential for generating a stable reference current, which is vital for biasing the ring oscillator on the supply chip and is also used on the core chip. The design guarantees that the reference current remains consistent despite fluctuations in process, voltage, and temperature (PVT).

\begin{equation}
\label{eq:bandgap_current_formula}
i\_ref = \frac{W_3}{W_2}\left(\frac{V_{\text{BE1}}}{R_2} + \frac{V_T \ln{8}}{R_3}\right)
\end{equation}

The formula for calculating the reference current $i_{ref}$ is outlined in Eq.(\ref{eq:bandgap_current_formula}), based on the work from \cite{razavi2001design}. This calculation incorporates the base-emitter voltage $V_{BE1}$ of transistor Q1, which has a negative temperature coefficient, and the thermal voltage $V_T$, which has a positive temperature coefficient. The resistors $R_2$ and $R_3$ used in the circuit are p+ poly resistors, which exhibits a negative temperature coefficient. Transistors M2 and M3 have widths $W_2$ and $W_3$, respectively. By precisely selecting the values of $R_2$ and $R_3$, the temperature coefficient of $i_{ref}$ is effectively compensated across a temperature range from $-20^\circ\text{C}$ to $65^\circ\text{C}$. Specifically, with $R_2 = 45.3\text{k}\Omega$ and $R_3 = 6.13\text{k}\Omega$, the reference current $i_{ref}$ is stabilized at 5$\mu$A at $36^\circ\text{C}$. Simulation results confirm that the variance of this current reference under various PVT conditions is maintained within $\pm 5ppm$.

\begin{figure*}[!htb]
\centering
\includegraphics[width=0.9\columnwidth]{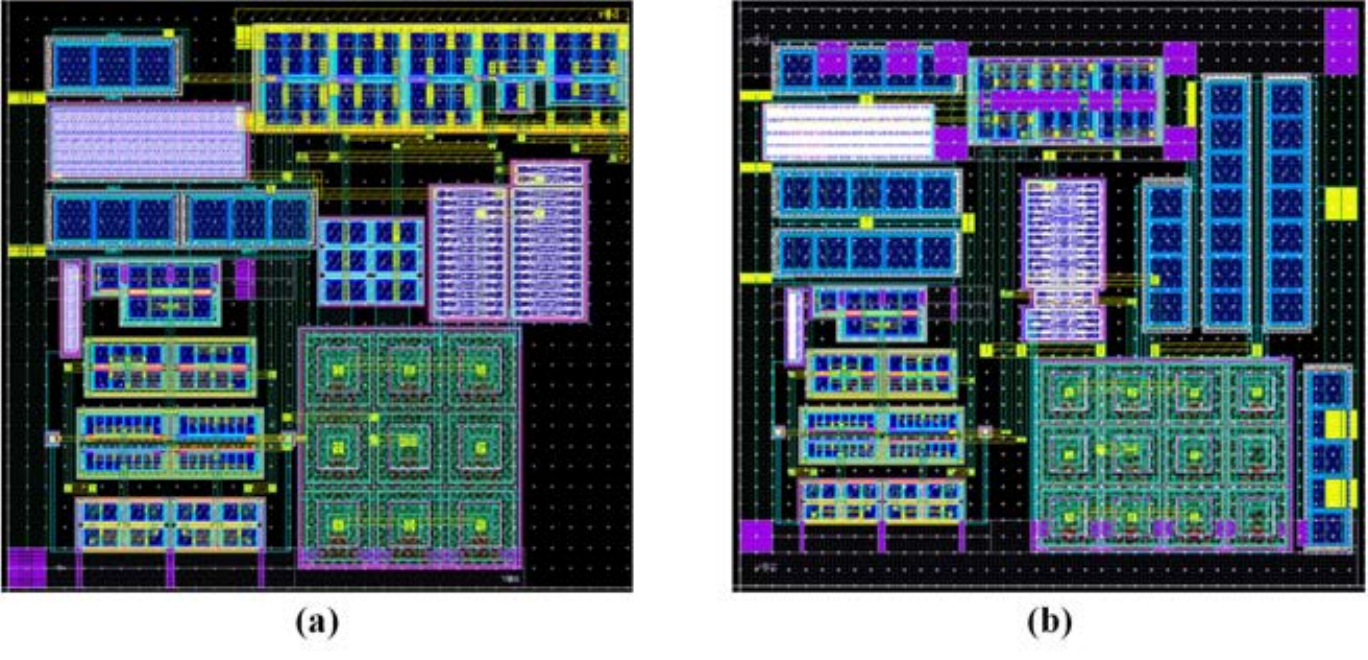}
\caption{Layout of voltage and current references.}
\label{ReferenceLayout}
\end{figure*}

Fig.~\ref{ReferenceLayout} (a) and (b) show the layout for the voltage and current reference, respectively. This will conclude the low-voltage parts of the design. In the next section, the high-voltage portion will be introduced in detail.

\clearpage

\subsection{High-voltage design}
The high-voltage design is an important part of the supply chip design. With the help of design of a $V_{DD}$-5V voltage generator, one novel solution to provide both 15V and 20V, is to only generate the 20V from the converter, and then derive the 15V reference from the 20V. This will reduce the die area as well as simplify the design complexity. In this chapter, firstly an analysis on the DC-DC converter has been done. Then, the feedback loop will be described in detail. That will conclude the high voltage design chapter.

\subsubsection{Multi-Stage Single-Output DC-DC Converter}
In contemporary electronics, the demand for handheld and portable devices is rapidly increasing. This includes a wide range of applications, from wearable biomedical systems to wireless sensors for the Internet of Things (IoT). As a result, efficient power management has become critically important. This chapter examines the complex world of DC-DC converters, focusing on the challenges of multi-stage multi-output (MSMO) structures and the innovative approach of tapered design.

At the core of this discussion is the goal of achieving low-voltage, power-efficient design methods. While reducing the supply voltage can decrease power consumption and device size, certain components within these systems need higher supply voltages to work effectively. Key examples are neural stimulators, RF power amplifiers, and bidirectional brain-machine interface (BMI) systems, all of which require higher voltages for their specific functions. To address these needs, the Dickson charge pump is a notable solution. Its design, which does not use inductors, is particularly suitable for on-chip implementation, meeting the requirements of modern semiconductor devices. Over time, various versions and designs of the Dickson charge pump have been created, each striving to improve power efficiency and overcome design challenges.

\begin{figure*}[!htb]
\centering
\includegraphics[width=1\columnwidth]{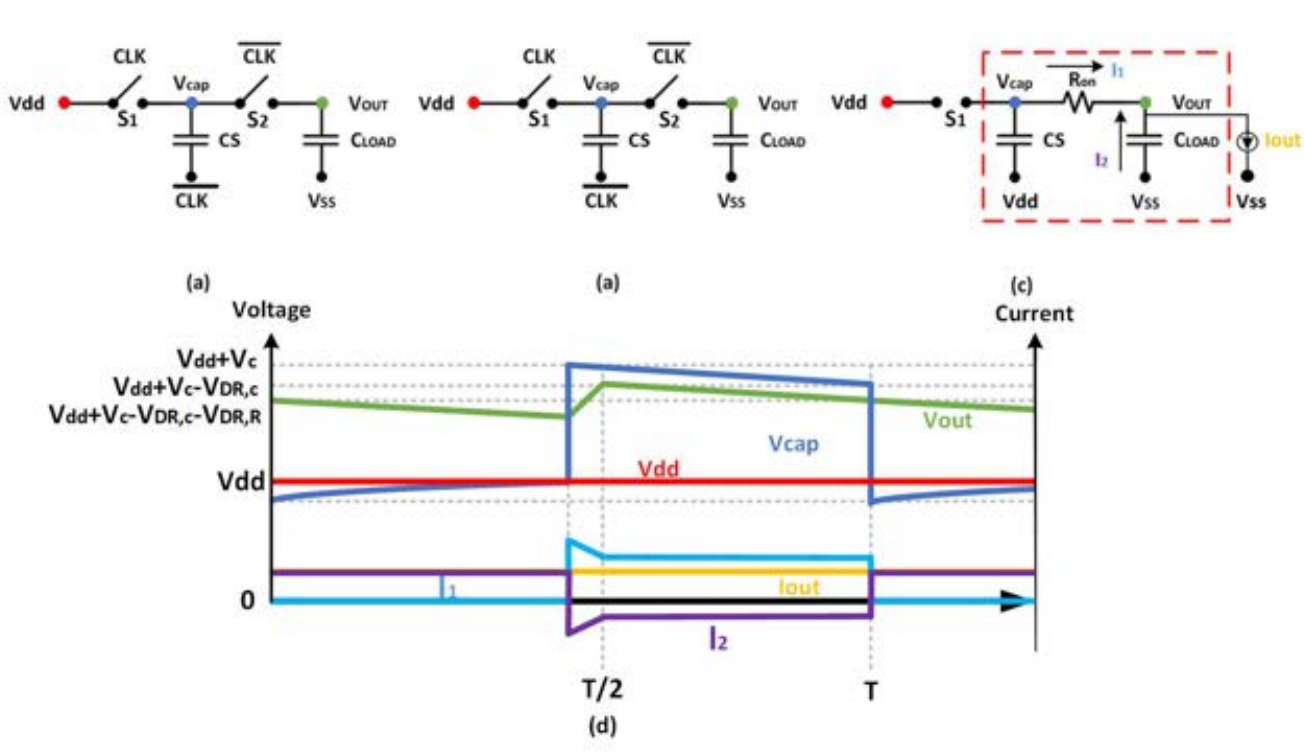}
\caption{Working Principle of single stage DC-DC converter with voltage and current waveforms for a period of operation}
\label{DCDCconv1}
\end{figure*}

The design of a DC-DC Converter (DDC) for specific uses requires consideration of several aspects, such as the output voltage stability under load and power efficiency metrics. Fig.~\ref{DCDCconv1} outlines the fundamental model of a single-stage charge-pump-based DDC with a fix dc input voltage, $\textit{V}_{dd}$. using an analysis similar to that in \cite{dcdcconv}, during the charging phase (see Fig.~\ref{DCDCconv1}(b)), the storage capacitor $C_{S}$ begins to charge via the equivalent resistance $R_{ON}$ of switch S1, setting the time constant for this phase. At the end of the charging phase, $V_{cap}$ reaches $V_C = (1 - \alpha) \textit{V}_{dd}$ (where $\alpha = \exp(-1/(2f_{CLK}R_{ON}C_S))$). In the discharge or pumping phase (see Fig.~\ref{DCDCconv1}(c)), $R_{\textit{on}}$ on from $S_{2}$ connects $C_{s}$ to the output node $V_{OUT}$, charging the $C_{load}$, and providing the output current. Choosing large $C_{load}$ keeps the $V_{out}$ variation minimal and it can be assumed that $V_{out}$ is nearly a dc voltage. Load of DC-DC converter is nearly a dc current, thus, we can assume that $I_{out}$ is nearly constant. For a big value of $C_{load}$, small variation of $V_{out}$ guarantees that $I_{out}$ has small variation, then the average value of $I_{out}$ can be used. For the rest of the chapter, $I_{out}$ is assumed to be a dc. Fig.~\ref{DCDCconv1} (d) shows the currents and voltages waveform for a single period of DC-DC converter. As it is shown, $I_{out}$ is assumed to be a dc waveform, and $V_{out}$ is assumed to has small variation around the average value. To calculate the average $V_{out}$, a calculation on the voltage drop in discharging phase is needed. By applying Kirchhoff's Current Law at $V_{out}$ node in discharging phase, it can be concluded that an average dc current equal to $I_{out}$ is going through $R_{on}$, which will results in a voltage drop equal to  $V_{DR,R} = R_{ON}I_{OUT}$ across switches resistor $R_{on}$. In addition, in the discharging phase, Voltage across $C_{S}$ change as $C_{s}$ is discharging. charge across $C_{s}$ is providing the current flowing into $R_{on}$, the voltage variation across $C_{s}$ can be approximated by $\frac{I_{OUT}}{2f_{CLK}C_S}$, resulting in an average voltage drop of $V_{DR,C} = \frac{I_{OUT}}{2f_{CLK}C_S}$ where the $f_{clk}$ is the clock frequency of the charge pump circuit. In \cite{dcdcconv,dcdcref1}, the steady-state output voltage is given by:
\begin{equation}
V_{OUT} = V_{CLK} + V_C - V_{DR,C} - V_{DR,R}
\end{equation}
where $V_{dd}$ denotes the supply voltage, $V_C$ represents the voltage across the capacitor at the end of charging.  and  $V_{DR,C}$ and $V_{DR,R}$ are the voltage drops across the capacitor and resistance, respectively.

To derive the efficiency equation based on the calculated $V_{out}$ a few assumptions is needed. Firstly, it is assumed that $1/f_{clk}>R_{on}C{eq}$ where $C_{eq}$ is the total capacitor value at $V_{cap}$ as shown in Fig.\ref{DCDCconv1}, Secondly, clock signal is assumed to have between zero and $V_{dd}$. Lastly, efficiency has been calculated for a single stage DC-DC converter where $C_{load}$ is assumed to be at least 10 times bigger than $C_{s}$ resulting in dc values for $I_{out}$, and $V_{out}$. With these assumptions, the efficiency $\eta$ in steady-state is given by \cite{dcdcconv}: 
\begin{equation}
\eta =\frac{P_{out}}{P_{t}+P_{d}+P_{supply}}= \frac{\overline{V_{\text{out}} I_{\text{out}}}
}{2R_{ON} I_{OUT}^2 + C_{P, eq} f_{CLK} V_{DD}^2 + 2V_{DD} I_{OUT} }
\end{equation}

where $P_{t}$ is thermal power consumption and $P_{d}$ indicates dynamic power consumption. Here, $C_{P, eq}$ is the switch's equivalent parasitic capacitance. For each set of design parameters, there is an optimum $I_{OUT}$ where the efficiency reaches its maximum. The optimum $I_{OUT}$ can be calculated as follows: Based on (4.2), $V_{OUT}$ can be written as a linear function of $I_{OUT}$ ($V_{OUT} = K - \beta I_{OUT}$), where $K = V_{CLK} + V_{C}$ and $\beta = \frac{1}{2f_{CLK}C_S} + R_{ON}$. Then by calculating the derivative of efficiency with respect to $I_{OUT}$, the optimum point can be derived.

\begin{equation}
\eta' = \frac{(K-2\beta I_{OUT})D-(4R_{ON} I_{OUT}-2V_{DD})(KI_{OUT}-\beta I_{OUT}^2)}{D^2}
\end{equation}

where D=$2R_{ON} I_{OUT}^2 + C_{P, eq} f_{CLK} V_{DD}^2 + 2V_{DD} I_{OUT}$. Since the effect of each design parameters on the overall efficiency is desired rather than a purely mathematical format, the effect of each design parameter on the peak efficiency and corresponding optimum $I_{out}$ is studied.

\begin{figure*}[!htb]
\centering
\includegraphics[width=0.9\columnwidth]{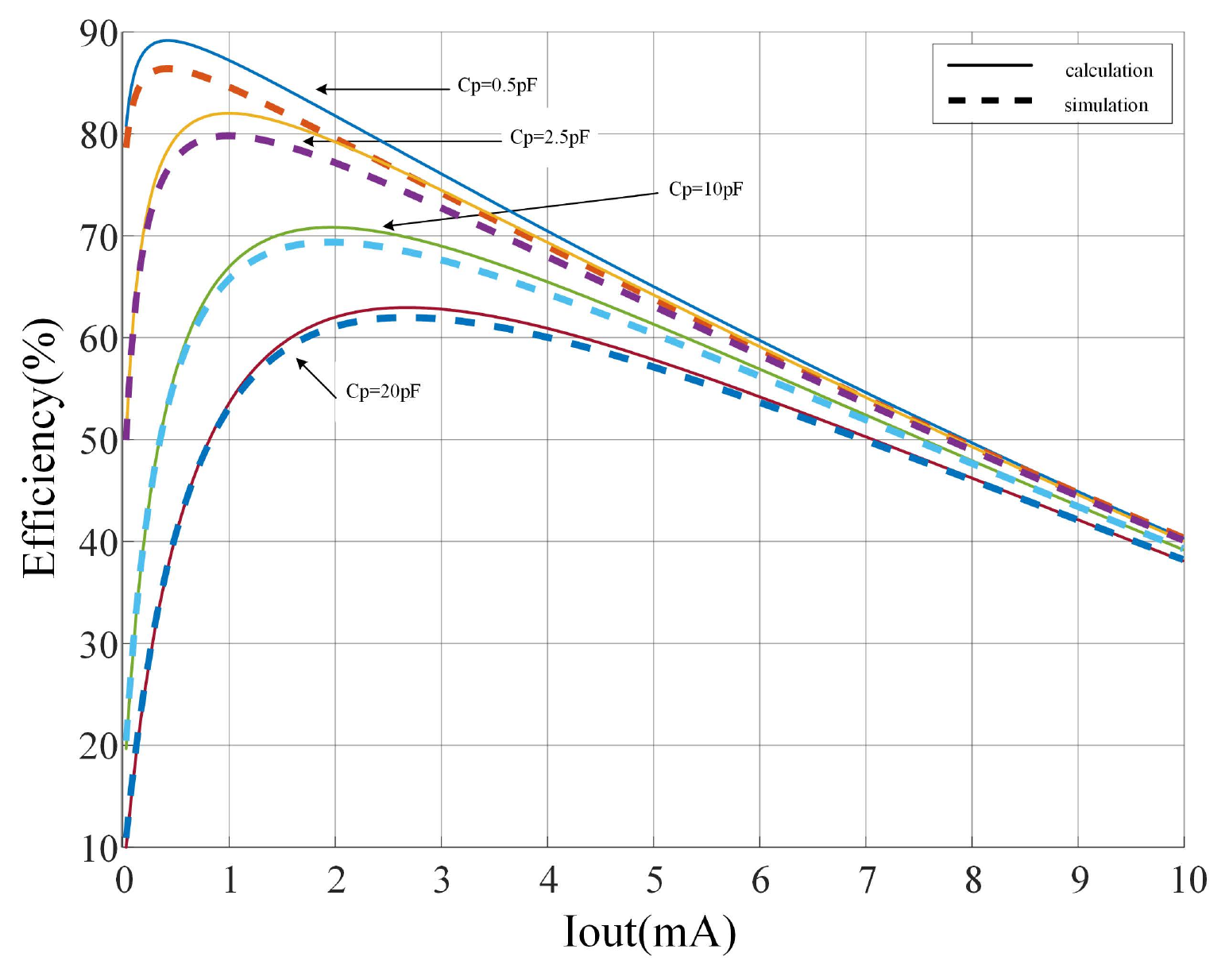}
\caption{effect of Cp on the efficiency}
\label{Cpdcdc}
\end{figure*}

As shown in Fig.~\ref{Cpdcdc}, with increasing the equivalent parasitic cap of switches, the overall efficiency will drop and the optimum $I_{out}$ will increase. Increasing in the parasitic cap, however, will flatten the curve which makes it more suitable for a higher range of current.

\begin{figure*}[!htb]
\centering
\includegraphics[width=0.9\columnwidth]{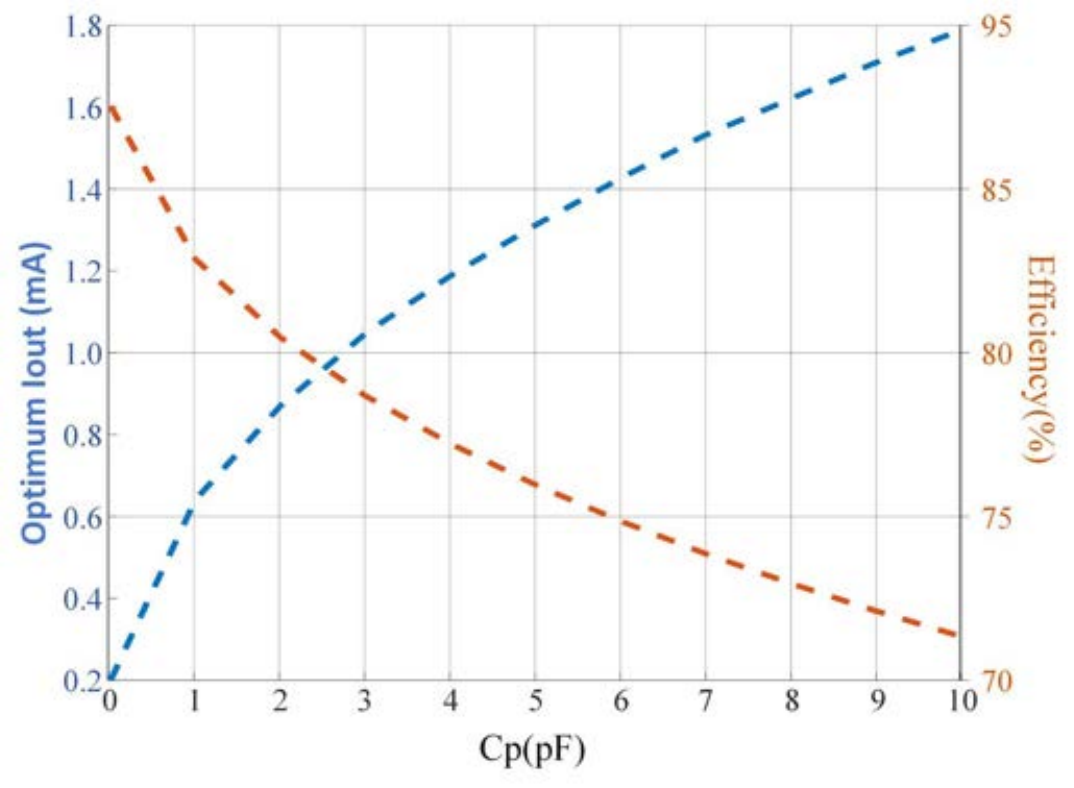}
\caption{Effect of $C_{p}$ on the optimum $I_{out}$ and its corresponding efficiency}
\label{Cpdcdc2}
\end{figure*}

This is illustrated in Fig.~\ref{Cpdcdc2}, where the optimum $I_{out}$ and its corresponding efficiency is shown as a function of the parasitic capacitance. In this simulation, the $R_{on}$ is set to 100 and the $C_{s}$ is set to 100 pF.

 \newpage

\begin{figure*}[!htb]
\centering
\includegraphics[width=0.9\columnwidth]{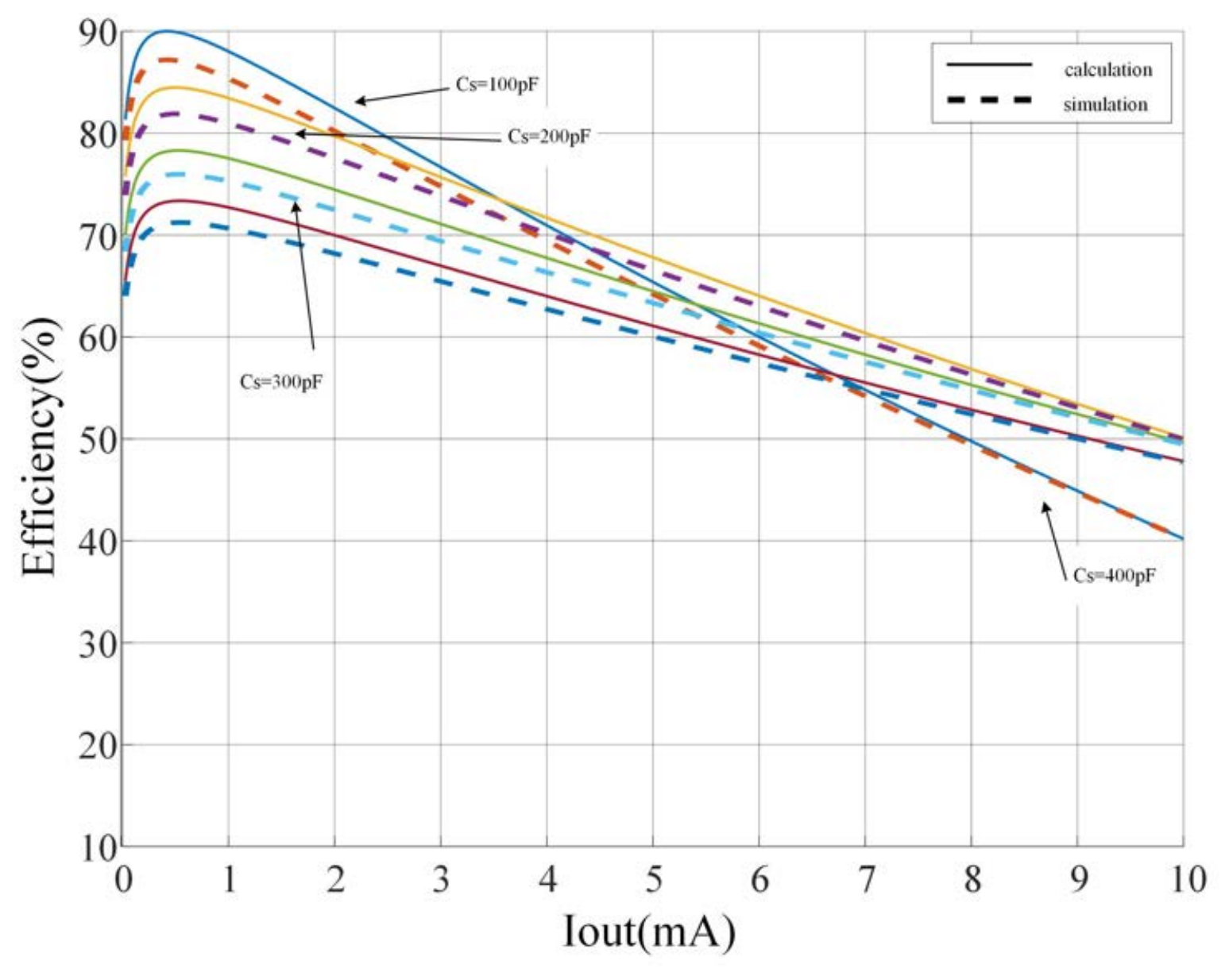}
\caption{effect of Cs on the efficiency}
\label{Csdcdc}
\end{figure*}

\begin{figure*}[!htb]
\centering
\includegraphics[width=0.9\columnwidth]{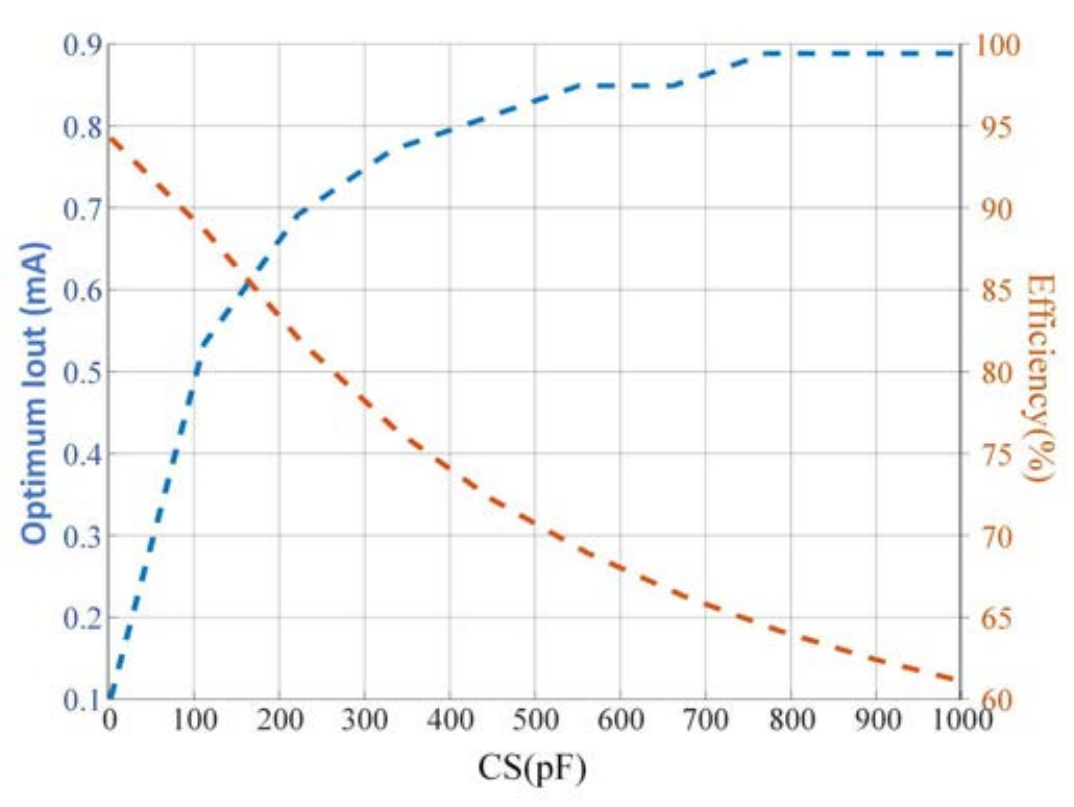}
\caption{effect of Cs on the optimum $I_{out}$ and its corresponding efficiency}
\label{Csdcdc2}
\end{figure*}

The effect of $C_{S}$ is shown in Fig.~\ref{Csdcdc}. Similar to $C_{p}$, by increasing the cap value, the efficiency at the optimum $I_{out}$ will decrease. There is a very close match between the calculation and simulation, which validates the model and calculations. In Fig.~\ref{Csdcdc2}, the optimum $I_{out}$ and its corresponding efficiency is shown as a function of $C_{S}$. For this simulation, $R_{on}$ is set to 100, and $C_{p}$ is set to 500fF, which is based on real simulation of current design. 

For the rest of this chapter it will be assumed that $V_{DD}$ is set to 5V and . The clock frequency, derived from the core chip specification, set to 13.56MHz.

 \newpage

\begin{figure*}[!htb]
\centering
\includegraphics[width=0.9\columnwidth]{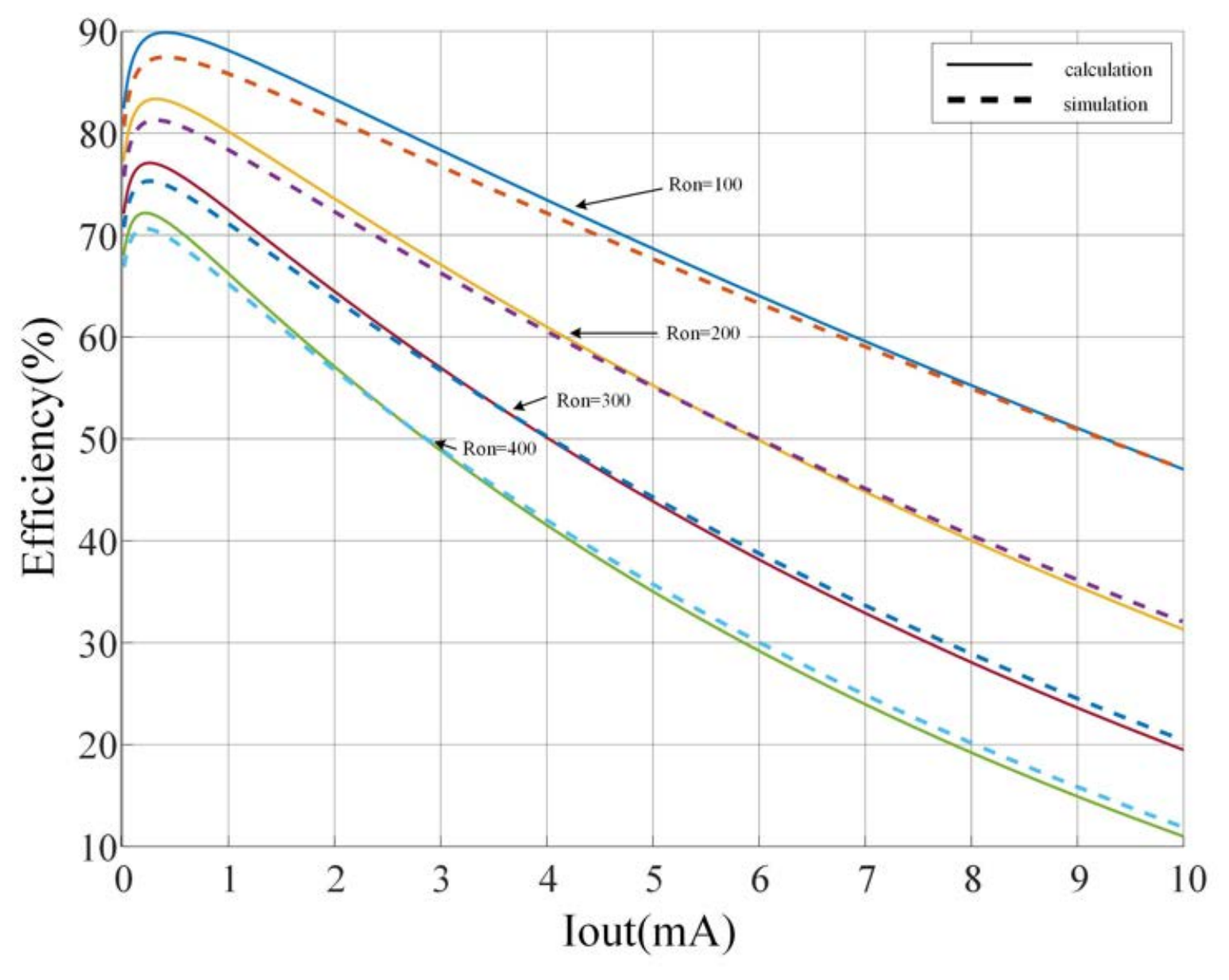}
\caption{effect of Ron on efficiency}
\label{Rondcdc}
\end{figure*}

 \newpage

\begin{figure*}[!htb]
\centering
\includegraphics[width=0.9\columnwidth]{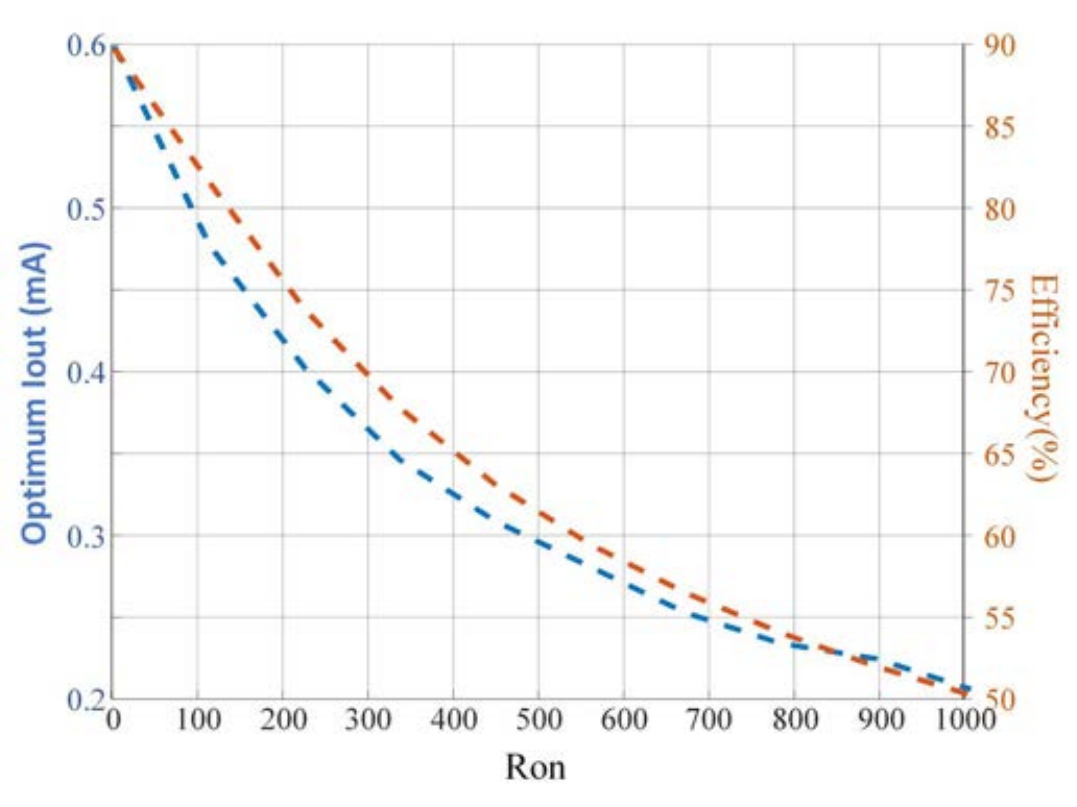}
\caption{effect of Ron on the optimum $I_{out}$ and its corresponding efficiency}
\label{R2dcdc}
\end{figure*}

 \newpage

The other important factor that need to be considered is $R_{on}$, which determines the size of switches, and has a significant impact on the efficiency. Fig.~\ref{Rondcdc} depicts the effect of $R_{on}$ on the efficiency. in a optimal design, $R_{on}$ is desired to be as small as possible,  however, small $R_{on}$ requires large switches resulting in higher parasitic capacitor. effect of $R_{on}$ on the optimum $I_{out}$ and corresponding efficiency is shown in Fig.~\ref{R2dcdc}.

\begin{figure*}[!htb]
\centering
\includegraphics[width=0.9\columnwidth]{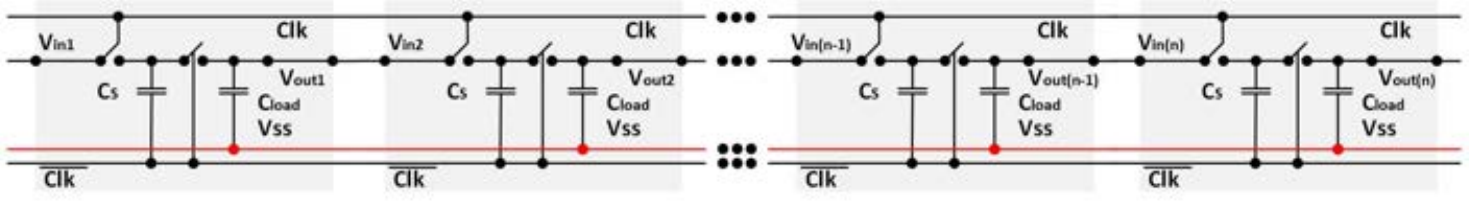}
\caption{Working Principle of Multi stages DC-DC converter}
\label{DCDCMstage2}
\end{figure*}

Fig.~\ref{DCDCMstage2} depicts a typical multi-stage single-output (MSSO) DDC configuration, where each block represents a stage of the charge pump circuit. As it is shown in \cite{dcdcconv,dcdcref1,dcdcref2}, An $N$-stage DDC consumes $(N+1) \times I_{OUT}$ from the supply to generate an output voltage of $(N+1) \times V_{DD}$. This statement which is verified with simulation and is derived from Kirchhoff current law and it is true when $I_{leak}$ for middle stages is equal to zero. $I_{leak}$ is defined as leakage current at the output of each stage of DC-DC converter, which ideally is zero. The detail is shown in Fig.\ref{nstageDCDC}. For this design, providing the required 20V under a heavy load as determined by the ear model is required, thus, a 7-stage DC-DC converter has been developed. The number of stages are chosen based on simulation results over the PVT. Considering relatively high value of $I_{out}$, each stage is boosting the input voltage by approximately around 3V. The switch size is selected based on two criteria: Firstly, to present low on-resistance, which is demonstrated to enhance the efficiency of the DC-DC converter and boost the output voltage. Secondly, to avoid reducing the output voltage due to the large parasitic capacitance of the switches. The capacitance value of the last stage is set significantly higher as an off-chip capacitor will be connected to this node, which ensures a reduction in the voltage and current ripple value. To ensure that the output voltage closely approximates 20V, the multi-stage DC-DC converter is incorporated within a simple feedback loop. The behavior of the feedback loop will be discussed in the following chapter. With some minor changes to the expression for the unit cell efficiency as was defined in \cite{dcdcconv}, the efficiency formula for a multi-stage DC-DC converter is given as:

\begin{figure*}[!htb]
\centering
\includegraphics[width=0.5\columnwidth]{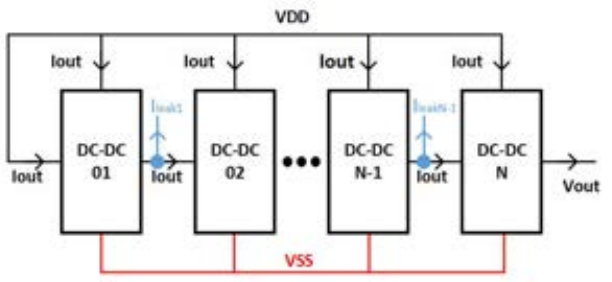}
\caption{Current flow for multi stage DC-DC converter- clock signals are not shown \cite{dcdcconv}}
\label{nstageDCDC}
\end{figure*}

\begin{equation}
\eta = \frac{\overline{V_{\text{out}} I_{\text{out}}}}{nC_{P,eq} f_{CLK} V_{DD}^2 + 2nR_{ON} I_{OUT}^2 + (n+1)V_{DD} I_{OUT}}
\end{equation}
where $\overline{V_{\text{out}} I_{\text{out}}}$ is the average $P_{out}$ of load connected to the output of last stage and 
\begin{equation}
V_{OUT} = V_{CLK} + nV_C - nV_{DR,C} - nV_{DR,R}
\end{equation}

\begin{figure*}[!htb]
\centering
\includegraphics[width=1\columnwidth]{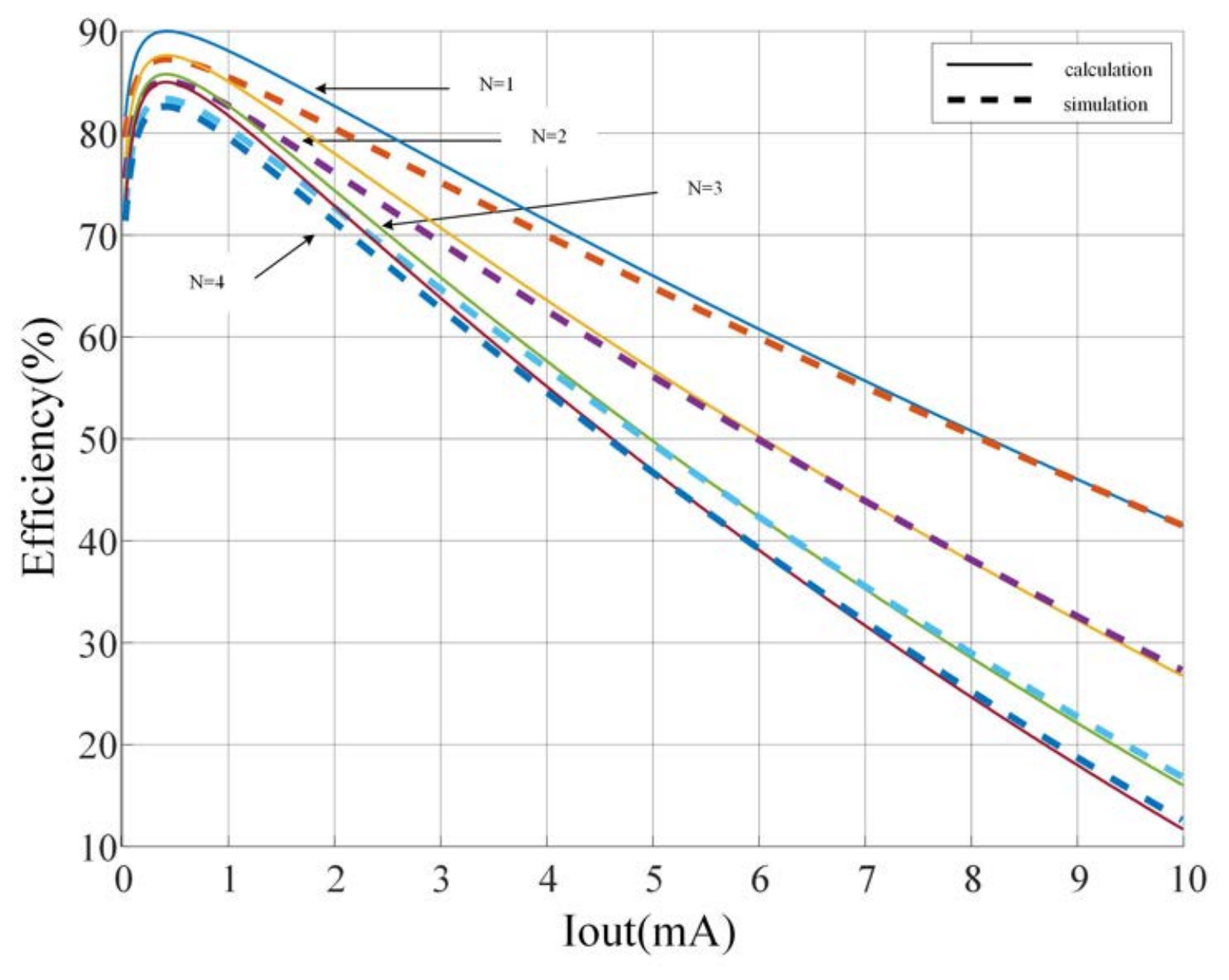}
\caption{Working Principle of multi-stage DCDC converter}
\label{Ndcdc}
\end{figure*}

Fig.~\ref{Ndcdc} shows the effect of number of stages (n) on the overall efficiency. as it is show, increasing the number of stage will not change the optimum $I_{out}$ but will lower the efficiency at the optimum $I_{out}$. 

\subsubsection{proposed DC-DC converter with on-chip feedback loop}

\begin{figure*}[!htb]
\centering
\includegraphics[width=0.9\columnwidth]{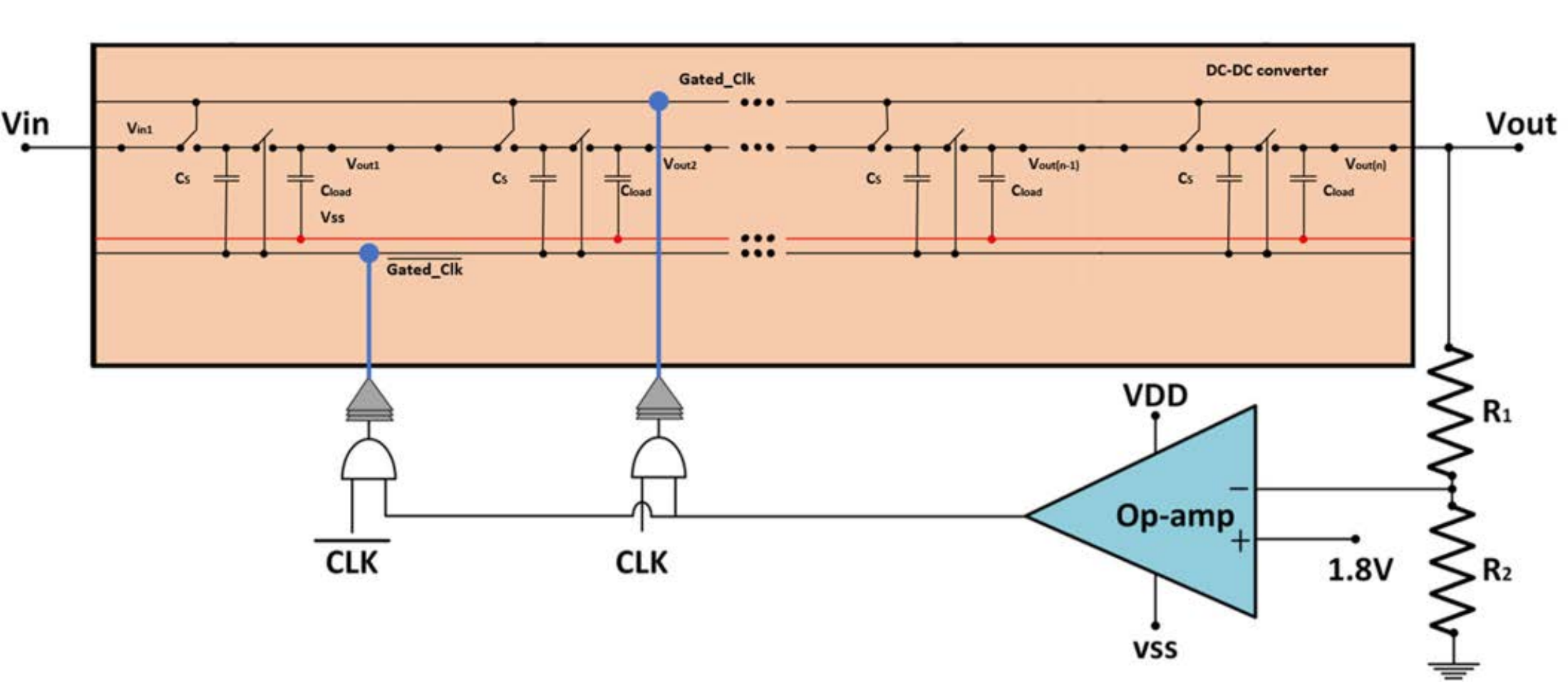}
\caption{Working Principle of proposed DC DC converter feedback loop}
\label{DCDCMstage}
\end{figure*}

As shown in the Fig.~\ref{DCDCMstage} , a simple feedback loop is implemented to ensure the output voltage remains close to the desired value, in this case, 20V. A portion of the output voltage is sampled by the \(\frac{R_2}{R_1 + R_2}\) ratio and then compared to a fixed voltage (in this case, 1.8V). The output signal determines whether the clk/clkb signal will be applied to the DC/DC converter.  As it is shown in Fig.\ref{DCDCMstage}, the polarity of the op-amp is chosen so If the output voltage exceeds 20V, then the output of the op-amp will be set to zero, setting the input clock of the DC/DC converter to zero. Under this condition, the output voltage will drop. This will continue until the output of the op-amp flips, which will turn on the multi-stage DC/DC converter. The values of \(R_1\) and \(R_2\) are set to \(R_2 = 1.1 \, k\Omega\) and \(R_1 = 10 \, k\Omega\).

\subsubsection{Design of the $V_{DD}$-5V voltage generator}

Recent advances in power electronics have led to the widespread adoption of monolithic power converters manufactured using high-voltage CMOS BCD technology. However, a notable limitation of this technology is the maximum permissible gate-to-source voltages, typically limited to 5.5V. To effectively operate the high-voltage PMOS transistors used as switches, their gates must be biased at a voltage that is no lower than the difference between the core chip's supply voltage (e.g., 20V) and a safe source-to-gate voltage (e.g., 5V, just below the maximum limit). This biasing requires no DC current from the voltage source, yet it must be precise and stable despite variations in process, voltage, and temperature (PVT). This subsection describes a circuit specifically designed to generate this essential bias voltage reliably for the core chip. It is worth mentioning that the maximum allowable voltage between each two terminals is 5V.

\begin{figure*}[!htb]
\centering
\includegraphics[width=0.75\columnwidth]{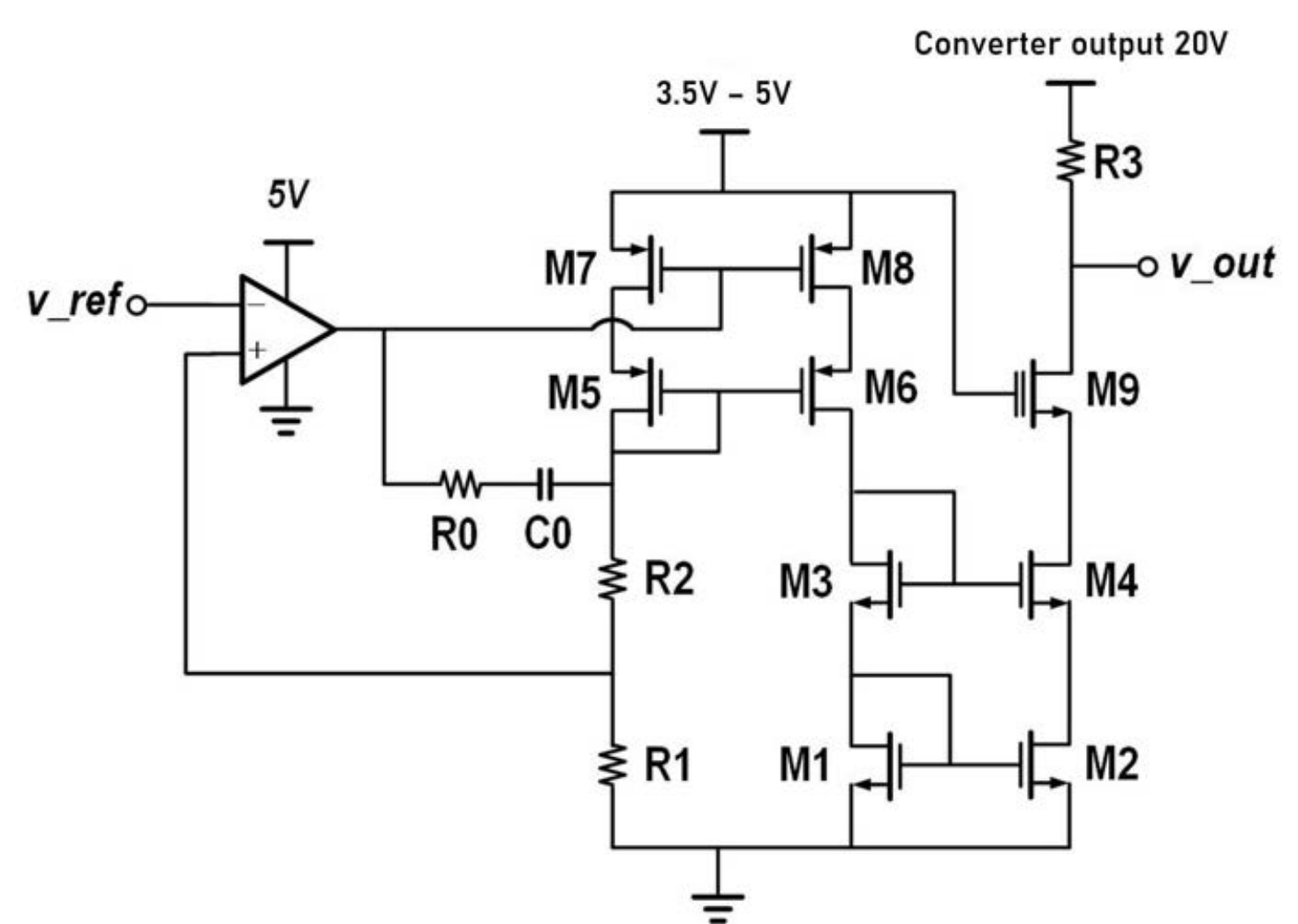}
\caption{Circuit design for the $V_{DD}-5V$ Voltage Source.}
\label{fig:Vdd-5V Generator}
\end{figure*}

Figure~\ref{fig:Vdd-5V Generator} illustrates the circuit designed to generate the specified bias voltage, where $v_{ref}d = 1.25V$ originates from a bandgap voltage reference. This circuit receives somewhere between 3.5V to 5V and 20V inputs, sourced from rectifier and boost DC-DC converter, respectively. In the schematic, transistors M1 through M4, along with M5 through M8, form two sets of unity-gain current mirrors. 
A Miller compensation network, consisting of components R0 and C0, is used to separate the poles of the system. This network positions the dominant pole at the output node of the operational amplifier, maintaining a phase margin consistently above 60 degrees across all PVT conditions.

\begin{equation}
\label{eq:vdd_minus5_equation}
v\_out = 20V - V\_{ref} \frac{R_{3}}{R_{1}}
\end{equation}

With $v_{ref} = 1.25V$ and $R_{3} = 4R_{1}$, $v_{out}$ is calculated to be 15V. If resistors R1 and R3 are precisely matched and positioned closely together in the physical layout, and the current mirrors function with high accuracy, the voltage drop across R3 remains consistently close to 5V relative to the 20V supply. This configuration ensures that $v_{out}$ stays approximately 5V below the supply voltage.

\begin{figure*}[!htb]
\centering
\includegraphics[width=0.9\columnwidth]{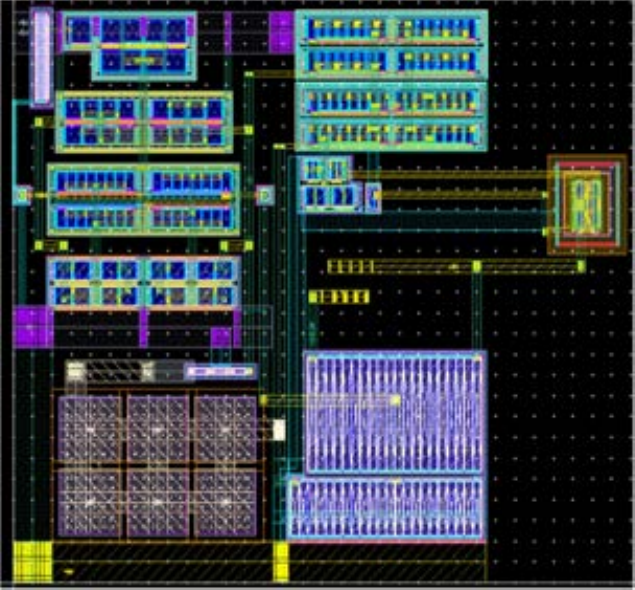}
\caption{Circuit Layout design for the $V_{DD}-5V$ Voltage Source.}
\label{fig:VOLTAGEFOLLOWER}
\end{figure*}

\clearpage

\subsection{Simulation results}

Figure~\ref{fig:VOLTAGEFOLLOWER} depict the simulation results for the post layout simulation of the chip. As it is shown, both high voltages rise when chip is powered on. The difference is shown in green which reach to its steady-state point in less than 20$\mu$s. This will guarantee that the difference between two high voltage nodes always stays within the desired range. As the 20V output voltage reaches its optimum point, which depends on $I_{out}$, there will be a swing around the optimum voltage. This is needed to be observed and chosen depending on the application. In this work, the off chip component is set to keep the output voltage always within 1 volt of desired voltage after reaching to its nominal voltage.

\begin{figure*}[!htb]
\centering
\includegraphics[width=0.9\columnwidth]{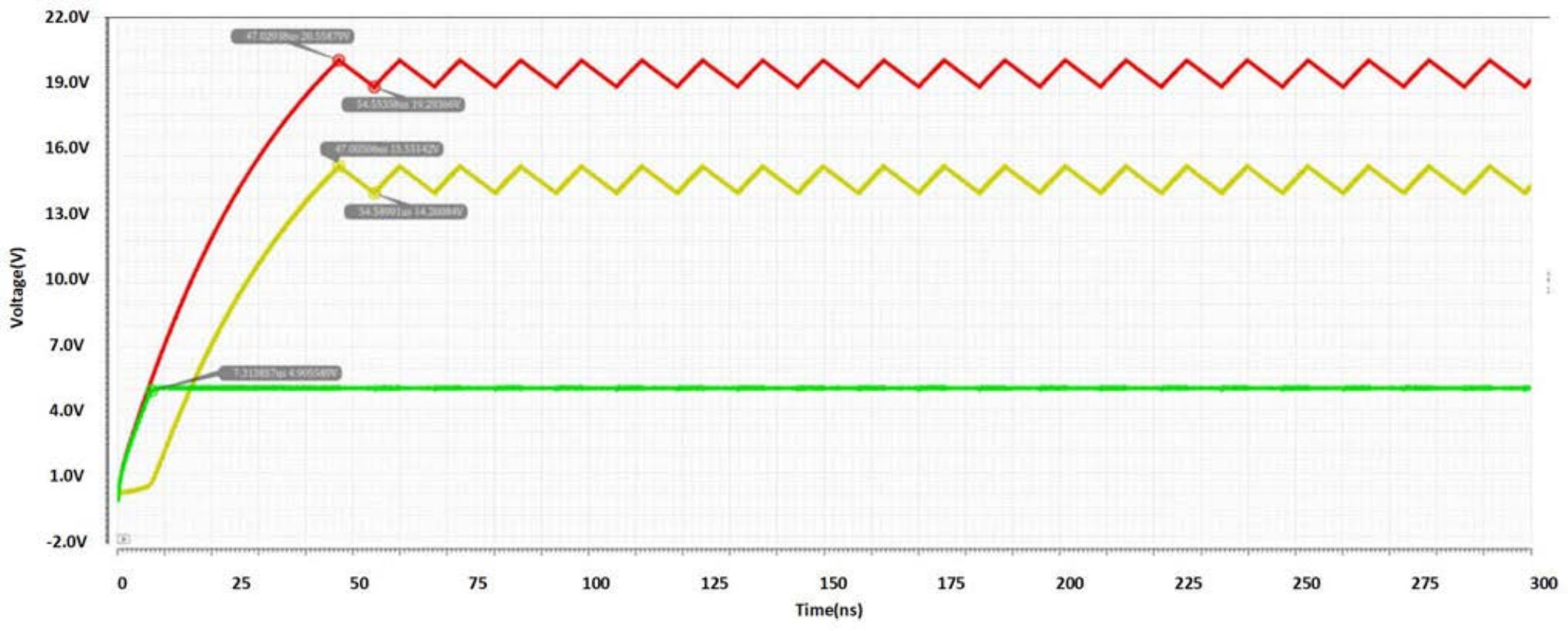}
\caption{Simulation results of DC-DC converter loop with feedback}
\label{fig:VOLTAGEFOLLOWER}
\end{figure*}

\begin{figure*}[!htb]
\centering
\includegraphics[width=0.9\columnwidth]{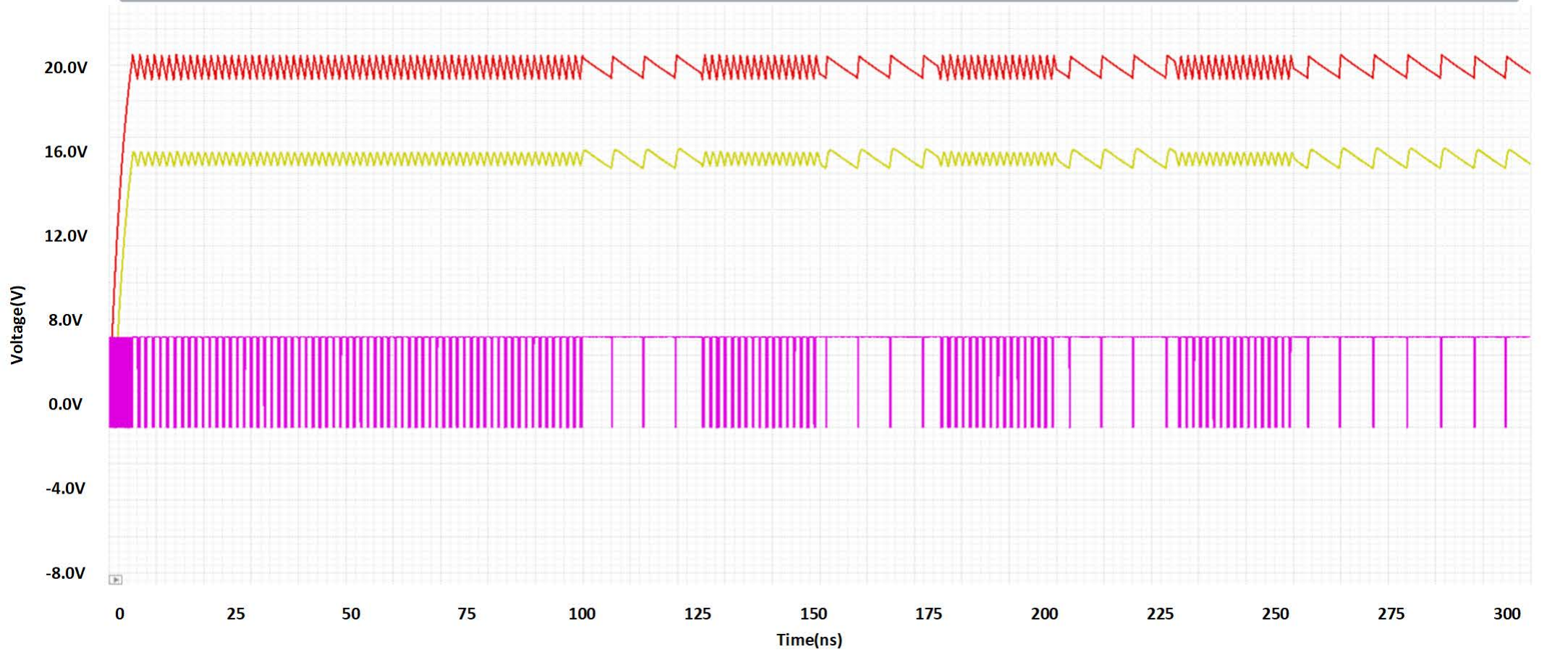}
\caption{Operation of feedback loop with corresponding clock.}
\label{fig:LOOPPERFORMANCE}
\end{figure*}

Figure~\ref{fig:LOOPPERFORMANCE} depicts the simulation results for the post layout simulation of the feedback loop in more detail for different load. As it is shown, in the first phase where the load is heavier, the output voltage tends to decrease more quickly when the dc-dc converter is off which leads to more frequent charging and discharging. In either case, the output ripple is purely the same as it is a function of the feedback loop parameters.

\begin{figure*}[!htb]
\centering
\includegraphics[width=0.9\columnwidth]{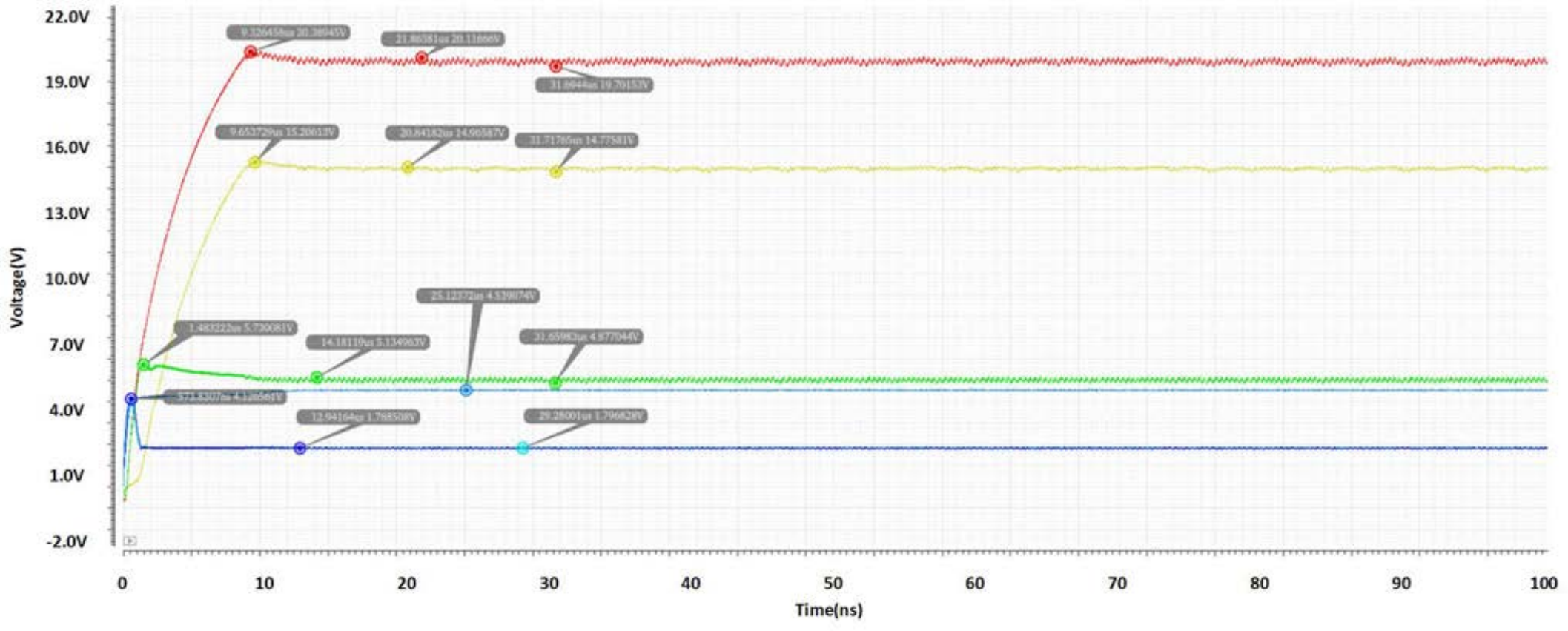}
\caption{Operation of High voltage and low Voltage blocks with 1nF off-chip capacitor connected to the output.}
\label{fig:SimulationResultsdcdc}
\end{figure*}

Figure~\ref{fig:SimulationResultsdcdc} depicts the simulation results for the post layout simulation of both high voltage and low voltage part. As shown, the chip is providing both high voltage and low voltages needed to power up the core chip.

\newpage

\section{Layout of the supply chip}

\begin{figure*}[!htb]
\centering
\includegraphics[width=0.8\columnwidth]{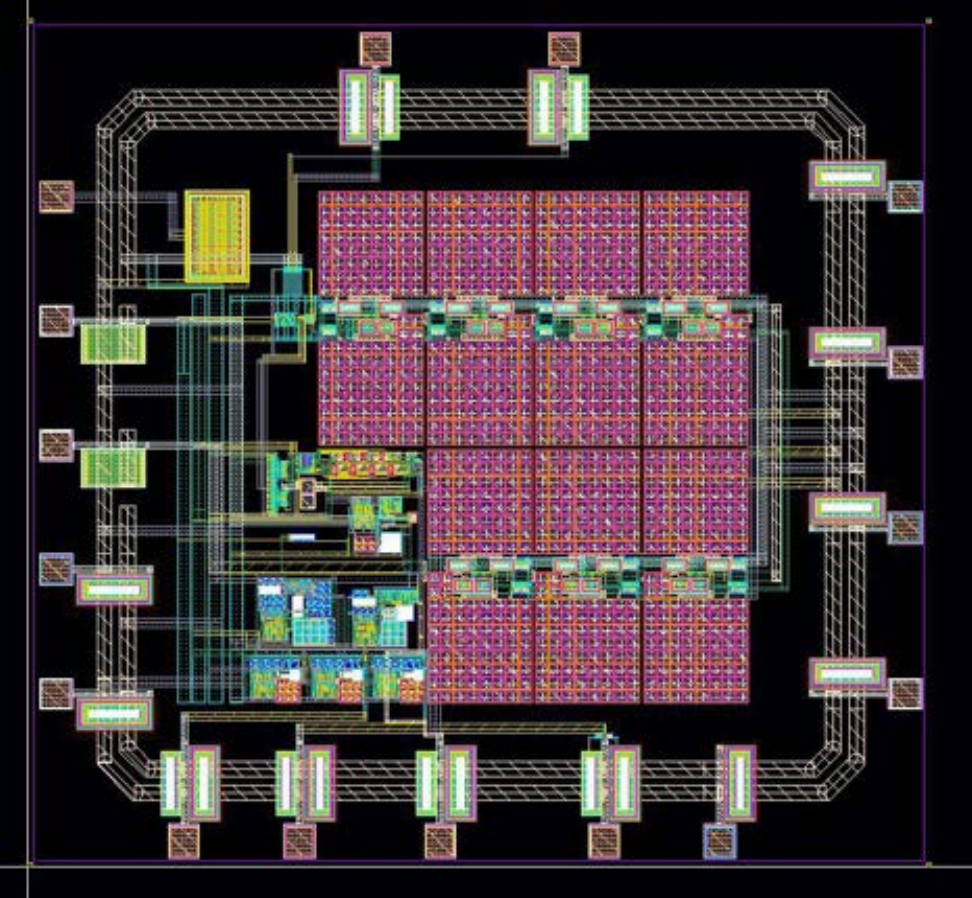}
\caption{Layout of supply chip.}
\label{Layout of supply chip}
\end{figure*}

Similar to the core chip, the supply chip has also been laid out manually using the Cadenced Virtuoso layout editor tool as shown in Fig.~\ref{Layout of supply chip}. Compliance checks for design rules, layout vs. schematic, antenna rule, as well as parasitic extraction, were all performed using the Mentor Graphics Calibre tools. With parasitic resistors and capacitors been included, post-layout simulation shows the chip functions as expected. the die of supply chip is fabricated successfully with size of 2.34mm by 2.33mm.

\chapter{Integrated Module prototype}

\begin{figure*}[!htb]
\centering
\includegraphics[width=1\columnwidth]{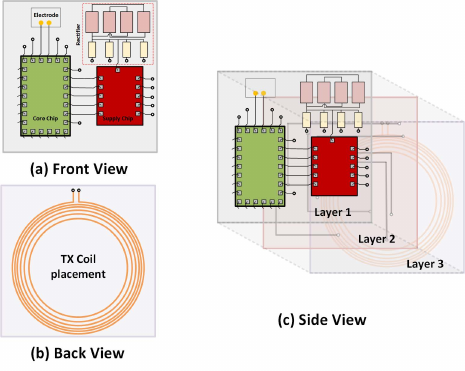}
\caption{Block diagram of supply chip.}
\label{Module}
\end{figure*}

To create a 3-D module with the correct form factor(4mmX4mmX3mm) including the supply chip and the core chip and the TX coil, a detailed PCB design is needed. One possible option is shown in Fig.\ref{Module}. Two chips are placed side by side with a small margin for routing. On the back side, the designed TX coil, with a diameter of 4mm, is implemented. To save space on the TX module, all the matching network has been put on the RX side. Based on initial measurements, we have verified that an off-chip rectifier is needed for the top module as the on-chip rectifier shows poor efficiency due to its structure. The four diode makes a full bridge rectifier are placed on the front side of the module as shown in Fig.\ref{Module}(a). The middle layer is used for the necessary routing between the chips and RX coil. The current module form factor is a cubic structure with a 8mmx8mmx4mm size. The larger than target form factor is due to the large die size of supply and core chip. This can be reduced in the future steps simply by changing the fabrication node. Based on initial calculations, moving to a smaller fabrication node, allows a reduction in the area by factor of 4. This will enable to have the final module size to be 4mmx4mmx4mm. 
\clearpage

For the coil design, a numerous simulation has been done to verify the matching network performance of the RX/TX coil.

 \chapter{Measurement}


\clearpage

\section{Measurement setup}

\begin{figure}[!htb]
\centering
\includegraphics[width=0.75\columnwidth]
{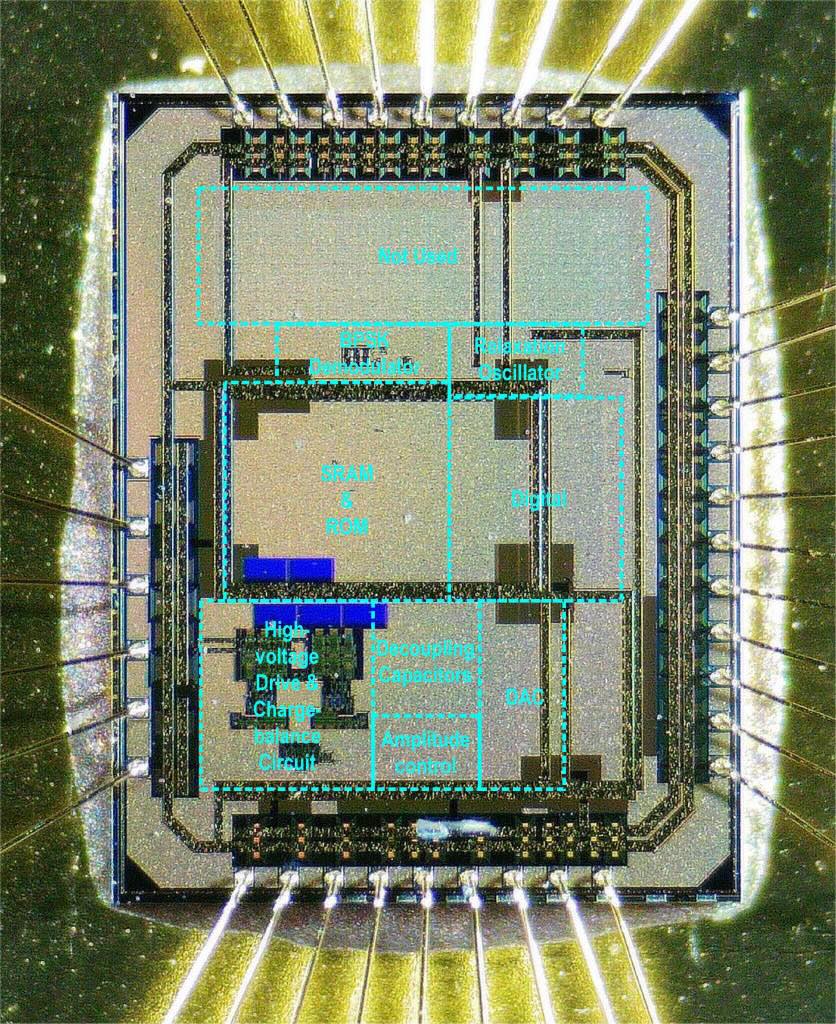}
\caption{Core chip die photo.}
\label{Core chip die photo}
\end{figure}

For this project, the two main integrated circuits have been fabricated, both using the TSMC 180nm BCD G2 process. A die photograph of the core chip, shown in Fig.~\ref{Core chip die photo}. Showing dimensions of a 2.34mm × 3mm. We note that there is an unused area near the top of the chip due to manufacturing requirements for the multi-project wafers. The digital portion of the chip - digital and memory- are placed in the center of the die, while all the high-voltage portion placed in the lower part of the die. In the future work, the core chip die area can be reduced if a smaller technology node is chosen for fabrication, that will potentially help meet the form factor specification.

\begin{figure}[!htb]
\centering
\includegraphics[width=0.75\columnwidth]
{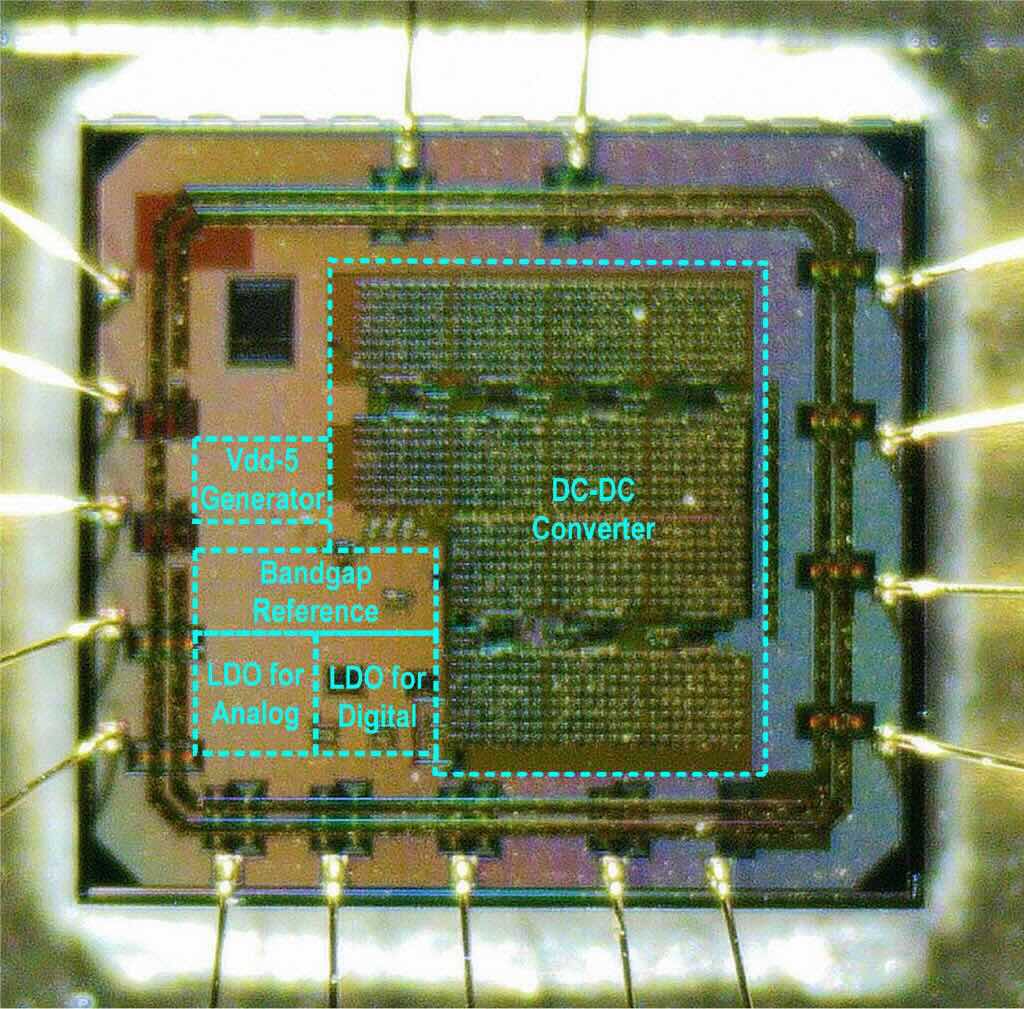}
\caption{Supply chip die photo.}
\label{Supply chip die photo}
\end{figure}

A die photo of supply chip, measuring 2.13mm x 2mm, is shown in Fig.~\ref{Supply chip die photo}. Due to the requirement for high  value of metal-insulator-metal (MIM) capacitors in the DC-DC converter consumes approximately 75\% the chip area. The reminder consists of low voltage circuitry, including - bandgap reference, LDO and Vdd-5 generator. Due to the multi-module nature of the design, the measurement has been done in a few separate steps. Each step will be discussed separately.

\begin{figure}[!ht]
\centering
\includegraphics[width=0.85\columnwidth]
{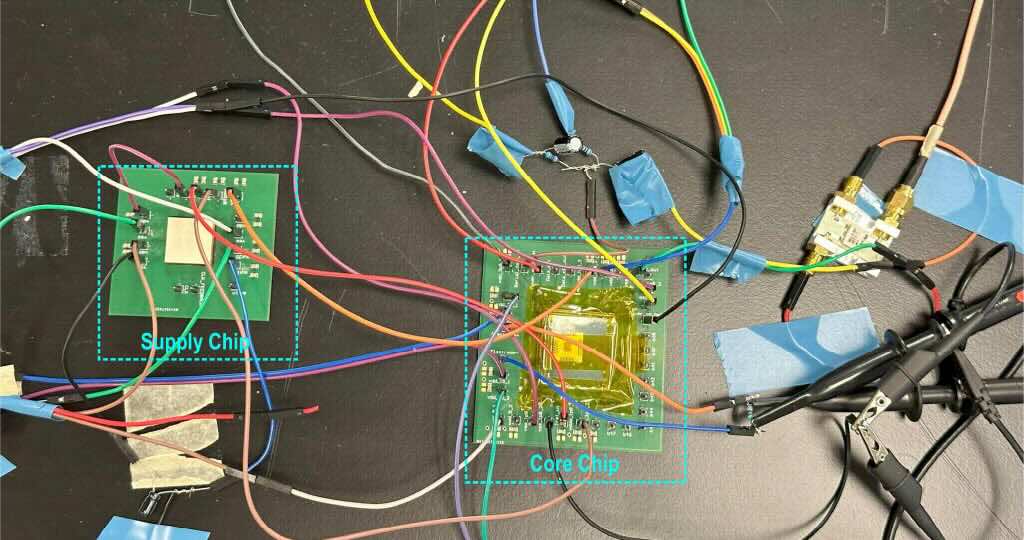}
\caption{Core chip supply chip connected.}
\label{Core chip supply chip connected}
\end{figure}

In the complete setup, the core chip plays the role of the load of the supply chip. In this measurement an equivalent resistive load of 15k has been connected as the supply chip load. Measurement had been successfully finished with relatively close match with simulated results.A Raspberry Pi device is used to transmit, to transmit the modulated signal to the core chip. A digital oscilloscope is utilized to probe the differential voltage waveform across the load. After successfully measuring each chip seperately, in the second step, the two printed-circuit boards (PCB) are connected together, using jump wires, to demonstrate a step closer to the final expected module as shown in Fig.~\ref{Core chip supply chip connected}. As mentioned earlier the purpose of the supply chip is to provide the 5-V, 15-V, 20-V high-voltage supplies, the analog 1.8-V supply, the digital 1.8-V supply, the 5$\mu$A reference current to the core chip. For the third and final part a few changes has been done towards the final single module board. First, instead of directly transmitting the modulated data to the supply chip using wire, a RX/TX pair of coils with corresponding matching network has been implemented. The matching network topology has been carefully selected based on both numerous simulation and specifications. Second, a single PCB with the required form factor has been designed which allows putting the chips close together while implementing the RX coil on the same PCB. The measurement results for the second and third phases is shown in fig.~\ref{Measurement results with pig-ear load}

The measurements were conducted using two types of loads designed to mimic human inner-ear tissue: The first load comprises a 3-k$\Omega$ resistor in series with a 100-nF capacitor, representing the human inner-ear model; the second load consists of a section of pig-ear tissue.

\clearpage

\section{Measurement results}


The total power consumption of the core chip at maximum stimulation mode is 33 mW, with the high-voltage drive and charge-balance block accounting for 99.5\%, the DAC for 0.4\%, and all digital circuits and other blocks for 0.1\%. During the measurement of charge-balance precision, a 3-k$\Omega$ resistor in series with a 100-nF capacitor served as the load. The chip was switched from stimulation mode to charge-balance mode, and a multi-meter measured no more than 0.1 mV across the load. To demonstrate the chip's ability to generate arbitrary current stimulation waveforms with tunable amplitudes, consecutive cycles, and duration of charge-balance time, several measurements were conducted. The oscilloscope captured the differential voltages across the loads.

\begin{figure}[!ht]
\centering
\includegraphics[width=0.85\columnwidth]
{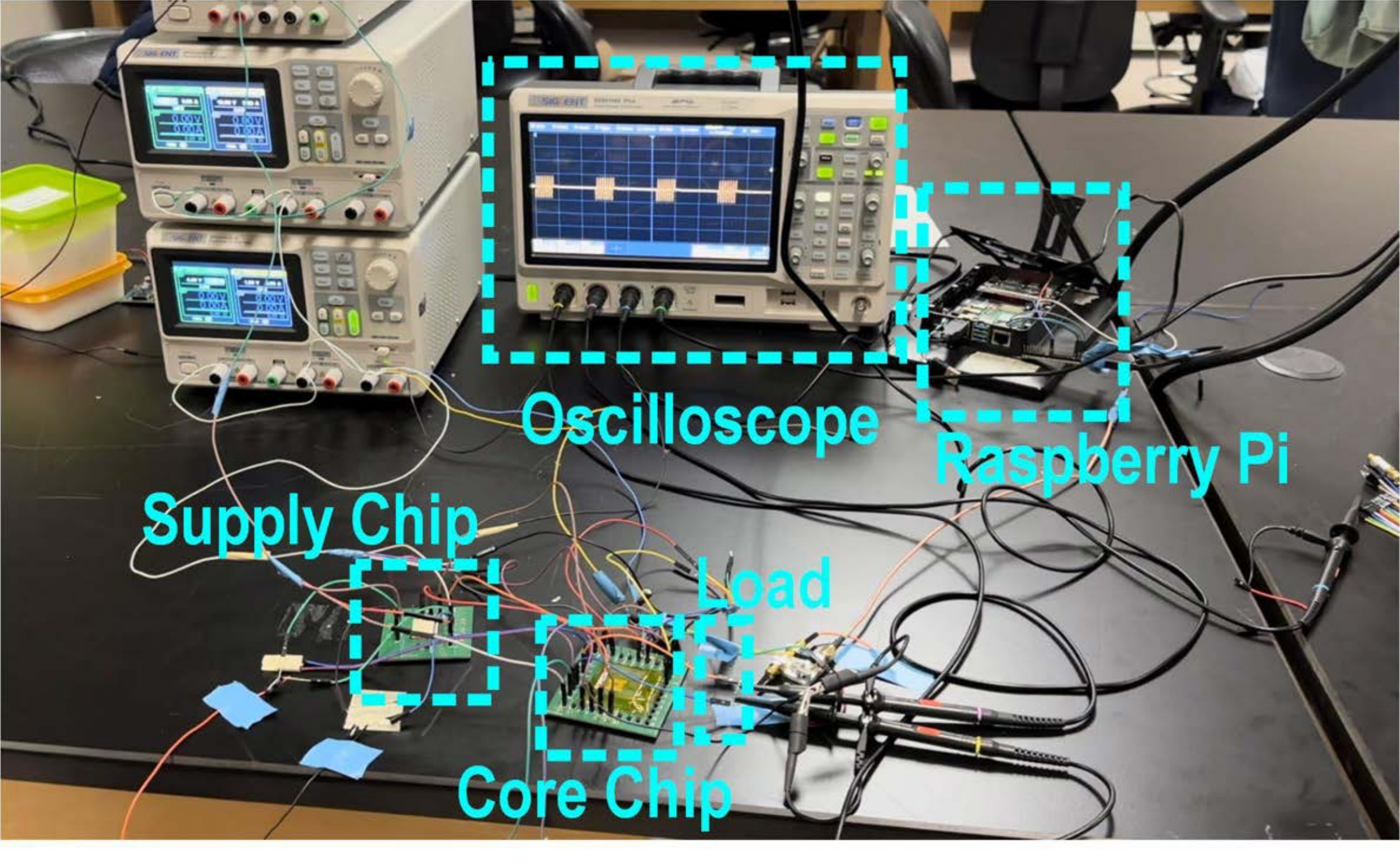}
\caption{Entire measurement setup.}
\label{Entire measurement setup}
\end{figure}

\begin{figure}[!ht]
\centering
\includegraphics[width=1.0\columnwidth]
{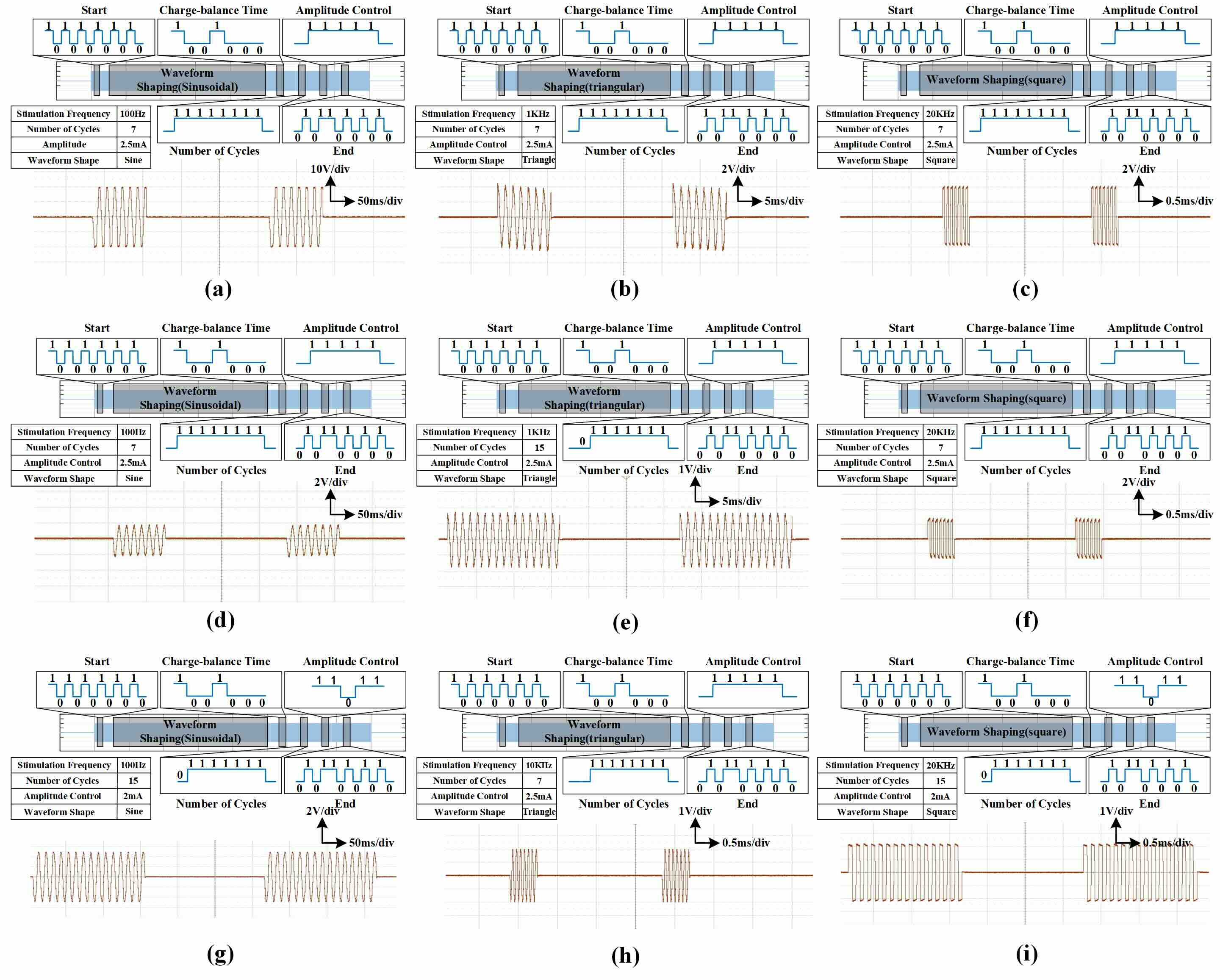}
\caption{Measurement results with RC load.}
\label{Measurement results with RC load}
\end{figure}

For the results shown in Fig.~\ref{Measurement results with RC load}, the load is the series-RC. The top digital signal waveform of each figure shows the associated input digital signals that encode the waveform parameters. The waveform specification is shown in the table associated with each figure. Observing Fig.~\ref{Measurement results with RC load} (a), (b), (c), we proved that a variety of wave shapes with different stimulation frequencies can be generated, including a sinusoidal waveform with 100Hz stimulation frequency, a triangular waveform with 1kHz stimulation frequency, and a square waveform with 20kHz stimulation frequency. The number of consecutive simulation cycles for all figures is 7 and the amplitude of all waveforms is 2.5mA. Comparing Fig.~\ref{Measurement results with RC load} (d) to Fig.~\ref{Measurement results with RC load} (a), the number of stimulation cycles is increased to 15 and the amplitude is reduced to 2mA. For stimulation waveform shown in Fig.~\ref{Measurement results with RC load} (e), its amplitude and stimulation frequency is same as that shown in  Fig.~\ref{Measurement results with RC load} (b); however, its charge-balance duration is twice of that shown in Fig.~\ref{Measurement results with RC load} (b). Comparing Fig.~\ref{Measurement results with RC load} (f) to Fig.~\ref{Measurement results with RC load} (c), its number of cycles is increased to 15 and its amplitude is reduced to 2mA.

\begin{figure}[!ht]
\centering
\includegraphics[width=0.8\columnwidth]
{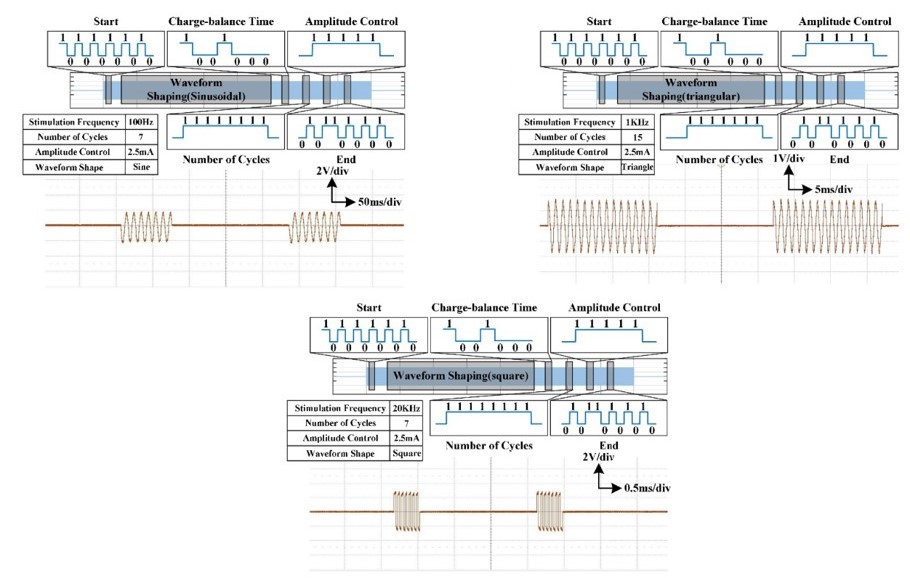}
\caption{Measurement results with pig-ear load.}
\label{Measurement results with pig-ear load}
\end{figure}

Fig.~\ref{Measurement results with pig-ear load} shows the measurements with pig-ear tissue load. In this figure, three representative waveforms are shown, namely, sinusoidal, triangular, square. With these measurements, we proved that the chip has enough driving capability to the real pig-ear tissue. Comparing Fig.~\ref{Measurement results with RC load} to Fig.~\ref{Measurement results with pig-ear load}, it can be observed that the impedance of the pig-ear tissue is lower than that of the human inner-ear series-RC model. These measured waveforms confirm the effectiveness of the proposed design under varying loads, the ability to program the current stimulation waveform, and the charge balance period. 

Comparison of this work with prior works is shown in Table~\ref{Comparision table}. The chip performs extremely well considering the charge-balance precision when comparing with prior works; this is as expected thanks to the novel high-voltage drive and charge-balance circuit.

A standalone evaluation of the supply chip was conducted to assess its performance. Seven assembled chips were measured at room temperature. The measurement indicates that the bandgap voltage reference, bandgap current reference, LDOs, and the $V_{DD}$-5V voltage generator all operate as expected. Furthermore, the voltage and current accuracy of all seven chips fall within the designed $\pm3\%$ variation.

\begin{table*}[!ht]
\begin{center}

{\small
\caption{\small Comparison of This Core Chip with Prior Works \normalsize}

\label{Comparision table}
\resizebox{\textwidth}{!}{\begin{tabular}{|c|c|c|c|c|c|c|c|c|c|}

\hline
&{Number of}&{Maximum}& {DAC}&{Charge }&{Charge }&{Voltage}&{On-Chip}&{Implemented}\\
{Reference}&{Channels}&{Stimulation}&{Resolution}&{Balance}& {Balance}&{Compliance}& {Control} &{On-chip}\\
&{(N)}&{ Current(mA)}&{(bits)}&{Method}&{Precision(mV)}&{(V)}&{Circuit}&{Memory}\\
\hline
\cite{ref5}&1&3&4&DCM+Passive&$\pm$9&12&No&No\\
\hline
\cite{ref6}&6&10&9&OR&$\pm20$&49&No&Yes\\
\hline
\cite{ref7}&1&5.12&9&IPCC+OR&$\pm20$&22&No&No\\
\hline
\cite{ref8}&8&0.775&5&CPI&$\pm50$&4&Yes&No\\
\hline
\cite{ref9}&1&0.2&-&No&-&1&No&No\\
\hline
\cite{dac}&16&12.75&8&TBCB&$\pm2$&40&No&No\\
\hline
This Work&1&1.25&8&Passive&$\pm0.1$&20&Yes&Yes\\
\hline
\hline

\end{tabular}
}
}
\end{center}
\end{table*}


 \chapter{Conclusion and Future Work}


\clearpage

\section{Conclusion}

Tinnitus, characterized by the perception of sound in the absence of external stimuli, poses a significant challenge in the field of auditory medicine. Traditional treatment methods often involve cumbersome equipment and may yield varying degrees of success. However, a paradigm shift is underway with the emergence of an integrated solution that promises to revolutionize tinnitus therapy. At the heart of this revolutionary approach lies an integrated solution designed to streamline tinnitus treatment. Unlike traditional methods that require specialized equipment and settings, this solution empowers clinicians to administer therapy directly within their office environments. By eliminating the need for extensive equipment and complex procedures, it enhances the accessibility of treatment and simplifies the therapeutic journey for patients.

Central to the effectiveness of this integrated solution is its comprehensive capabilities, validated through meticulous measurement and analysis. The solution boasts the ability to produce customizable arbitrary current stimulus waveforms with remarkable precision. Clinicians can tailor treatments to the specific needs of individual patients, adjusting parameters such as waveform amplitude and frequency with unparalleled flexibility. This customization ensures that therapy is optimized for each patient, maximizing the likelihood of successful outcomes.

The seamless and efficient approach offered by this integrated solution holds profound implications for patient care. By simplifying the treatment process and enhancing accessibility, it elevates the standard of care for individuals suffering from tinnitus. Patients no longer face the barriers associated with traditional treatment methods, such as undergoing complex procedures. Instead, they can receive therapy conveniently within their clinician's office, leading to improved adherence and overall satisfaction with treatment. The measurement results

\section{Future work}

The successful fabrication of our latest chips, following the taping-out process, serves as a testament to the validity of our innovative concept. Moving forward, we're set to propel our project to greater heights. We've devised a comprehensive plan aimed at refining our chip design, utilizing flip-chip pads, and embracing the full-block type-out service. Moreover, we're integrating specialized processes to manufacture the MCM board, which will serve as the cornerstone for integrating our chips.

Our vision entails mounting the two chips onto the MCM board, a pivotal step in ensuring seamless functionality and enhanced performance. To uphold the utmost standards of safety and reliability, we're committed to encapsulating the entire module in a specialized medical-graded silicon package. This meticulous approach underscores our unwavering commitment to delivering a product that not only meets stringent quality standards but also complies with regulatory requirements.

As we progress towards clinical validation, we embark on a crucial phase of our research journey. Live animal experiments, specifically conducted with pigs, will furnish invaluable insights and data to validate the efficacy and safety of our device. These experiments serve as a vital precursor to human trials, guiding us in refining our approach and surmounting potential challenges.

Buoyed by promising results from animal experiments, our ultimate aim is to transition to human trials and seek approval from the Food and Drug Administration (FDA). This milestone signifies the culmination of years of dedication, research, and collaboration. By navigating the regulatory landscape and adhering to stringent standards, we're poised to introduce our transformative technology to the market, ultimately enhancing the lives of patients grappling with the debilitating effects of our target condition.

In essence, our strategic roadmap charts a comprehensive path towards advancing our technology from conception to commercialization. Through meticulous planning, rigorous testing, and an unyielding commitment to excellence, we're steadfast in our conviction to revolutionize the field and make a tangible impact on patient care.




\chapter{Ultra Low Noise VCO - Introduction}

A Voltage-Controlled Oscillator (VCO) is a fundamental component in numerous electronic systems. it generates a periodic signal whose frequency can be adjusted by changing the voltage applied to its control input. This makes VCOs essential components in a wide array of applications, from simple audio synthesizers to sophisticated communication systems, radar, and electronic instrumentation. The core functionality of a VCO lies in its ability to convert voltage variations into changes in oscillation frequency. This characteristic enables the VCO to be a pivotal element in frequency modulation (FM) and phase modulation (PM), where the frequency or phase of the output signal is varied in accordance with the input voltage, respectively. Consequently, VCOs play a crucial role in analog signal processing and have also been widely adopted in digital systems.

In telecommunications, VCOs are employed in Phase-Locked Loops (PLLs), clock recovery, and demodulation in both analog and digital communication systems. PLLs leverage the voltage-to-frequency conversion property of VCOs to maintain a constant phase relationship between the output and a reference signal, which is vital for stable frequency generation and signal demodulation. Radar systems also benefit from the rapid frequency tuning capabilities of VCOs, especially in Frequency-Modulated Continuous-Wave (FMCW) radar, where the VCO is used to generate a signal whose frequency varies with time, enabling precise distance and velocity measurements of targets.

The design and implementation of VCOs can vary widely, ranging from Ring oscillators, LC circuits, where the oscillation frequency is determined by the inductance (L) and capacitance (C) values, to more complex designs involving digital techniques and Direct Digital Synthesis (DDS). Despite their versatility, VCOs are not without challenges. Design considerations must address factors such as temperature stability, power consumption, phase noise, and linearity to ensure that the VCO's performance meets the application's requirements. The versatility and utility of VCOs across numerous applications, from signal generation in synthesizers to frequency modulation in telecommunications, hinge on their specific characteristics and performance parameters. This chapter delves into the crucial specifications of VCOs, providing insights into how these parameters influence their functionality, application suitability, and overall performance.

\textbf{Frequency Range and Tuning range:} The frequency range of a VCO is one of its most defining characteristics, indicating the minimum and maximum frequencies the oscillator can generate. This range is vital for applications where a wide bandwidth is necessary, such as in frequency synthesizers and agile communication systems. The choice of a VCO for a particular application depends heavily on its ability to cover the required operational frequencies. 

\begin{figure}
    \centering
    \includegraphics[width=1.0\columnwidth,height=0.4\columnwidth]{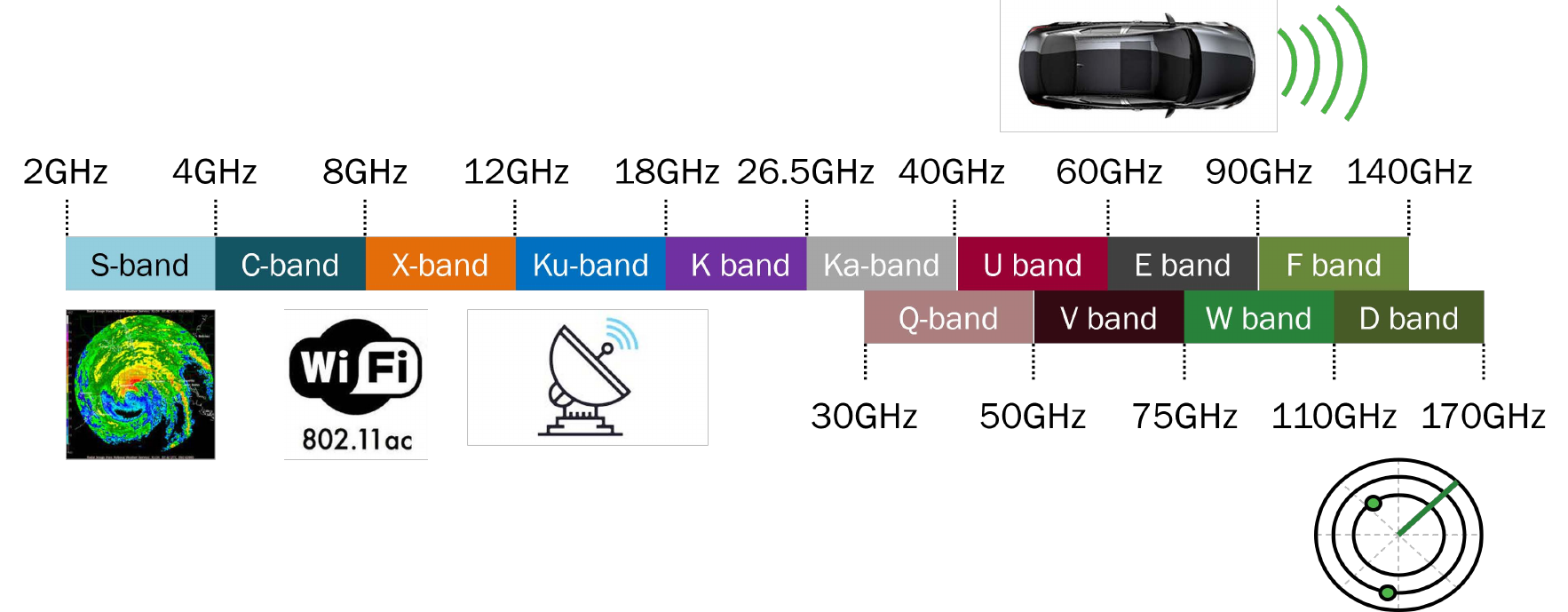}
    \caption{Overview of Frequency Bands and Their Respective Applications}
    \label{FrequencyBand}
\end{figure}

Fig.\ref{FrequencyBand} presents a spectrum of frequency bands ranging from the S-band to the D-band, each with its associated applications. The S-band is widely used for weather radar and some communications satellites, providing clarity and penetration through the atmosphere. Moving up the spectrum, the C-band is typically employed for satellite television and long-distance radio telecommunications. The X-band serves radar applications, including air traffic control, weather monitoring, and military uses. The Ku-band is favored for satellite communications, notably in direct-broadcast satellite television, as well as in some radar technologies. In the realm of deep space satellite communications, the K-band is a key player, where high data rates are required. The Ka-band is becoming increasingly popular for satellite communications, offering higher bandwidths, thus supporting faster data transmission rates. The U band, although less commonly referenced in public resources, finds its niche in specialized applications. The E band and F band are gaining momentum in high-capacity radio telecommunications, as well as in experimental and future communication technologies. The Q-band, V band, and W band are primarily used for research and development in scientific and military applications due to their high-frequency characteristics. Finally, the D band is at the frontier of wireless communications research, where the highest frequencies on this chart reside. The applications of the D band are still being explored, with potential uses in ultra-high-speed wireless broadband services and advanced radar systems. Each band's unique properties make it suitable for specific applications, shaping the way we use and interact with technology across various domains.

The tuning range of a Voltage-Controlled Oscillator (VCO) is a key characteristic that significantly impacts its utility in various electronic systems, where frequency agility and range are critical. This feature enables the VCO to modify its frequency output in accordance with changes in control voltage, making it an essential component in diverse applications. Notably, the VCO's tuning range is crucial in Phase-Locked Loops (PLLs) and Frequency-Modulated Continuous-Wave (FMCW) radar systems. In PLLs, a broad tuning range is vital for maintaining lock across a wide frequency spectrum, ensuring stable and accurate frequency synthesis. In the context of FMCW radar systems, the tuning range of the VCO directly affects the bandwidth (\(B\)) of the chirp signal, which is fundamental to the system's range resolution (\( \Delta R \)). This is expressed in the following formula:

\begin{equation}
\Delta R = \frac{c}{2B}
\end{equation}

where \(c\) is the speed of light in a vacuum, approximately \(3 \times 10^8\) meters per second, and \(B\) is the bandwidth of the frequency sweep facilitated by the VCO's tuning range. An extensive tuning range, enabling a larger bandwidth, permits finer resolution, i.e., smaller values of \( \Delta R \), which is crucial for distinguishing between closely spaced objects in the radar's surveillance area. The tuning range of a Voltage-Controlled Oscillator (VCO) in an FMCW radar system is intrinsically linked to the system's operational capabilities, including the maximum unambiguous range. A broader tuning range allows for a wider frequency sweep in the chirp signal, which is pivotal in extending the radar's operational envelope and enhancing its ability to accurately measure long distances. This aspect of the VCO's performance is critical in defining the radar's maximum unambiguous range, which is the furthest distance at which the system can reliably distinguish a return signal from a target. The relationship between the chirp duration and the maximum unambiguous range is encapsulated in the following formula:

\begin{equation}
R_{\text{max}} = \frac{c \cdot T_{\text{chirp}}}{2}
\end{equation}

where \( c \) represents the speed of light in a vacuum, approximately \( 3 \times 10^8 \) meters per second, and \( T_{\text{chirp}} \) is the duration of the chirp signal. This equation underscores the importance of the chirp duration, which is directly influenced by the VCO's tuning range, in determining the radar's ability to measure distances accurately, by calculating the time it takes for the radar signal to travel to the target and back.

\textbf{Tuning Sensitivity (Kv):} Tuning sensitivity, often denoted as Kv, represents the change in output frequency for a given change in control voltage, typically expressed in MHz/Volt. This parameter is crucial for understanding how precisely the VCO can be controlled. High sensitivity VCOs are beneficial in applications requiring fine frequency adjustments, whereas lower sensitivity may be preferable for broader, less precise control.

The VCO gain (\(K_{VCO}\)) plays a pivotal role in the functionality and efficiency of Phase-Locked Loops (PLLs), significantly impacting the system's response to variations in the input signal and control voltages. A higher \(K_{VCO}\) enhances the VCO's sensitivity to control voltage changes, facilitating a quicker lock to the input signal, which can improve the locking speed and potentially expand the locking range. However, this increased sensitivity also introduces trade-offs. On the positive side, a high \(K_{VCO}\) can lead to better system responsiveness and a broader operational bandwidth, thus offering greater versatility across different frequencies. Conversely, it may also escalate phase noise and induce jitter, especially in scenarios with volatile control voltages or in noise-prone environments. Therefore, the selection of \(K_{VCO}\) requires a deliberate balance, as the benefits of rapid locking and extensive bandwidth need to be carefully weighed against the drawbacks of heightened noise susceptibility and possible signal instability.

The VCO gain \(K_{VCO}\) is defined as the change in output frequency \(\Delta f\) per unit change in control voltage \(\Delta V\):
\begin{equation}
K_{VCO} = \frac{\Delta f}{\Delta V} \quad \text{(in Hz/Volt)}
\end{equation}

The PLL bandwidth \(\omega_{BW}\) can be approximated for a simple second-order loop as:
\begin{equation}
\omega_{BW} \approx \frac{1}{2\pi} \sqrt{\frac{K_{PD} \cdot K_{VCO} \cdot K_{F}}{N \cdot L}}
\end{equation}
where \(K_{PD}\) is the phase detector gain (in V/radian), \(K_{F}\) is the loop filter transfer function, \(N\) is the division ratio in the feedback path, and \(L\) is the loop filter inductance (in Henries) for an analog PLL.

The locking range (or capture range, \(\Delta \omega_{C}\)) for a basic PLL can be simplified as:
\begin{equation}
\Delta \omega_{C} = 2\pi \cdot K_{VCO} \cdot V_{C_{\text{max}}}
\end{equation}
where \(V_{C_{\text{max}}}\) is the maximum control voltage for the VCO.

\textbf{Linearity:} The linearity of a VCO refers to the consistency of its tuning sensitivity across the entire operating range. Ideal VCOs exhibit perfect linearity, where a linear increase in control voltage results in a proportional linear increase in frequency. However, real-world VCOs often exhibit some non-linearity, which can introduce distortion in frequency modulation applications, making linearity a critical parameter in high-fidelity communication systems.

\textbf{Power Consumption:} The power consumption of a VCO is a critical consideration in battery-operated or power-sensitive applications. Lower power VCOs are preferred in portable devices and remote sensors to extend battery life and reduce thermal effects, which can influence other components' performance and reliability.

\textbf{Temperature Stability:} Temperature stability indicates how well the VCO maintains its frequency over the operating temperature range. Temperature variations can cause significant frequency drift in VCOs, affecting the performance of temperature-sensitive applications like precision instrumentation and timekeeping devices. VCOs designed for such applications often incorporate temperature compensation mechanisms to mitigate these effects.

\textbf{Phase Noise:} Phase noise is a measure of the signal purity or the stability of the VCO's output frequency, representing the frequency fluctuations that can cause jitter in the output signal. Low phase noise is essential in applications such as radar systems, precision measurement equipment, and digital communication systems, where signal integrity and clarity are paramount. The phase noise performance can significantly impact the system's overall noise characteristics and data integrity.

Phase noise (PN) in Voltage-Controlled Oscillators (VCOs) arises from various noise sources, each contributing to the instability and spectral impurity of the VCO output. The intrinsic noise of active devices, such as thermal noise characterized by the formula $S_{\text{thermal}} = 4kTR$ where $k$ is Boltzmann's constant, $T$ is the absolute temperature, and $R$ is the resistance, and flicker noise, which follows a $1/f$ dependency, significantly influence PN. Supply and substrate noise, represented by voltage fluctuations $\Delta V$, modulate the VCO frequency, thus contributing to PN as $\Delta f / \Delta V$. Load noise, including noise from interfaced circuitry, and resonator noise from the VCO's tank circuit elements, like varactors and inductors, also play crucial roles. Moreover, the modulation of parasitic elements within the VCO circuit can translate and mix noise components into the VCO's operating frequency, further impacting PN. Addressing these noise contributions demands a holistic approach that encompasses careful device selection, biasing optimization, and the implementation of circuit techniques aimed at minimizing the sensitivity of the VCO's output phase to these noise sources.

Implications of Phase Noise Modeling
The Leeson Model and other phase noise theories provide crucial insights for VCO design and application:

\textbf{Design Optimization:} By understanding the dependencies in the phase noise model, designers can optimize oscillator circuit components, biasing conditions, and loop filter designs to minimize phase noise.
System Performance Analysis: Phase noise models enable the prediction of system-level performance metrics, such as bit error rate (BER) in digital communication systems or radar range resolution.
Component Selection: Accurate phase noise modeling assists in selecting the appropriate VCOs for specific applications, balancing performance requirements with cost and power consumption constraints.
Advanced Modeling Techniques
While the Leeson Model offers a foundational understanding of phase noise, advanced modeling techniques address the complexities of modern VCO designs, including:

Nonlinear Dynamic Models: These models account for the nonlinear behavior of oscillator circuits, providing a more accurate representation of phase noise under large signal conditions.
Time-Domain Simulations: Techniques like phase noise simulation in the time domain allow for a detailed analysis of noise sources and their impact on oscillator performance.
Empirical Models: Based on measured data, these models can accurately predict phase noise performance for specific VCO designs and operating conditions.

Passive components, such as capacitors, transmission lines, and inductors, play a pivotal role in the design and operation of Voltage-Controlled Oscillators (VCOs). While their primary function is to determine the oscillation frequency and shape the oscillator's output waveform, these components also significantly influence the phase noise characteristics of VCOs. Understanding the effects of these passive elements on phase noise is crucial for optimizing VCO design for minimal phase noise, which is essential for high-performance applications in communications, radar, and precision measurement systems. This chapter explores the interactions between passive components and phase noise in VCOs, highlighting the mechanisms through which capacitors, transmission lines, and inductors contribute to phase noise and discussing strategies to mitigate their impact.

Capacitors and Phase Noise
Capacitors in VCOs are primarily used for setting the oscillation frequency in conjunction with inductors (in LC oscillators) and for filtering purposes. However, they also introduce noise sources that can degrade phase noise performance:

Dielectric Absorption and Loss: The dielectric material in capacitors exhibits absorption and loss phenomena, which introduce noise and contribute to phase jitter. Low-loss dielectric materials can mitigate this effect.
Voltage Coefficient: Some capacitors exhibit a voltage coefficient, where the capacitance value changes with the applied voltage, leading to frequency modulation and increased phase noise.
Thermal Noise: The thermal noise generated by the resistive components of capacitors can couple into the oscillator loop, contributing to phase noise.
Mitigation strategies include the careful selection of capacitor types with low dielectric losses, minimal voltage coefficient, and proper placement to minimize noise coupling into the VCO loop.

Transmission Lines and Phase Noise
Transmission lines in VCO circuits, used for impedance matching and signal routing, can also affect phase noise:

Losses and Dispersion: Transmission line losses and dispersion can lead to signal attenuation and phase distortion, respectively, which can modulate the oscillator's frequency and increase phase noise.
Reflections: Impedance mismatches can cause reflections, resulting in standing waves that modulate the oscillator's amplitude and phase, contributing to phase noise.
Designing with well-characterized transmission lines, ensuring proper impedance matching, and using low-loss materials can help reduce their adverse effects on phase noise.

Inductors and Phase Noise
Inductors are key components in determining the oscillation frequency and quality factor (Q) of LC oscillators, with their properties directly influencing phase noise:

Quality Factor (Q): The Q factor of an inductor is a critical parameter, as it determines the oscillator's selectivity and phase noise. Higher Q inductors lead to lower phase noise by reducing the bandwidth over which noise can contribute to phase jitter.
Magnetic Losses: Losses in the magnetic core material (if used) can introduce noise and degrade the Q factor, increasing phase noise.
Skin Effect: At high frequencies, the skin effect increases the effective resistance of the inductor, reducing Q and worsening phase noise.
Selecting high-Q inductors, minimizing core losses (or using air-core designs), and designing for minimal skin effect are effective strategies for reducing phase noise contributions from inductors.

Design Considerations for Minimizing Phase Noise
The design and selection of passive components in VCOs require careful consideration to minimize their adverse effects on phase noise:

Component Selection: Choosing components with characteristics conducive to low phase noise, such as high-Q inductors, low-loss capacitors, and low-resistance transmission lines.
Circuit Layout: Optimizing the layout to reduce coupling of noise sources into the oscillator loop, using shielding and grounding techniques to minimize interference.
Thermal Management: Designing for stable thermal conditions to reduce fluctuations in component values due to temperature changes, which can modulate the oscillation frequency and increase phase noise.

\chapter{Ultra Low Noise VCO - Flicker noise reduction technique}

Flicker noise, also recognized as \(1/f\) noise, is a pervasive noise source in electronic devices, manifesting a power spectral density (PSD) that exhibits an inverse relationship with frequency. The physical origins of flicker noise are complex and multifaceted, often linked to the dynamic interactions of charge carriers with defects or impurities within the semiconductor material. These interactions lead to temporal fluctuations in the device's conductivity, predominantly observable at lower frequencies. The PSD of flicker noise is typically expressed as:

\begin{equation}
S(f) = \frac{K}{f^\alpha}
\end{equation}

where \(S(f)\) is the power spectral density at frequency \(f\), \(K\) is a proportionality constant that depends on the specific device and its operating conditions, and \(\alpha\) is a parameter close to 1, varying slightly depending on the material and the nature of the fluctuations. This equation highlights the characteristic \(1/f\) dependency, underscoring the significance of flicker noise in shaping the noise profile of electronic circuits, especially in the low-frequency regime.

Flicker noise, or \(1/f\) noise, in NMOS transistors is a critical factor affecting low-frequency operation and stability. It originates from the fluctuation of carrier mobility due to traps at the silicon-oxide interface and in the oxide near the interface. The power spectral density (PSD) of flicker noise in NMOS devices is given by:

\begin{equation}
S_{V_{f}} = \frac{K_{F} \cdot W}{L \cdot f \cdot C_{ox}^2}
\end{equation}

where \(S_{V_{f}}\) represents the PSD of the flicker noise voltage, \(K_{F}\) is the flicker noise coefficient, \(W\) and \(L\) are the width and length of the NMOS channel, \(f\) is the frequency, and \(C_{ox}\) is the gate oxide capacitance per unit area. The inverse proportionality to frequency (\(1/f\)) underscores the dominance of flicker noise at lower frequencies. Comparatively, PMOS transistors exhibit similar flicker noise characteristics; however, the flicker noise coefficient (\(K_{F}\)) often differs due to the distinct mobility and trapping mechanisms of holes in the PMOS devices. Generally, \(K_{F}\) for PMOS is higher than that for NMOS, making PMOS devices more susceptible to flicker noise. This difference is attributed to the higher energy and deeper trap levels associated with holes, which lead to more significant carrier number fluctuations and, consequently, increased noise levels.

In essence, while both NMOS and PMOS transistors are affected by flicker noise, the magnitude and impact may vary due to the inherent material and physical differences between electrons and holes as charge carriers in these devices.

In \cite{Tailcurrent}, a tail current-shaping technique in LC-VCOs aimed at improving phase noise by dynamically adjusting the tail current. This method increases the tail current during the oscillator's output voltage peaks, where phase noise sensitivity is lowest, and decreases it during zero crossings, where sensitivity is highest. This approach not only reduces the phase noise contributions of active devices but also increases the oscillation amplitude, leading to better DC to RF conversion efficiency without additional power consumption or the need for extra noisy active devices. The technique's effectiveness is supported by extensive analysis and experimental results, showcasing its potential in enhancing VCO performance.

In \cite{AMFM},The document explores the AM-to-FM conversion mechanisms in MOS LC-VCOs, focusing on how the voltage-dependent capacitors of active devices contribute to this conversion. It identifies an optimal oscillation amplitude where the effective differential capacitance versus amplitude has a minimum, leading to minimal AM-to-FM noise conversion and improved phase noise performance. This optimal amplitude, close to the MOS threshold voltage, is crucial for achieving better VCO operation, verified through simulations and measurements of a 2 GHz differential NMOS LC-VCO.

In \cite{flickerNoiseN}, The document introduces a novel CMOS LC VCO topology aimed at reducing flicker noise by addressing the intrinsic noise sources in the tail transistor. By employing a memory-reduced tail transistor approach, the design mitigates flicker noise by minimizing the long-term memory effects associated with carrier trapping in oxide states. This technique effectively reduces the flicker noise, leading to enhanced phase noise performance in the VCO, as demonstrated through both simulation and practical implementation results presented in the document.

In \cite{flickerNoiseN}, The document presents a technique to mitigate phase noise in LC oscillators by incorporating a filtering approach that specifically targets the noise generated by the current source. This method involves using a narrowband circuit to suppress noise frequencies emanating from the current source, effectively rendering it "noiseless" to the oscillator. By strategically placing a large capacitor in parallel with the current source and an inductor between the current source and the oscillator's tail, the design achieves a high impedance at critical frequencies. This noise filter setup significantly reduces phase noise by minimizing the impact of low-frequency noise components that are typically problematic in oscillator designs.

Incorporating a second harmonic resonator in the source of a VCO can effectively lower flicker noise by leveraging the noise shaping phenomenon. This approach utilizes the resonator to selectively enhance the impedance at the second harmonic frequency of the oscillator's output signal. By doing so, it increases the VCO's sensitivity to its own second harmonic while reducing its sensitivity to low-frequency flicker noise. This results in a form of "noise filtering" that attenuates the flicker noise components before they can modulate the oscillator's output signal. The enhancement of the second harmonic can also contribute to a cleaner, more stable oscillation frequency by providing a secondary feedback path that helps maintain the oscillation at the desired frequency. This technique, therefore, not only suppresses flicker noise but can also improve the overall spectral purity and phase noise performance of the VCO.

\chapter{Ultra Low Noise VCO - Proposed topology}

\section{Introduction}

Thanks to the recent advancements in Si/SiGe transistor technologies, interest in mm-wave and sub/THz sensing and ranging applications, in particular frequency modulated continuous wave (FMCW) radars and pulsed radars, are rapidly growing\cite{fmcwapp,fmcwapp1,FMCWRADAR,yahya,aghasi2020terahertz,aghasi2017power,aghasi2016design,aghasi2020millimeter,aghasi2023broadband,han201525,alesheikh2024electronically,aghasi2019fully,maktoomi2022sub,han2015sige}. Moving to  mm-wave frequencies allows to extend the chirp bandwidth of an FMCW radar to improve the range resolution, i.e, the capability of the radar to distinguish close-by objects in the line of sight direction but at different distances \cite{bandwidth,FMCWCHIRP}. The resolution enhancement at higher frequencies comes at the expense of deteriorating phase noise\cite{vadim}, RF power, and efficiency\cite{mmwavepower} as the quality factor of passive components as well as power gain of active devices drops by increasing frequency \cite{power}. The phase noise of VCO impact the signal-to-noise ratio of the intermediate frequency components in an FMCW radar and the output power determines the detectable range of radar\cite{radarnoise,radarrange}.
Therefore, enhancing the PN and power efficiency of mm-wave VCOs incorporated inside FMCW radars is critical \cite{aghasi202349,aghasi202376}.

The PN reduction of CMOS VCOs should capture the effect of both thermal noise sources and transistor Flicker noise \cite{aghasi2024single}. As CMOS technology scales, especially at lower offset frequencies, $1/f$ flicker noise effect becomes more pronounced\cite{siri,flickerthermal},\cite{1/fscale}. 
Different theories to improve noise behavior in $1/f^3$ region of PN have been developed \cite{shahmohammadi,flickcernoiseT,hajimiri,abidiR,implicit,CICC2022}. Conventional cross-coupled VCOs tend to suppress harmonic components by employing a fundamental resonator at the drain terminal which is prone to two major issues: (a) frequency drift due to Groszkowski theory {\cite{Groszkowski}} (b) introducing asymmetry to the output perform which increases the dc value of effective ISF ($\Gamma_{eff}$), i.e., the product of noise modulation factor (NMF) and ISF ($\Gamma$) \cite{hajimiri}. This is due to the fact that 
all the higher harmonic components flow into a capacitive load and thus unbalance the reactive energy and increase the waveform asymmetry \cite{flickcernoiseT}. It is shown in \cite{hajimiri} that the dc term of ISF, $\Gamma_{dc}$ determines how strong  $1/f$ noise will upconvert to $1/f^3$ PN. By adding a common mode resonator at $2f_{osc}$, it is shown that $\Gamma_{dc}$ can be reduced where the fundamental and second harmonic voltage components are assumed in-phase\cite{shahmohammadi,implicit,dmresonator}. 

\begin{figure*}[h!]
      \centering \includegraphics[width=1\textwidth]{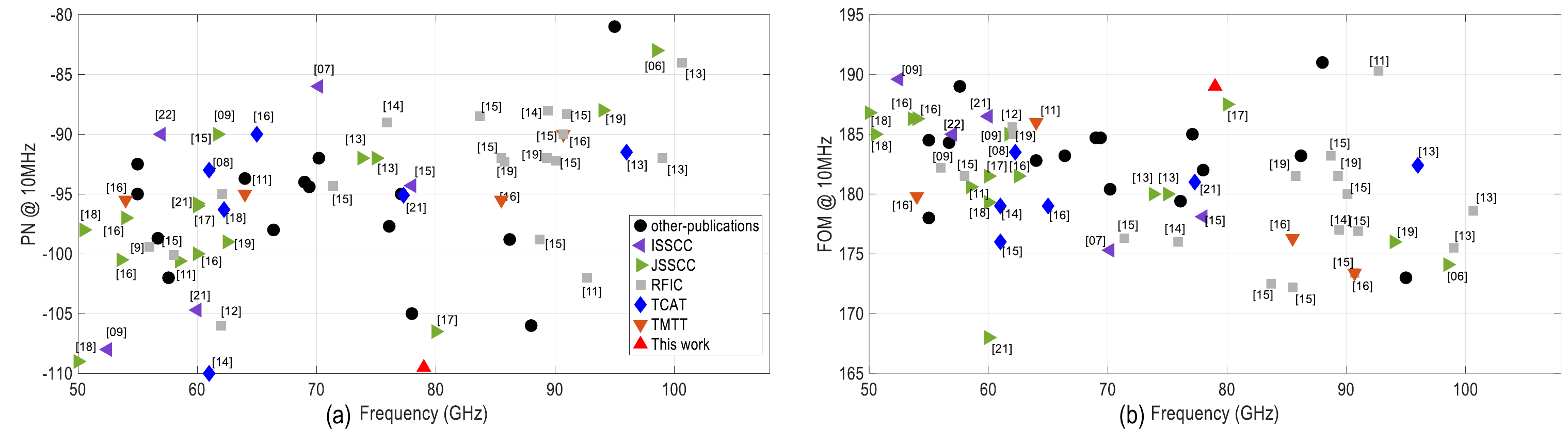}
      \vspace{-0.25in}
     \caption{(a) PN, and (b) FoM at 1MHz offset frequency for recent state-of-art works over past two decades. For all the works published after 2000, the year of publication is shown inside brackets. For more meaningful conclusion, only VCOs in CMOS/BiCMOS technologies are included.}
     \label{stateofart}
\end{figure*}

At higher frequency offsets, typically larger than 1MHz, thermal noise becomes the dominant contributing source to the phase noise. Improving oscillator phase noise usually will be accomplished by increasing quality factor, which includes transformer-based designs \cite{transformbased}--\cite{ALIMEDI}, ladder circuits \cite{Yazdi}, and coplanar waveguides \cite{hajimiriCPW}, or increasing the voltage swing. Based on \cite{hajimiriISF},  for effective ISF ($\Gamma_{eff}$), the root mean square value should be reduced to lower the thermal noise contribution to output PN. In \cite{babaie1}, the authors proposed a transformer-based soft-clipping class F oscillator to effectively lower tail current noise contribution while maintaining high voltage swing. 
Despite a clear understanding of NMF ($\alpha$) and ISF ($\Gamma$) at radio frequencies, the behavior of these functions in mm-wave oscillators is not well studied. 

To enhance the power efficiency of mm-wave oscillators, device-centric approaches have been pursued by prior art \cite{amirahmad,khodam,reference1,reference2,reference3}. In \cite{amirahmad, khodam} the passive embedding of the transistor is optimized to enhance the output power efficinecy above 10\%. The design in \cite{amirahmad} has around 10\% of frequency tuning while the design in \cite{khodam} is fixed frequency. In \cite{reference1} a 20 GHz VCO with \textit{4}-th harmonic power extraction at 80 GHz with efficiency of 0.7\% is presented. Similarly in \cite{reference2} a VCO with third harmonic power generation efficiency of 2.8\% at 57.8 GHz is proposed. The digitally controlled oscillator (DCO) in \cite{reference3} is part of a fractional-\textit{N} digital frequency synthesizer at 57.5-67.2 GHz with a DC-to-RF efficiency of 3.2\%. Despite the excellent power efficiency, strong majority of these works have a PN FoM below 180 dBc/Hz.

Fig.\ref{stateofart} summarizes the status quo in terms of PN at 1MHz offset frequency (Fig. \ref{stateofart}(a)) and peak FoM (Fig. \ref{stateofart}(b)). Despite significant PN improvements over the past few years, there are still few works with lower than -100 dBc/Hz PN at 1MHz offset frequency which indicates that there is still room for improvement of VCO PN along with other advances in the design of FMCW radars. In this work, we characterize the effect of oscillator voltage waveforms on the PN and generated harmonic power. By finding the desired relative phase and amplitude of the fundamental and second harmonic voltage components across transistor terminals, we propose a circuit configuration that achieves the desired phase noise suppression while maintaining a reasonable DC-to-RF efficiency.
The rest of this papers is organized as follows. In Section \ref{sec:2}, we study the super-harmonic VCO behavior and demonstrate how the dc and rms values of effective ISF can be reduced in the presence of a harmonically rich output waveform. Moreover, we study the effect of proposed architecture on second harmonic power generation. In Section \ref{sec:3}, the details of the proposed VCO, the design procedure, and layout considerations are presented. In Section\ref{sec:4}, the measurement results are presented and compared against simulation results. The paper is concluded in Section \ref{sec:5}.

\section {Waveform Synthesis to improve pn and harmonic power}
\label{sec:2}

\begin{figure*}[h!]
      \centering \includegraphics[width=1\textwidth]{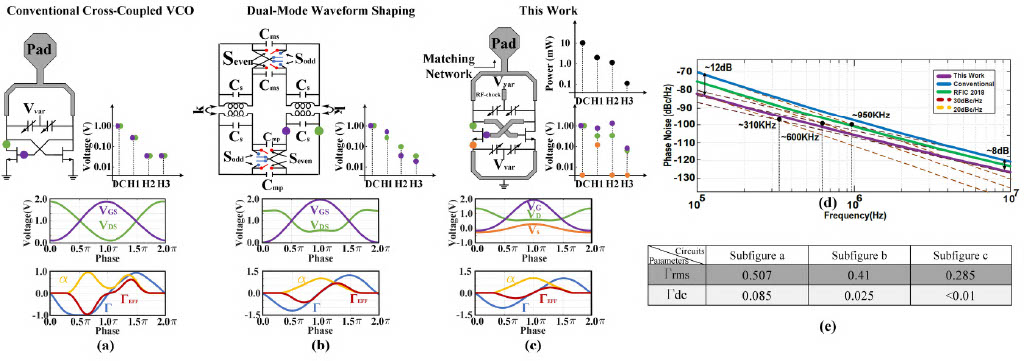}
      \vspace{-0.25in}
\caption{a) Simplified cross-coupled schematics b) dual-mode waveform shaping schematics, c) proposed schematics and d) corresponding PN, (e) Comparison of dc and RMS values of ISF among the illustrated circuits. }
\label{compare}
\end{figure*}

\begin{figure*}[h!]
      \centering \includegraphics[width=1\columnwidth,height=5cm]{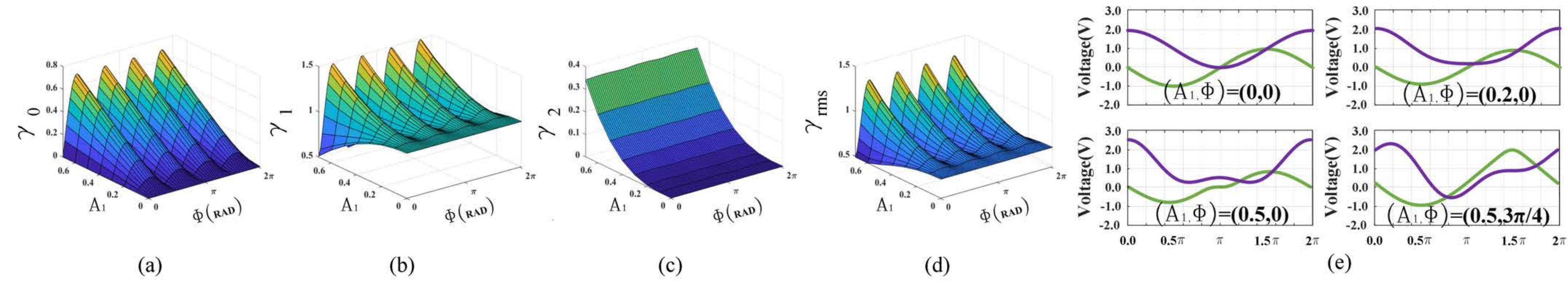}
      \vspace{-0.15in}
    \caption{Simulated (a) $\gamma_{0}$, (b) $\gamma_{1}$, (c) $\gamma_{2}$, (d) ${rms}$ of $\Gamma$ in presence of second harmonic as a function of $A_{2}$,$\Phi_{2}$, (e) examples of $\Gamma$,$V_{ds}$ as a function of $A_{2}$,$\Phi_{2}$.}
\label{ISF1}
\end{figure*}

In this section, we characterize the PN and harmonic power based on the constituent harmonic terms of the voltage waveforms across the device terminals in a super-harmonic VCO. The PN analysis is based on ISF theory in \cite{hajimiri} and for the power analysis we adopt the models from \cite{OmeedMomeniR, aghasi20170}. To obtain intuition on the effectiveness of the proposed design, we compare our proposed oscillator with two other configurations of cross-coupled oscillators in Fig. \ref{compare}. In Fig. \ref{compare}(a), the schematic of a conventional 
cross-coupled VCO is shown where the drain-source and gate-source fundamental voltages are out of phase. Harmonic current component tends to flow into a capacitive path making tank’s reactive energy unbalanced. This results in a non-zero dc and relatively high rms value for the $\Gamma_{eff}$ of transistor noise. In Fig. \ref{compare}(b), the even/odd mode switching is combined with coupled transformers. By shaping the waveform of the gate terminal voltage, the dc and rms value of $\Gamma_{eff}$ in this design are reduced. In this work (Fig. \ref{compare}(c)), we propose a modified cross-coupled oscillator which further reduces the dc, and rms value of $\Gamma_{eff}$. The values for dc and rms of ISF for these circuits are provided in Fig. \ref{compare}(e). For all these designs, terminals voltage waveforms along with the corresponding $\Gamma$,NMF or $\alpha$, and $\Gamma_{eff}$ are also shown in Fig. \ref{compare} (a-c). Moreover, in the proposed design second harmonic power is maintained inside the oscillator critical current loop and reasonably high harmonic power can be obtained. The comparison of phase noise profiles in Fig. \ref{compare}(d) shows an effective PN reduction by up to 6 dB and phase noise corner frequency reduction by more than 290 kHz compared to the other two designs.

\subsection{Impact of Fundamental and Harmonic Voltages on PN}
For a VCO with non-negligible harmonic components, the arbitrary drain-to-source waveform (excluding dc) is written as $V_{ds}=\sum_{n=1}^{\infty} A_{n}\cos(n\theta+\phi_n)$ where $\theta=\omega t$ is the angular frequency. $A_{n}$ and $\phi_{n}$ are the \textit{n}-th harmonic amplitude and phase, respectively. Similar to prior art of super-harmonic VCO designs with assumption on the dominance of fundamental and second harmonic components, in the proposed VCO, we neglect the effect of  harmonic components with index larger than 2 \cite{OmeedMomeniR,shahmohammadi}. The $\Gamma$ estimation based on time-domain derivatives of the voltage waveform shown in \cite{hajimiri} is calculated for the assumed $V_{ds}$ and due to its periodicity can be written as \cite{ISFFOURIER}:
\vspace{-0.05in}
\begin{equation*}
\Gamma=\sum_{n=0}^{\infty} \gamma_{nc}\cos(n\theta)+\gamma_{ns}\sin(n\theta)=\gamma_{n}\cos[n\theta+\tan^{-1}(\gamma_{ns}/\gamma_{nc})],
    \label{f1}
\end{equation*}

where $\gamma_{nc}$,  $\gamma_{ns}$, and $\gamma_{n}$ are real-valued coefficients of the Fourier expansion for $\Gamma$. To minimize the effect of cyclo-stationary noise contribution from the active device, effect of $\alpha$ should also be observed. As shown in \cite{shahmohammadi}, the $\alpha$ of a transistor noise can be derived by calculating the periodic $G_m$ of the device and normalizing it to the maximum value. Due to the periodic nature of $G_m$ in an oscillator, the time-domain waveform of $\alpha$ can be written as:

\begin{equation}
\alpha=\sum_{m=0}^{\infty} \zeta_{mc}\cos(m\theta)+\zeta_{ms}\sin(m\theta)=\zeta_{m}\cos[m\theta+\tan^{-1}(\zeta_{ms}/\zeta_{mc})],
    \label{f1}
\end{equation}

where $\zeta_{mc}$, $\zeta_{ms}$, and $\zeta_{m}$ are also real-valued coefficients. From an analytical perspective, $\Gamma_{eff}$ which is $\Gamma\times\alpha$ can be written as:
\begin{equation} \Gamma_{eff}=\sum_{n=0}^{\infty} \sum_{m=0}^{\infty}\gamma_{n}\zeta_{m}\frac{\cos[(m+n)\theta+\Tilde{\phi}_{m,n}]+\cos[(m-n)\theta+\Tilde{\phi}_{m,-n}]}{2}
    \label{3}
    \small
\end{equation}

where $\Tilde{\phi}_{m,n}=tan^{-1}(\zeta_{ms}/\zeta_{mc})+tan^{-1}(\gamma_{ns}/\gamma_{nc})$, and $\Tilde{\phi}_{m,-n}=tan^{-1}(\zeta_{ms}/\zeta_{mc})-tan^{-1}(\gamma_{ns}/\gamma_{nc})$. According to the Parseval theorem \cite{Parseval}, to minimize the rms value of a periodic function, the summation of squared Fourier coefficients are minimized. Thus, we can expect the minimum of $\Gamma_{eff,rms}$ is achieved under condition(s) when collectively the individual Fourier coefficients of $\Gamma_{eff}$ are small values. Based on (\ref{3}), for each Fourier coefficient of $\Gamma_{eff}$, the product of multiple coefficients from $\Gamma$ and $\alpha$ should be summed. To minimize the magnitude of first three Fourier coefficients of $\Gamma_{eff,rms}$ which are the most dominant in this second-harmonic VCO, the Cauchy-Schwarz inequality \cite{Cauchy} allows us to go after minimizing the Fourier coefficients of $\Gamma$ and $\alpha$ up to the 2nd-order terms. In the first step, we plot the Fourier coefficients of $\Gamma$ as well as $\Gamma_{rms}$ in terms of $A_2$ and $\phi_2$ when $A_1=1$ and $\phi_1=0$, shown in Fig. \ref{ISF1} (a-d). Despite the direct proportionality of $\gamma_2$ with $A_2$ (Fig. \ref{ISF1}(c)), $\gamma_0$ and $\gamma_1$ which normally have larger magnitude than $\gamma_2$ are sensitive to both $A_2$ and $\phi_2$ (Fig. \ref{ISF1}(a,b)) which ultimately impacts the variations of $\Gamma_{rms}$ (Fig. \ref{ISF1} (d)). As shown in Fig. \ref{ISF1} (a,d), $\Gamma_{0}$ and $\Gamma_{rms}$ reach their minimum values of 0 and 0.22 when $\phi_2=n\pi$ for integer $n$, respectively. Under $\phi_2=n\pi$, $\Gamma_{0}$ stays at minimum for all values of $A_2$ While $\Gamma_{rms}$ reaches the minimum at $A_2=$0.7. The time-domain variations of $\Gamma$ by changing the composition of $A_2$ and $\phi_2$ for $V_{ds}$ when $A_1=1$ and $\phi_1=0$, are shown in Fig. \ref{ISF1}(e). 

As the next step, we analyze the $\alpha$ Fourier coefficients inside the oscillator when the voltage relationship between $V_{ds}$ and $V_{gs}$ is varied. Conventionally, the drain and gate of an NMOS pair inside a cross-coupled oscillator are cross-connected forcing the fundamental voltage at the gate and drain terminals of each transistor to be out of phase. To break this relationship by introducing a transmission line between the gate and drain, $R_{v,n}=V_{g,n}/V_{d,n}$, which is the gate-to-drain voltage ratio at the \textit{n}-th harmonic, will obtain a non-unity magnitude and excess phase shift compared to conventional cross-coupled.
\begin{figure*}[h!]
\centering
\includegraphics[width=1\textwidth,height=5cm]{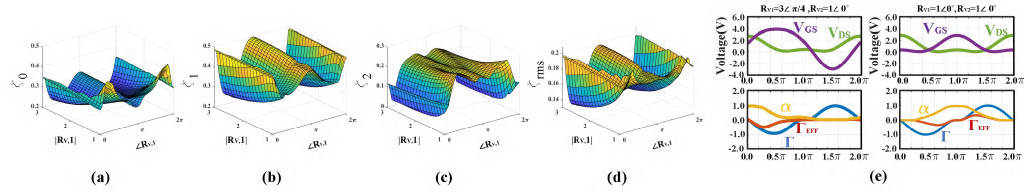}
\vspace{-0.25in}
\caption{Simulated (a) $\zeta_{0}$, (b) $\zeta_{1}$, (c) $\zeta_{2}$, (d) $rms$ of $\alpha$ in presence of second harmonic as a function of $R_{V,1}$, (e) examples of $\Gamma$, $\alpha$, $\Gamma_{eff}$ as a function of $R_{V,1}$. }
\label{nmf}
\end{figure*}
\begin{figure*}[h!]
\centering
\includegraphics[width=1\textwidth,height=5cm]{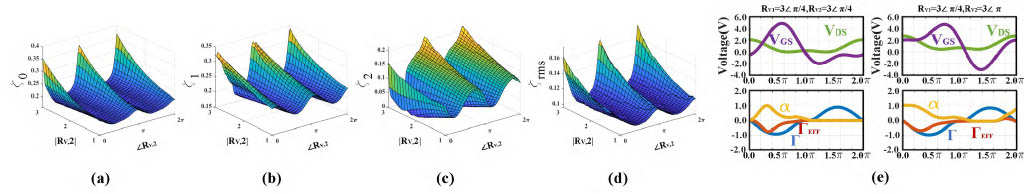}
\vspace{-0.25in}
\caption{Simulated (a) $\zeta_{0}$, (b) $\zeta_{1}$, (c) $\zeta_{2}$, (d) $rms$ of $\alpha$ in presence of second harmonic as a function of $R_{V,2}$, (e) examples of $\Gamma$, $\alpha$, $\Gamma_{eff}$ as a function of $R_{V,2}$. }
\label{nmf2}
\end{figure*}
For desired $V_{ds}$ derived from Fig. \ref{ISF1} which minimizes $\Gamma_0$ and $\Gamma_{rms}$, $\alpha$ Fourier coefficients ($\zeta_m$) are calculated  in terms of phase and magnitude of $R_{v,1}$, and $R_{v,2}$, shown in Fig. \ref{nmf} and Fig. \ref{nmf2}, respectively. The practical range of magnitude for $R_{v,1}$ and $R_{v,2}$ is bounded to 3 for the circuit configuration in this design. Instead of formulating the global minima for each coefficient, we attempt to find one of the practical scenarios under which these coefficients are sufficiently small. To identify this scenario, we take two steps: (1) assuming an invariant $R_{v,2}$ and minimizing the coefficients with respect to the magnitude and phase of $R_{v,1}$, happening at $R_{v,1,opt}$ (2) Setting $R_{v,1}=R_{v,1,opt}$ and minimizing the coefficients with respect to the magnitude and phase of $R_{v,2}$, happening at $R_{v,2,opt}$. The presented simulation results are based on a transistor with total width of 40$\mu$m and channel length of 65nm. In Fig. \ref{nmf}, $R_{v,2}$ is assumed $1\angle0$\textdegree. Despite the fact that each $\zeta_n$ dependency on $R_{v,1}$ is different, by observing rms value dependency on $R_{v,1}$ it can be concluded that $R_{v,1}$=(3,$\pi/4$) results in the smallest magnitude of $\alpha_{rms}$. By looking at the waveforms for this case and another case of $R_{v,1}=(1,0)$ in Fig. \ref{nmf}(e), it can be concluded that the larger magnitude of $R_{v,1}$ and the phase of $\pi/4$ when the $V_{ds}$ is kept invariant, results in transistor being in the cut-off or triode regions (corresponding to $\alpha\simeq 0$) for a longer portion of period. 

In Fig .\ref{nmf2}, dependency of $\alpha$ on $R_{v,2}$ is shown when $R_{v,1}=R_{v,1,opt}$. For $R_{v,2}=$3$\angle45$\textdegree, $\alpha_{rms}$ reaches its minimum. The waveforms in Fig. \ref{nmf2}(e) confirm that when $R_{v,2}$ is set to these optimum values, similar to Fig. \ref{nmf}(e), the saturation region portion of the period decreases, resulting in smaller $\alpha_{rms}$. From a design perspective, under the optimal value of $R_{v,1}$ the gate-source voltage swing is readily large and further adjustments to the magnitude of $R_{v,2}$ does not change the $\zeta_m$ coefficients significantly, as one can conclude by comparing Fig. \ref{nmf}(e) and Fig. \ref{nmf2}(e). As a summary of this subsection, we discussed the desired conditions on transistor terminal voltage waveforms in terms of $(A_{2},\phi_{2},R_{V,1},R_{V,2})$ to reduce $\Gamma_{eff,0}$, $\Gamma_{eff,rms}$, and subsequently translation of transistor noise into PN. 
Before applying these conditions in the VCO design in Section III, in the next subsection, we synthesize the voltage waveforms to maintain a reasonable harmonic power (and power efficiency).


\subsection{Impact of Voltage Waveforms on Harmonic Power}

The conditions to reduce PN from the previous section do not necessarily ensure a high harmonic power in a super-harmonic VCO. This is in contrast with fundamental oscillators where a higher DC-to-RF efficiency and lower thermal noise PN happen concurrently \cite{banks}. Without loss of generality, we assume that for drain-to-source voltage, ($A_{2opt},\Phi_{2opt}$) is chosen to minimize the $\Gamma_{rms}$. Based on \cite{OmeedMomeniR}:

\begin{equation}
\vec{I}_{d}^{\hspace{0.05in} 2f_{0}}=
      \vec{I}_{d,a}^{\hspace{0.05in} 2f_{0}}+\vec{I}_{d,t}^{\hspace{0.05in} 2f_{0}}=
    G_{21,a}^{2f_{0}}.
    \vec{V}_{gs}^{\hspace{0.05in} 2f_{0}}+
    G_{22,t}^{2f_{0}}.
    \vec{V}_{ds}^{ \hspace{0.05in} 2f_{0}},
    \label{f1}
\end{equation}

where indices $a$ and $t$ denote the operation in active and triode regions, $I_{d}$ is the transistor current, and $G_{21}^{2f_{0}}$, $G_{22}^{2f_{0}}$ show the real part of Y parameter of the 2-port device model. Moreover, $\vec{V}_{gs}^{2f_{0}}$ can be written as: $\vec{V}_{gs}^{2f_{0}}=\vec{R}_{v,2}.\vec{V}_{ds}^{2f_{0}}$, then second harmonic current based on \cite{OmeedMomeniR} should be written as:

\begin{equation}
\vec{I}_{d}^{\hspace{0.05in} 2f_{0}}=
      \vec{I}_{d,a}^{\hspace{0.05in} 2f_{0}}+\vec{I}_{d,t}^{\hspace{0.05in} 2f_{0}}=
    (G_{21,a}^{2f_{0}}.\vec{R_{v,2}}+
    G_{22,t}^{2f_{0}})
        \vec{V}_{ds}^{2f_{0}}.
    \label{eq:5}
\end{equation}

\begin{figure}[h!]
\centering
\includegraphics[width=1\textwidth]{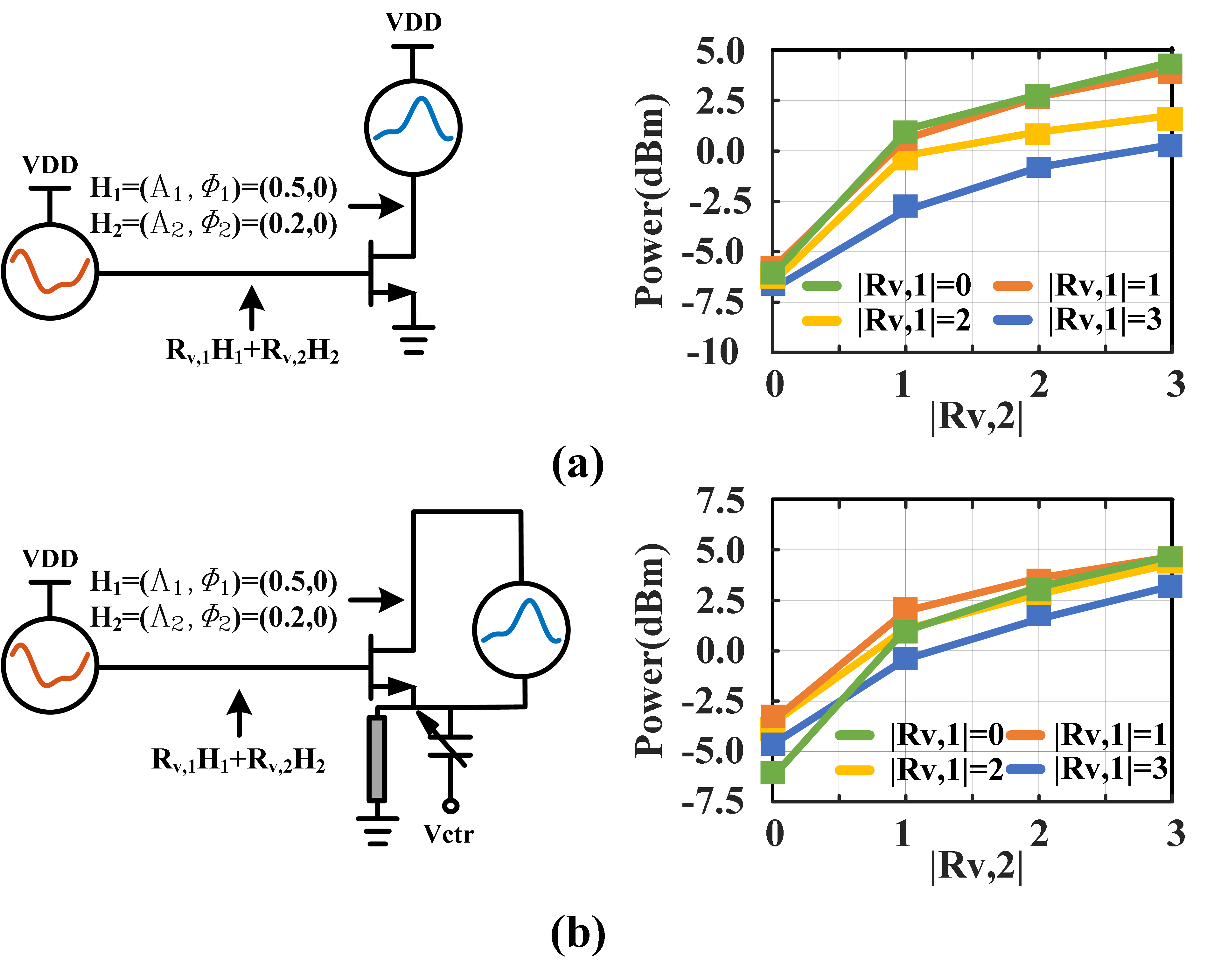}
\vspace{-0.2in}
\caption{(a) schematic of test bench for the simulation of harmonic power as a function of $R_{v,2}$, (b) schematic of test bench for the simulation of harmonic power as a function of $R_{v,2}$ in the presence of source degeneration.} 
\label{Power2}
\end{figure}

 However, according to the Volterra-Weiner  non-linear model of transisotrs in \cite{aghasi20170}, generated second harmonic current is not only a function of second harmonic voltages and instead is a combination of fundamental and harmonic components. 
 Fig. \ref{Power2}(a) depicts effect of $R_{v,2}$ on generated harmonic power demonstrating that maximizing the harmonic power happens for maximum magnitude of $R_{V,2}$ if the relative phase of $V_{gs}^{2f_{0}}$ and  $V_{ds}^{2f_{0}}$ is small. As shown in Fig. \ref{Power2}(a), different harmonic power values for a fixed $R_{v,2}$ happen $R_{v,1}$ obtains various values. More interestingly, in Fig. \ref{Power2} (b), harmonic power for a transistor degenerated with a fundamental resonator at source terminal (similar to this design) is plotted, demonstrating effective increase in harmonic power without impacting the second harmonic voltages (the resonator behaves effectively like a short circuit at the second harmonic). These observations mandate the inclusion of nonlinear terms in (\ref{eq:5}), i.e.,

\begin{equation}
\vec{I}_{d}^{\hspace{0.05in} 2f_{0}}=
    (G_{21,a}^{2f_{0}}.\vec{R_{v,2}}+
    G_{22,t}^{2f_{0}})
        \vec{V}_{ds}^{2f_{0}}+
        M_{12}\vec{V}_{gs}^{f_{0}^2}+
        N_{12}\vec{V}_{ds}^{f_{0}^2}
    \label{eq:6}
\end{equation}

Where, $M_{12}$ and $N_{12}$ coefficients represent the
$2_{nd}$ order nonlinear translation of $V_{gs}^{f0}$, and $V_{ds}^{f0}$ to $\vec{I}_{d}^{\hspace{0.05in} 2f_{0}}$\cite{aghasi20170}. Hence, generated harmonic power should be written as:

\begin{equation}
{P}_{ds}^{\hspace{0.05in} 2f_{0}}=
      \left({G}_{21,a}^{\hspace{0.05in} 2f_{0}} + \vec{R}_{v,2}{G}_{22,t}^{ 2f_{0}}\right)\vec{V}_{ds}^{ 2f_{0}^2} +
      \left({M}_{12}\vec{R}_{v,1} + {N}_{12}\right)\vec{V}_{ds}^{ f_{0}^2}\vec{V}_{ds}^{ 2f_{0}}.
    \label{eq:7}
\end{equation}

According to (\ref{eq:7}) and simulation results in Fig. \ref{Power2}, both $R_{v,1}$ and $R_{v,2}$ impact the magnitude of generated harmonic power. The impact of $R_{v,1}$ and $R_{v,2}$ on the power is not similar with their impact on $\Gamma$ and $\alpha$. More specifically, for the proposed cross-coupled oscillator, larger $R_{v,1}$ and $R_{v,2}$ will lead to reduced PN if they hold a certain phase relationship. However, according to Fig. \ref{Power2}, increasing the magnitude of $R_{v,1}$ will degrade the harmonic power and larger magnitude of $R_{v,2}$ is critical for harmonic power generation. 
It is noteworthy that the nonlinear harmonic modeling can be useful for signal processing in communication systems where the baseband analog processing prior to analog-to-digital converters can become more power efficient by deployment of polynomial generation circuits \cite{aghasi2022capacity,aghasi2022mimo,aghasi2022quantifying}.

\section{Design Procedure}
\label{sec:3}

In this section, we present a new architecture for cross-coupled oscillators that allows PN reduction while maintaining a reasonably high harmonic generation efficiency.
In this work, a super-harmonic VCO is designed where the fundamental frequency is set to 40GHz, employing an NMOS pair with total width of 40$\mu$m ($20\times 2\mu$m) and channel length of 65nm.

\subsection{ISF Assumptions}
\begin{figure}[h!]
\centering
\includegraphics[width=1\textwidth]{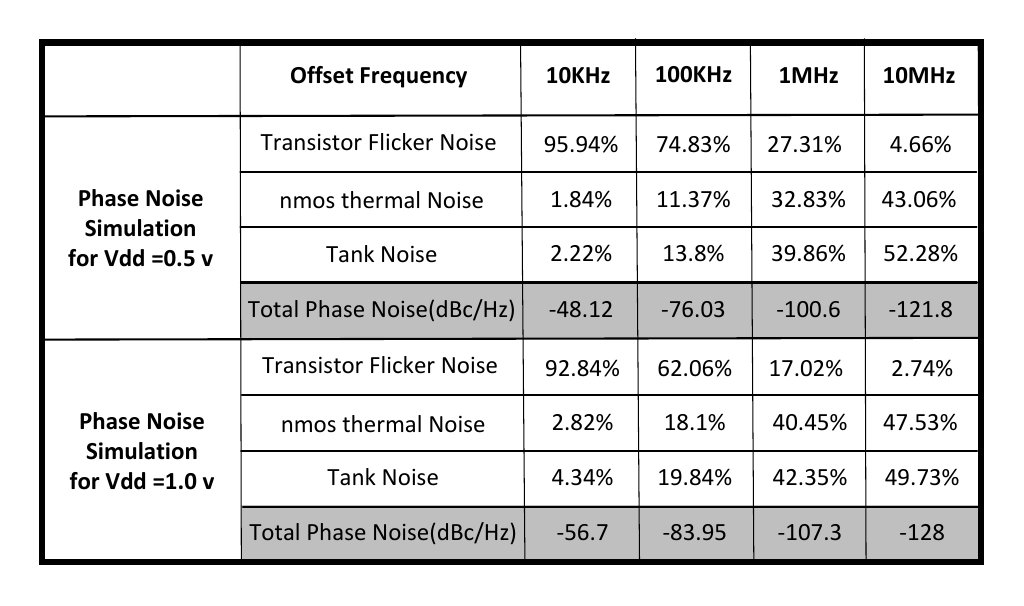}
\vspace{-0.25in}
\caption{Simulated noise contribution for $V_{dd}=0.5$, and $V_{dd}=1.00$.} 
\label{noiseperc}
\end{figure}

The noise contributions from various sources for the proposed VCO based on Cadence Virtuoso simulations, are presented in Figure \ref{noiseperc}. At 100 kHz offset, flicker noise from the transistors is the dominant noise source. However, at higher offset frequencies, the flicker noise contribution decreases significantly, becoming less than 5.0\% at 10MHz offset frequency. According to Fig. \ref{noiseperc}, the total transistor noise, i.e., thermal and flicker noise, are the most significant contributors to the phase noise for offset frequencies up to 1MHz and by approaching 10 MHz offset, their effect becomes less pronounced than the tank noise. Therefore, in the following analysis, we focus on monitoring the variations of time-domain voltage waveforms across the active device terminals as the primary independent variable that impacts both thermal noise and flicker noise.

\subsection{Effect of Feedback Transmission line}

For a transmission line with characteristic impedance of $Z_{0}$, electrical length of $l$ connected to the gate terminal with reflection coefficient of $\tilde{\Gamma}_{G,n}$\footnote{To avoid confusion with ISF symbol, we denote reflection coefficients with $\tilde\Gamma$ in this paper.} at the \textit{n}-th harmonic frequency, the corresponding $R_{v,n}$ can be calculated as:

\begin{equation}
    R_{v,n} = \frac{1 + \tilde{\Gamma}_{G,n}}{e^{j\beta_{n} l} + \tilde{\Gamma}_{G,n} e^{-j\beta_{n} l}}
    \label{f1}
\end{equation}

To derive the desired $l$ that maximizes the magnitude of $R_{v,n}$ to reduce the PN (according to Figs \ref{nmf},\ref{nmf2}) as a function of complex $\tilde{\Gamma}_{G,n}=$  $\tilde{\Gamma}_{G,n,r}+i\tilde{\Gamma}_{G,n,i}$, the magnitude of $R_{v,n}$ can be written as: 
\begin{equation}
    |R_{v,n}| = \frac{\sqrt{(1+\tilde{\Gamma}_{G,n,r})^2 + \tilde{\Gamma}_{G,n,i}^2}}{|e^{j\beta_{n} l} + (\tilde{\Gamma}_{G,n,r} + i\tilde{\Gamma}_{G,n,i}) e^{-j\beta_{n} l}|}
    \label{maxr}
\end{equation}

the maximum of $|R_{v,n}|$ for feasible values of $l$ happens when the nominator multiplied by the partial derivative of the denominator (with respect to $l$) in (\ref{maxr}) becomes zero. The nominator only becomes zero when $\tilde{\Gamma}_{G,n}=-1$, corresponding to a short-circuit, which cannot be the case for a cross coupled oscillator. To set the partial derivative of denominator to zero, we replace $e^{j\beta_{n} l}$ with $cos{\beta_{n} l}+isin{\beta_{n} l}$ in the denominator which results in $l$ to be:  

\begin{equation}
    l=\frac{\lambda_{n}}{4\pi} tan^{-1}{\frac{\tilde{\Gamma}_{G,n,i}}{\tilde{\Gamma}_{G,n,r}}}
    \label{max}
\end{equation}

\begin{figure}[!htb]
\centering
\includegraphics[width=1\textwidth]{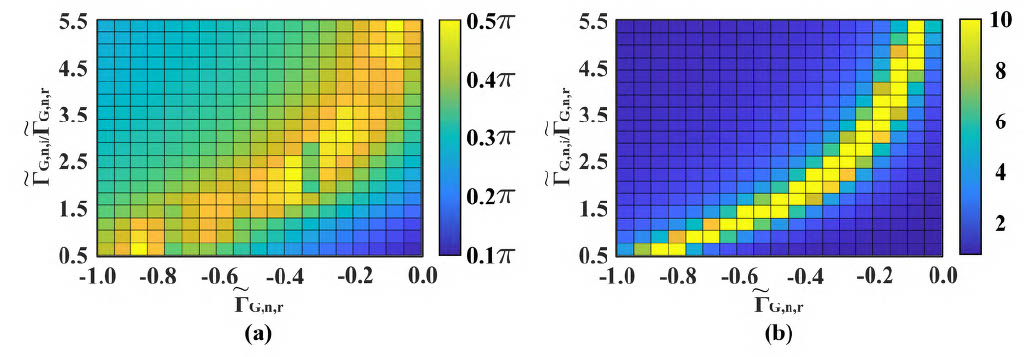}
\vspace{-0.25in}
\caption{(a) Phase of $R_{v,n}$ as a function of $\frac{\tilde{\Gamma}_{G,n,i}}{\tilde{\Gamma}_{G,n,r}}$ and $\tilde{\Gamma}_{G,n,r}$ (b) Magnitude of $R_{v,n}$ as a function of $\frac{\tilde{\Gamma}_{G,n,i}}{\tilde{\Gamma}_{G,n,r}}$ and $\tilde{\Gamma}_{G,n,r}$.}
\label{TLGformula}
\end{figure}

\begin{figure*}[h!]
\centering
\includegraphics[width=1\textwidth]{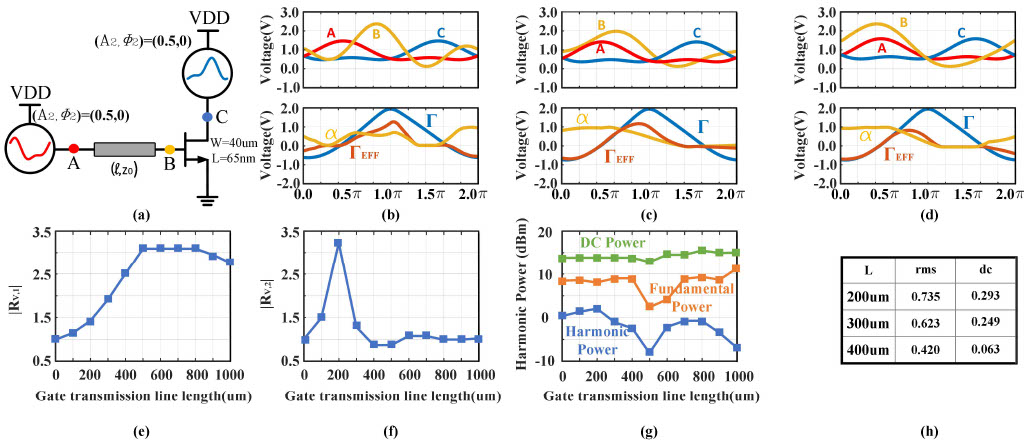}
\vspace{-0.15in}
\caption{a) Simulation setup for feedback transmission line. $\alpha$, $\Gamma$, and $\Gamma_{eff}$ in addition to terminal voltage waveform for b) $l=200\mu m$, c) $l=300\mu m$ d) $l=400\mu m$ e) fundamental drain-to-gate voltage gain ($R_{v,1}$) f) harmonic drain-to-gate voltage gain ($R_{v,2}$) g) DC power consumption, generated fundamental and harmonic power (dBm) h) dc, and rms value for corresponding $\Gamma_{eff}$waveforms.}
\label{tlgselection}
\end{figure*}

Fig. \ref{TLGformula} shows magnitude and phase of $R_{v,n}$ as a function of $\tilde{\Gamma}_{G,n,r}$ and $\frac{\tilde{\Gamma}_{G,n,i}}{\tilde{\Gamma}_{G,n,r}}$. It is evident from Fig. \ref{TLGformula}(b) that a small subset of values for $\tilde{\Gamma}_{G,n,r}$ and $\frac{\tilde{\Gamma}_{G,n,i}}{\tilde{\Gamma}_{G,n,r}}$ leads to a large magnitude of $R_{v,n}$.
To examine the range of values for $\tilde{\Gamma}_{G,n,r}$ and $\frac{\tilde{\Gamma}_{G,n,i}}{\tilde{\Gamma}_{G,n,r}}$, we fix the transistor size to an NMOS with size of (40$\mu$m/65nm). For values of $V_{dd}$ between 0.5 to 1V, $\tilde{\Gamma}_{G,1,r}$, and $\tilde{\Gamma}_{G,1,i}$ are calculated by Cadence simulations which exhibit relatively insignificant variations of $\pm$30m, and $\pm$40m around the mean values of -157m and -656.7m, respectively as the large signal gate terminal impedance is calculated as 18.61-44.7i$\Omega$ corresponding to a 78fF capacitor in series with a 18.61 $\Omega$ resistor. By selecting the mean values for $\tilde{\Gamma}_{G,1,r}$ and $\tilde{\Gamma}_{G,1,i}$, for fundamental oscillation frequency of 40 GHz, the desired $l$ = 430$\mu$m which maximizes the magnitude of $|R_{v,1}| = 3.2$ with a phase of $32^{\circ}$, results in $R_{v,2} = 1.2 \angle 65^{\circ}$. According to Fig. \ref{Power2} this value of $R_{v,2}$ does not lead to a high harmonic power. This is another example of non-identical PN and harmonic power enhancement conditions. It is noteworthy that one can also boost the generated power of the VCO by replacing  $R_{v,n}$ from (\ref{f1}) in (\ref{eq:6}) and (\ref{eq:7}). The optimum points of power are not the focus of this work and it suffices to show that the selected $TL_G$  allows efficient high-power second harmonic generation. We will discuss the power optimization in a future article.

\begin{figure}[!htb]
\vspace{-0.15in}
\centering
\includegraphics[width=1\textwidth]{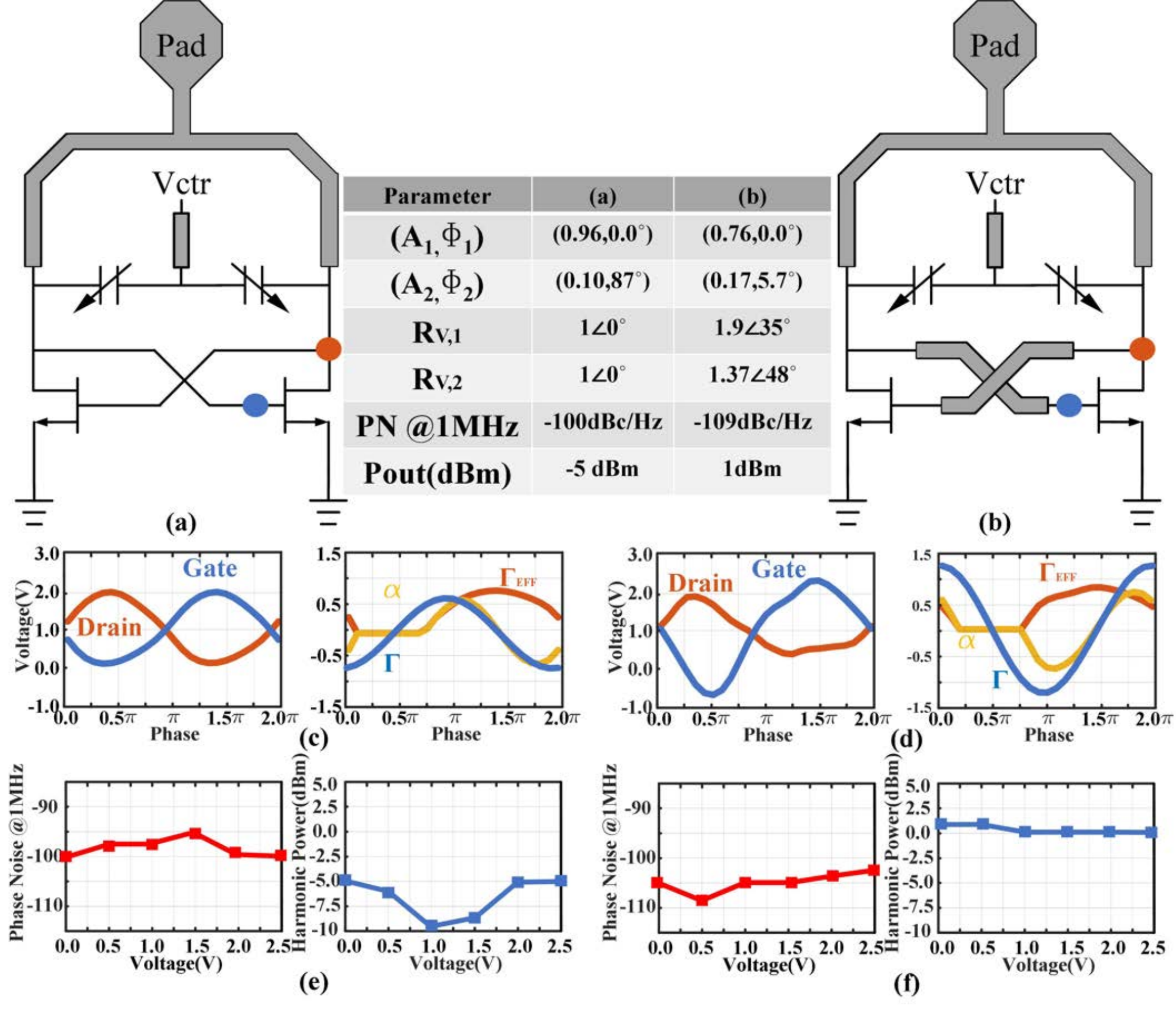}
\vspace{-0.25in}
\caption{(a) schematic of conventional cross-coupled VCO (b) with feedback transmission line (c)-(d) voltage waveforms at different terminals with corresponding $\alpha$,$\Gamma$,$\Gamma_{eff}$ and (e)-(f) PN and harmonic power for $V_{dd}=1.0$.} 
\label{gate}
\end{figure}

Fig. \ref{tlgselection} (a) shows the simulation setup demonstrating effect of $l$ on $\alpha$ and hence, $\Gamma_{eff}$. $V_{ds}$ is derived from Section \ref{sec:2} such that  $\Gamma_{rms}$ can become as small as possible. In Figs \ref{tlgselection} (b)-(d) the terminal voltage waveforms as well as corresponding $\Gamma$, $\alpha$, and $\Gamma_{eff}$ waveforms for three various $TL_G$ length values (200$\mu$m, 300$\mu$m and 400$\mu$m) are illustrated. For $l=$400$\mu$m, the dc and rms values of $\Gamma_{eff}$ reach the lowest values, which is very close to the value obtained from (\ref{max}). Fig. \ref{tlgselection} (e-f) depict $|R_{v,1}|$ and $|R_{v,2}|$ as a function of $l$. The transmission line length that yields the highest $|R_{v,2}|$, resulting in highest harmonic power generation, happens at a length of $200\mu$m as it is shown in Fig .\ref{tlgselection} (g). It is noteworthy that the generated fundamental power does not closely follow $R_{v,1}|$ in Fig. \ref{tlgselection} since the phase relationship of the drain voltage and current also change by varying $l$.  In Fig. \ref{tlgselection} (h), $\Gamma_{dc}$, and $\Gamma_{rms}$ for the three cases in Fig. \ref{tlgselection} (b-d) is shown. According to the simulation results in Fig. 
\ref{tlgselection}, reducing the $\Gamma_{eff}$ requires a transmission line length other than $200\mu$m, in this case $l=400\mu$m. This creates a trade-off between reducing PN and generating harmonic power.

To  understand the effect of  $TL_G$ on phase noise improvement, 
a simulation was conducted by comparing two cross-coupled oscillators with and without $TL_G$, as shown in Fig. \ref{gate}. To ensure fairness, power consumption and operation frequency of the two designs are sufficiently close. The corresponding terminal voltage waveforms and $\Gamma$, $\alpha$, and $\Gamma_{eff}$ for each case are shown in Fig. \ref{gate}. The dc and rms values of $\Gamma_{eff}$ for the VCO with $TL_G$ length of 400 $\mu$m compared to the design without $TL_G$ are reduced by more than 45\% and 20\%, respectively. The PN values at 1 MHz offset frequency for the two VCOs confirm the PN superiority of the VCO with $TL_G$. 
Moreover, in Fig. \ref{gate}, the generated harmonic power at the drain terminals for both VCOs are compared and it is evident that addition of $TL_G$ boosts the power by more than 7 dB across the tuning range.

\subsection{Effect of Source Resonator}

\begin{figure}[!htb]
\centering
\vspace{-0.15in}
\includegraphics[width=1\textwidth]{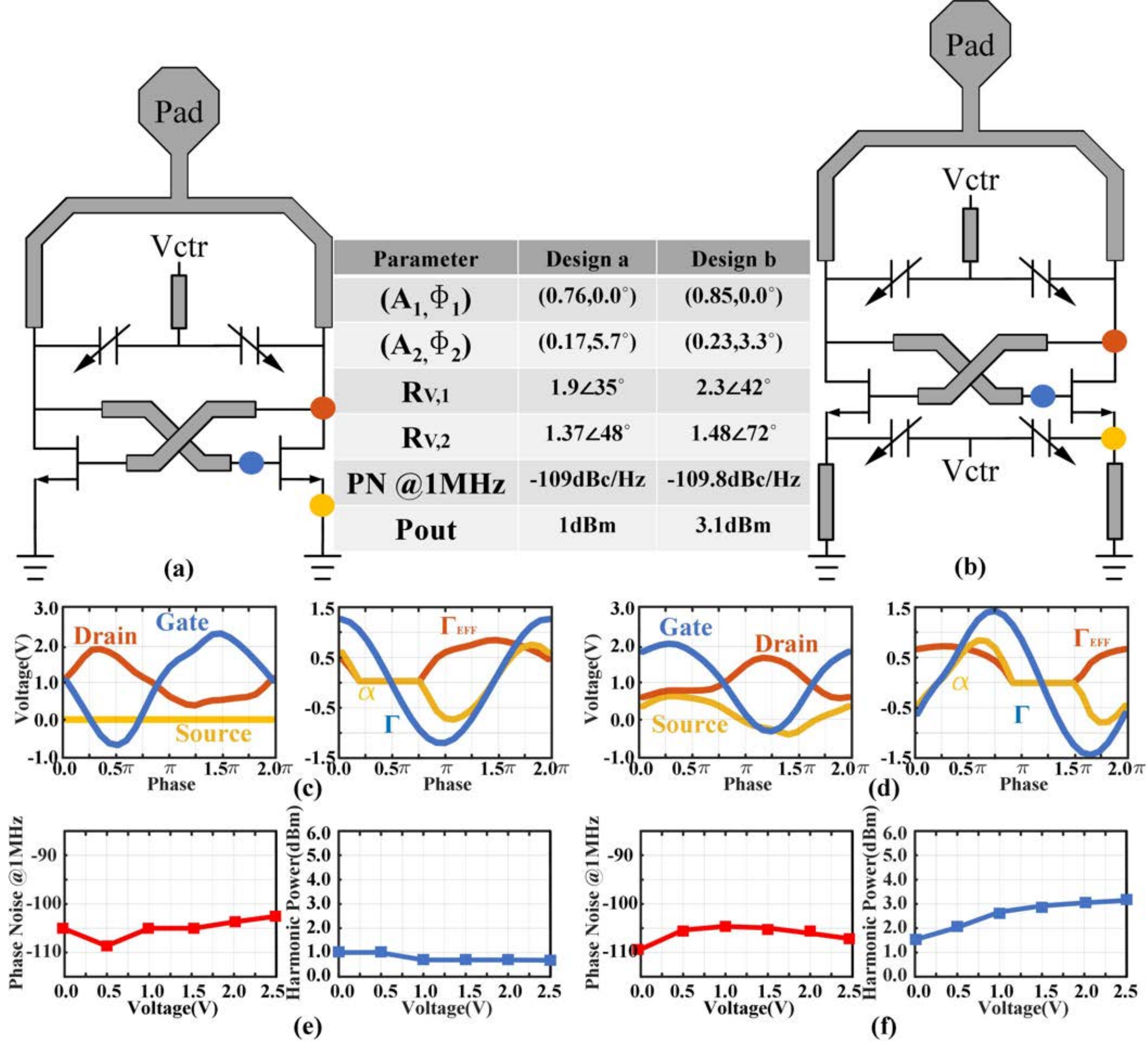}

\vspace{-0.15in}
\caption{Schematic of two cross-coupled VCOs (a) without and (b) with source degeneration, (c,d) voltage waveforms at different terminals with corresponding $\alpha$,$\Gamma$,$\Gamma_{eff}$ and (e)-(f) PN and harmonic power for $V_{dd}=1.0$.} 
\label{source}
\end{figure}

Previous subsection was focused on the impact of $TL_G$ on PN and harmonic power generation. Prior art of mm-wave cross-coupled VCOs have assumed a direct connection of the transistors' source terminals  to ground to satisfy the stringent Barkhausen criterion conditions with respect to $G_m$ at the higher frequencies \cite{mm1,mm2,mm3}. However, as highlighted in \cite{omeedsource}, the implementation of a source resonator to increase the deep triode region during the period can enhance the generated harmonic power. Fig. \ref{source} depicts a comparison betweentwo VCOs with $TL_G$ feedback transmission line, one with grounded source and one with a fundamental resonator at the source terminal. At vicinity of $f_{0}$, nmos Y parameters and subsequently $\Gamma$ at the gate terminal change by the source generation.
 Fig .\ref{source} (e)-(f) depicts the PN  at 1MHz offset frequency and generated harmonic power for both designs. Despite almost similar PN performance, the VCO with degeneration can generate more than 1mW (0dBm) of harmonic power with a harmonic power efficiency of 5\%. This represents a 30\% improvement compared to the VCO without source degeneration.

\vspace{-0.25in}
\subsection{Matching Network Design and Layout Considerations}

\begin{figure}[h!]
\centering
\includegraphics[width=1\textwidth]{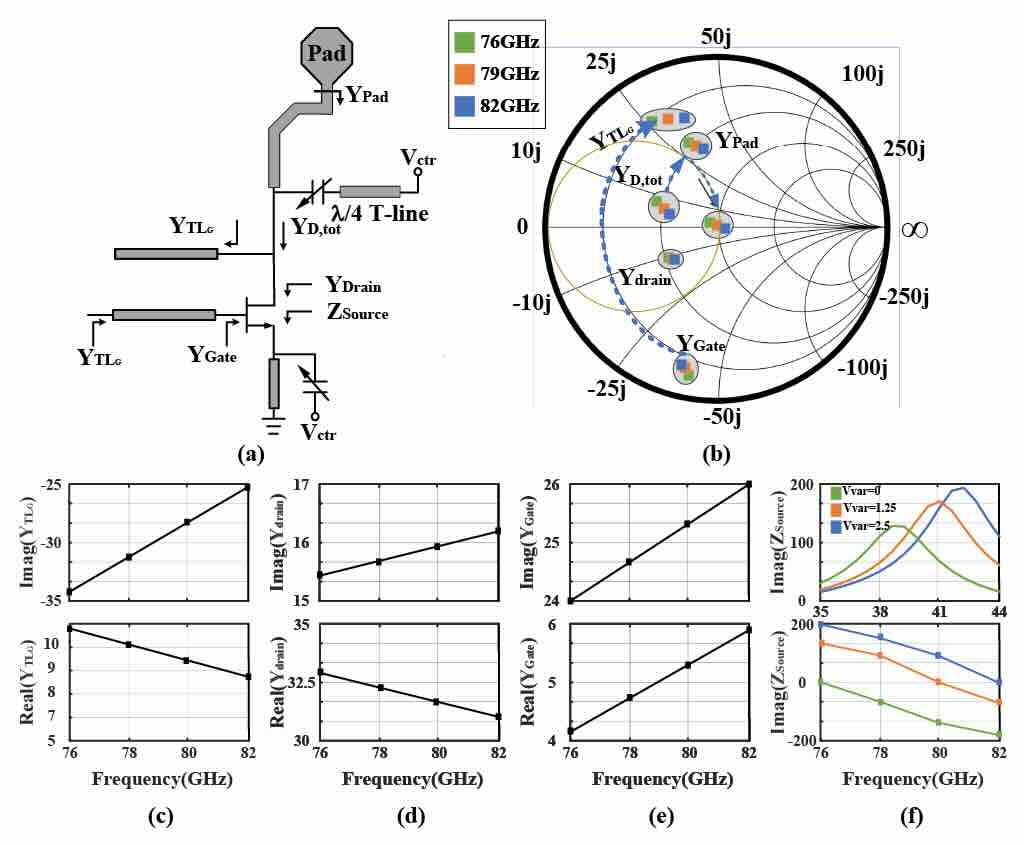}
\vspace{-0.25in}
\caption{(a) schematic of matching network structure, (b) smith chart with equivalent impedance transformation.  Impedance in vicinity of fundamental/second harmonic operation at (c) transformed gate impedance, (d) drain, (e) gate and (f) source.} 
\label{matching}
\end{figure}
Design of matching network and impedance transformation is described in Fig. \ref{matching}. Fig. \ref{matching} (a) shows a half-circuit schematic of proposed VCO including the matching network. The corresponding impedance transformation is depicted in the smith chart of Fig. \ref{matching} (b). In Fig. \ref{matching} (c)-(e) each terminal impedance at vicinity of second harmonic frequency is shown.
In order to increase harmonic power extraction, matching network should transform the $Z_{drain}$ to 50 $\Omega$. By incorporating an RF choke to avoid harmonic power leakage to ground, the drain varactor capacitive impedance will not been seen in equivalent half-circuit schematic. The observed admittance at drain $Y_{drain}$ combined with transformed admittance at gate $Y_{TLG}$ represent the total admittance at the drain and follow $TL_D$ and a 100 $\mu$m t-line before they reach the pad. The impedance transformation from $Z_{drain}$ to 50$\Omega$ is shown in Fig. \ref{matching} (b). In Fig. \ref{matching} (f) source terminal impedance is shown which demonstrates a high impedance at vicinity of $f_{0}$ while showing a very low impedance at $2f_{0}$.

\begin{figure}[t!]
\centering
\includegraphics[width=1\textwidth]{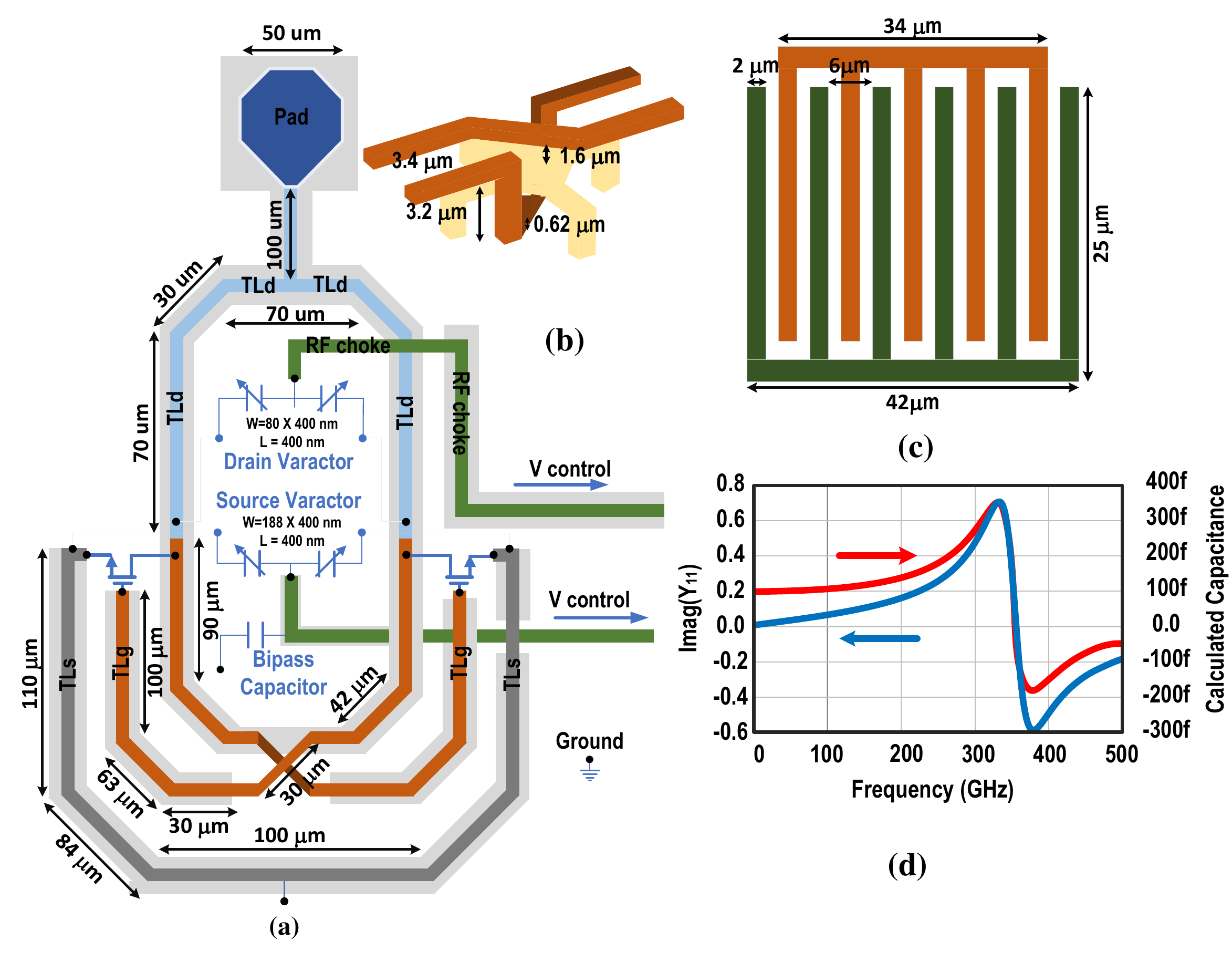}
\vspace{-0.20in}
\caption{a) Detailed figure of proposed harmonic VCO's layout b) metal layer in the cross-section layer c) fringing MOM capacitor d) Imaginary of proposed MOM capacitor Y parameter (blue), and equivalent capacitor as a function of frequency.(red)}
\label{layoutdetail}
\end{figure}

In Fig. \ref{layoutdetail}(a), the layout configuration of the proposed VCO is presented. A pair of NMOS transistors with a size of, W=20$\times$2$\mu$m, and L=65nm are used. 
According to Figs. \ref{tlgselection},\ref{gate}, \ref{source}, a 370$\mu$m transmission line as $TL_G$ is incorporated.  
We utilize grounded coplanar waveguide (GCPW) configuration for transmission lines to obtain a high isolation \cite{GCPW}. 
The 240$\mu$m transmission line combined with the varactor sized (w=188$\times$400nm, and length of 400nm) realize the tunable fundamental resonator at the source and the 135$\mu$m transmission line $TL_D$ in combination with a varactor sized (w=80$\times$400nm, and length of 400nm) create the fundamental resonator at the drain. 
In contrast to the varactors at the drain, the ones at the source are directly terminated to an ac ground. As shown in Fig. \ref{layoutdetail}(b), to minimize the coupling among the two $TL_G$ lines at the cross-section point, we keep the signal of one line at $M_9$ and transition the signal of the other line to a sandwich of $M_4$ and $M_5$ (to make it thicker) while a shielding $M_7$ layer lies between the signal of the two lines. By this design the cross-coupling of the lines is below -15 dB. 
The PDK metal-insulator-metal (MIM) capacitors cannot satisfy the requirement on self-resonance frequency (SRF)$\geq$ 100 GHz; hence, we have designed a fringe capacitor to ensure that the varactors at the source are terminated to an AC ground. The implemented capacitor is a sandwich from $M_4$ to $M_9$ which exhibits more than 100fF capacitance per unit (2 units connected in parallel to get more than 230 fF). The SRF of the capacitor shown in Fig. \ref{layoutdetail}(d) is above 250 GHz.

\clearpage
\section{MEASUREMENT RESULTS}
\label{sec:4}
\begin{figure}[h]
\centering
\includegraphics[width=1\textwidth]{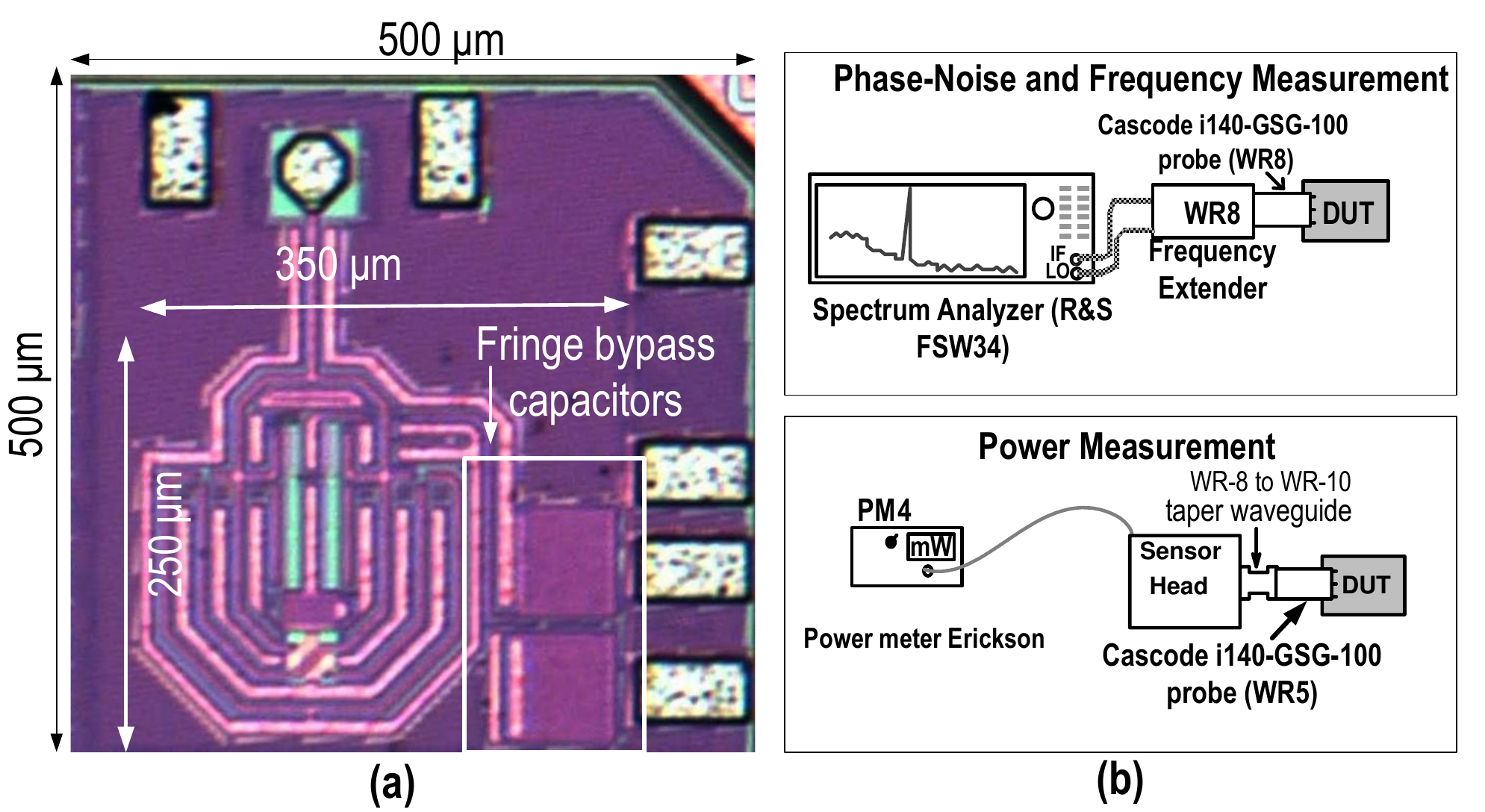}
\vspace{-0.25in}
\caption{ a) Die photograph of the chip, b) measurement setup for power extraction and PN measurement.}
\label{die}
\end{figure}

The proposed 76-82 GHz compact second-harmonic VCO is implemented and fabricated in 65nm bulk CMOS technology.
The die photograph is shown in Fig. \ref{die}(a), occupying a small area of (0.25$\times$0.35 $mm^{2}$ for core chip, and 0.5$\times$0.5 $mm^{2}$ with inclusion of pads), confirming its compact design. The setup for measuring PN, tuning range, and output harmonic power are also shown in Figure \ref{die}(b). 

\begin{figure}[h]
\centering
\includegraphics[width=0.75\textwidth,height=15cm]{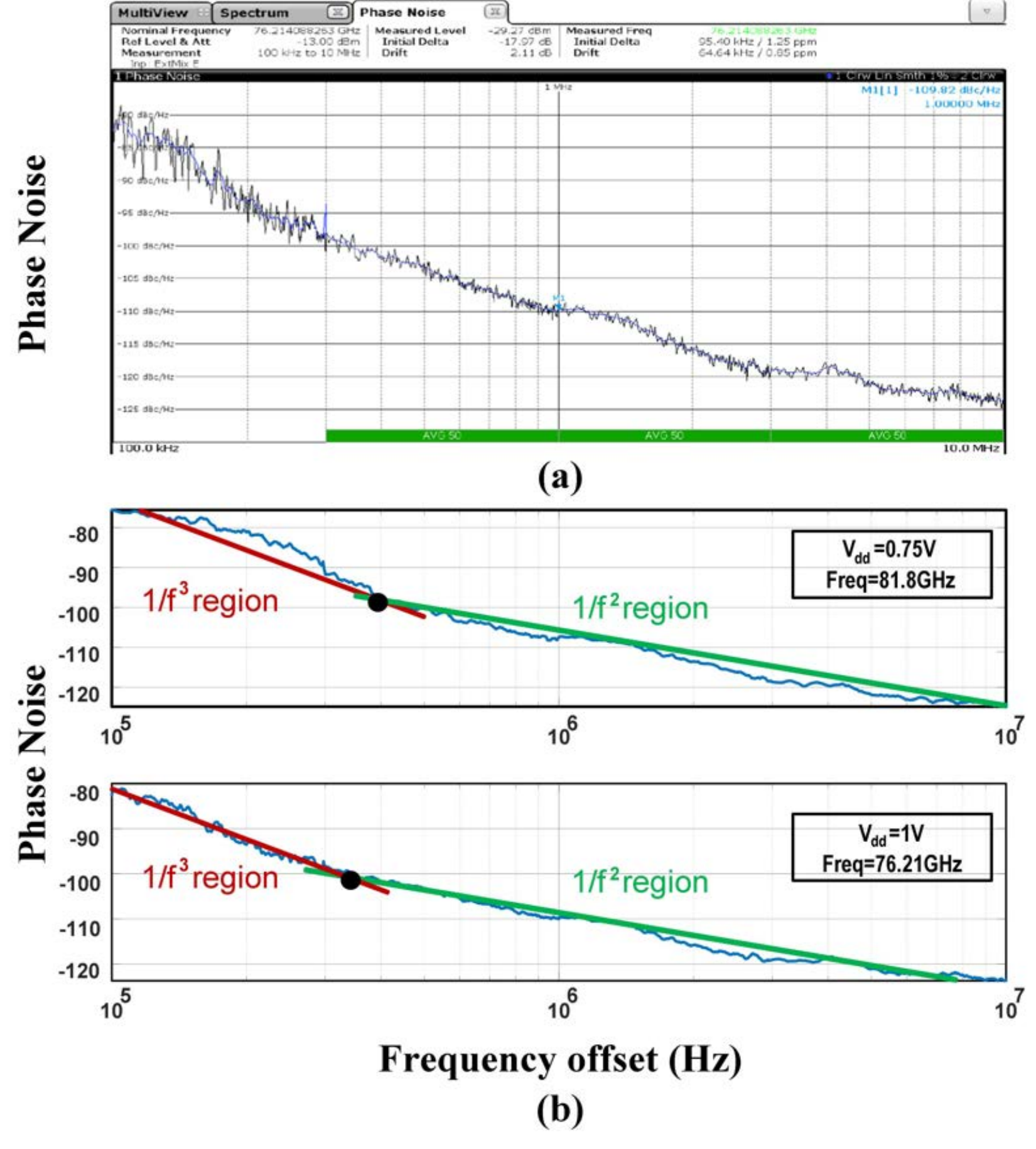}
\vspace{-0.15in}
\caption{a) Measured PN profile at 76 GHz, b) comparison of measured PN versus offset frequencies at two different frequencies of 81.8GHz (top) and 76.21 GHz (bottom).}
\label{flicker}
\end{figure}

\begin{figure*}[t!]
\centering
\includegraphics[width=1\textwidth]{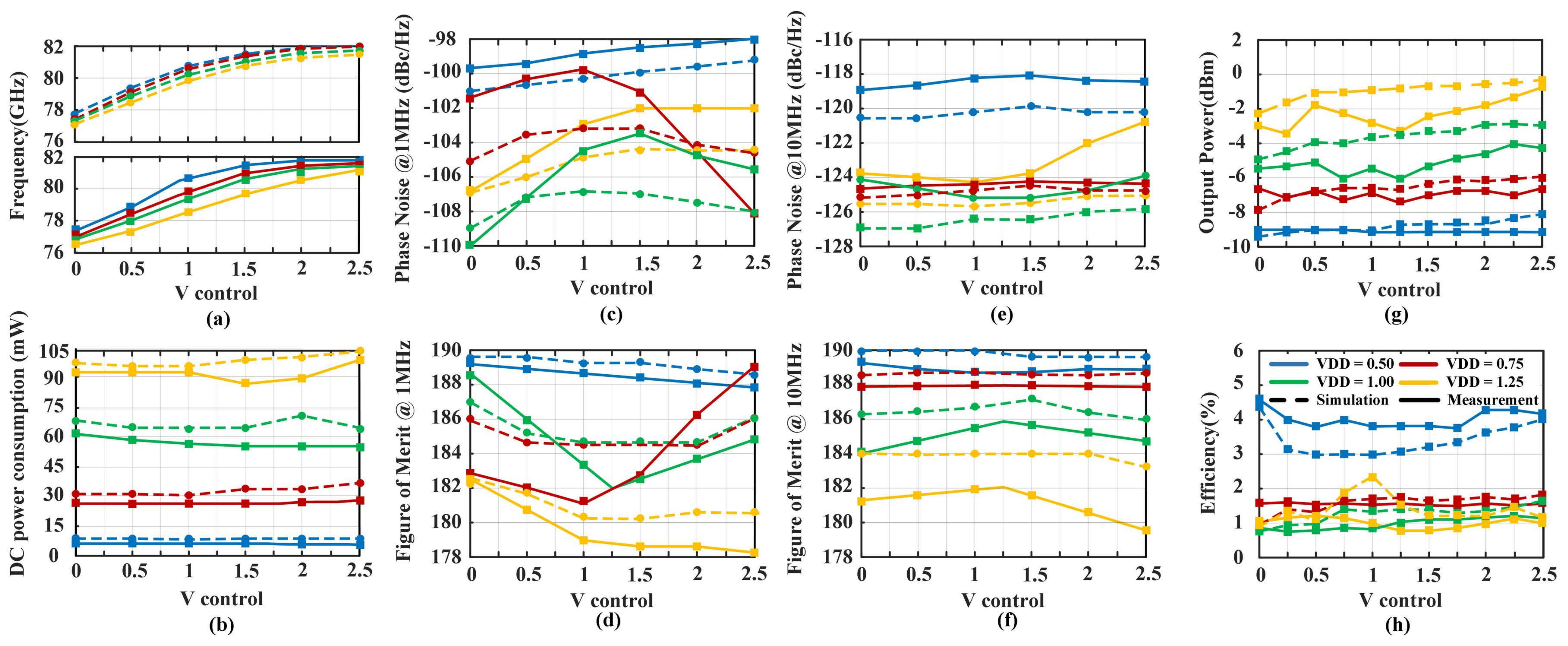}
\vspace{-0.20in}
\caption{ Measured and simulation result of (a) tuning range, (b) DC power consumption, (c,e) PN, (d,f) FoM at offset frequencies of  10MHz,1MHz, and (g,h) harmonic power/efficiency.}
\label{results}
\end{figure*}

A cascode i140-GSG-100 probe (WR8) and a WR8 frequency extender is used to guide the RF signal to a R$\&$S FSW34 spectrum analyzer. As shown in Fig. \ref{flicker}, the estimated $1/f^3$ PN corner frequency is $\approx$ 350 and $\approx$ 400 KHz at 76.21 and 81.8 GHz. respectively. Mode switching technique used in \cite{corner1}, effectively lower the corner frequency to 680KHz, and 710KHz at 43.43GHz, and 46.03 respectively. Also, in \cite{corner2}, circular triple-coupled transformer is used to lower the corner frequency to 1.3MHz, and 800KHz at 60.4GHz and 52.5GHz respectively. Compared to these works and other prior art, this VCO obtains the lowest corner frequency reported for a harmonic VCO at W frequency band.  Fig. \ref{results} depicts the measurement results as well as simulation results for phase noise and output power. As shown in Fig. \ref{results}(a), when both source and drain varactors are tuned by an identical control voltage $V_{tune}$, the proposed VCO demonstrates a $[76.14, 81.83]$ GHz oscillation equivalent to slightly higher than 7$\%$ tuning range. It is noteworthy that for various power supply scenarios, the effective voltage across the varactor will move within different regions of the \textit{C}-\textit{V} characteristic and leads to slightly nonlinear profile for smaller $V_{dd}$ values and at the same time moves the frequency to higher values due to reduction of the effective capacitance.

 Figs. \ref{results} (b) shows the measured DC power consumption over tuning range for various values of $V_{dd}$, demonstrating a nonlinear growth of power consumption with respect to $V_{dd}$. Moreover, measured results for PN and FoM at different frequency offsets, (1MHz, and 10MHz),  over tuning range for various values of $V_{dd}$ are also shown in Fig. \ref{results} (c,e) and (d,f) which outperform the prior art (as shown in Table 1) by more than 5 dB. A minimum PN of -109.8 dBc/Hz at 1MHz, and -125 dBc/Hz at 10MHz offset frequencies have been achieved. 
Higher offset PN values deviate from the 20 dB/dec trend due to elevation of noise floor in the spectrum analyzer by using harmonic mixers across an extended bandwidth. A 188.75 dBc/Hz FoM is reported for $V_{dd}$=0.5 V with PN of -119 dBc/Hz at 10MHz offset. At the same $V_{dd}$, a PN of -108.3 dBc/and FoM of 189.33dBc/Hz  is reported at 1MHz offset frequency. Measured output RF power with corresponding DC-to-RF efficiency for different values of $V_{dd}$ are displayed in Fig. \ref{results} (g)-(h). With a low power consumption at $V_{dd}=0.5$, the VCO proves to be an efficient power source with a peak DC-to-RF efficiency of 4.6\%. As the $V_{dd}$ value increases, the output power of the VCO also increases, reaching its peak value of -0.6 dBm at $V_{dd}=$ 1.25V. The measurement results are summarized in Table I, where a comparison with prior art is presented. The proposed VCO shows a significant improvement in the FoM, exceeding 4 dBc/Hz, compared to other state-of-the-art works. Despite the close match between simulation and measurement results, the slight mismatch between them in Fig. \ref{results} is attributed to the inevitable process variation (PVT).

\begin{table*}
\centering
{

\caption{\small Comparison of VCO with Prior Arts \normalsize{}}
\label{tb:JTHz_COMP_OSC2}
\resizebox{1\textwidth}{!}{  
\begin{tabular}{|c|c|c|c|c|c|c|c|c|c|c|c|c|c|c|c|}

\hline
{Reference} &   {Technology}  &   {$f_{0}$}   &   {Tuning }   &     {$P_{DC}$}   &   {RF Power}   &    \multicolumn{3}{|c|}{Phase Noise [dBc/Hz]}  &  \multicolumn{3}{|c|}{FoM [dBc/Hz]}  &  {1/f corner}\\
 &   {CMOS}  &   {[GHz]}   &   {[\%]}   &     {[mW]}   &   {[dBm]}   &    100KHz & 1MHz &10MHz  &  100KHz & 1MHz &10MHz   &  {[KHz]}\\
\hline
\cite{REF1} &  65nm   &  100.65  &  5.2  &  12 \textasciitilde\ 21  &  -5 \textasciitilde\ -2  &  N/A  &  N/A  &  -112.1  &  N/A  &  N/A  &  178.6  &  N/A\\ 
\hline 
\cite{REF2} &  65nm   &  85.75  &  8.3  &  8.5  &  1  &  -69.1  &  -91.8  &  -111.6  &  178.8  &  181.5  &  181.3  &  N/A\\ 
\hline
\cite{REF3} &  28nm   &  78  &  12  &  20.7  &  -4  &  N/A  &  -100  &  -121  &  N/A  &  182  &  182  &  N/A\\ 
\hline
\cite{REF4} &  28nm   &  73.75  &  9.8  &  35.6  &  N/A  &  N/A  &  -93.5  &  -117.7  &  N/A  &  176.3  &  179.4  &  2000\\ 
\hline
\cite{REF4} &  65nm   &  94.8  &  27  &  12  &  N/A  &  N/A  &  -90  &  -110  &  N/A  &  177.5  &  177.5  &  N/A\\ 
\hline
\cite{corner1} &  65nm   &  46.75  &  16.5  &  20.9  &  N/A  &  N/A & -106.1  &  -126.1    &  N/A  &  186.6  &  186.6  &  680\\ 
\hline
\cite{corner2} &  65nm   &  56.4  &  14.2  &  22.5  &  0  &  -75.2  &  -104.7  &  -124.7  &  N/A  &  177.8  &  186.5  &  800\\ 
\hline
This work ($V_{dd}=0.5$)&  65nm   &  79  &  7.2  &  3.95  &  -9.3  &  N/A  &  \textbf{-99.2}  &  \textbf{-119.1}  &  N/A  &  \textbf{189.3}  &  \textbf{188.8}  &  \textbf{350}\\ 

\hline
This work ($V_{dd}=1.0$)&  65nm   &  79  &  7.2  &  102.70  &  \textbf{-4(-0.6)}  &  \textbf{-75.4}  &  \textbf{-109.8}  &  \textbf{-125.4}  &  \textbf{175.8}  &  \textbf{188.3}  &  \textbf{185.8}  &  \textbf{400}\\ 

\hline

\end{tabular}
}
\tiny $^\dagger$ Maximum output power is achieved for $V_{dd}=1.25$V \\
\tiny $^\dagger$ FoM = $|\text{PN}| + 20\log_{10}\left(\frac{f_{0}}{\Delta f}\right) - 10\log_{10}\left(\frac{P_{DC}}{1\text{mW}}\right)$ \\}
\end{table*}

\section{Conclusion}
\label{sec:5}
A new design methodology by leveraging the harmonic components to improve the PN and power efficiency of mm-wave VCOs was presented in this work. By synthesizing the voltage waveforms across the transistor terminals, the impact of fundamental and harmonic voltage components that lead to manipulation of noise impulse sensitivity function and power efficiency were characterized.  A compact varactor-based super-harmonic oscillator with 7\% tuning range (76.14 to 81.83 GHz) has been presented, analyzed, and fabricated in a 65nm CMOS technology. The proposed compact VCO achieves a peak FoM of 189.3 dBc/Hz, and minimum PN of -109.8 dBc/Hz at 1 MHz offset frequency. Moreover, the VCO maintains a reasonable DC-to-RF efficiency of 4.6\% and generates -0.6 dBm of peak harmonic power. The design methodology and circuit configuration presented in this work can pave the path towards more reliable design of next generation mm-wave radars, communication links, and imaging systems.


\clearpage
\phantomsection

\bibliographystyle{unsrt}
\bibliography{thesis}

\captionsetup[figure]{list=no}
\captionsetup[table]{list=no}

\begin{appendices}

\end{appendices}

\end{document}